\documentclass[fleqn,usenatbib]{mnras}

\usepackage{newtxtext,newtxmath}

\usepackage[T1]{fontenc}

\DeclareRobustCommand{\VAN}[3]{#2}
\let\VANthebibliography\thebibliography
\def\thebibliography{\DeclareRobustCommand{\VAN}[3]{##3}\VANthebibliography}

\usepackage{graphicx}	
\usepackage{amsmath}	

\usepackage{color}
 \usepackage{ulem}

\title[end-to-end simulation for KN emission from BH-NS merger]{Three dimensional end-to-end simulation for kilonova emission from a black-hole neutron-star merger} 
\author[K. Kawaguchi et al.]{
Kyohei Kawaguchi,$^{1,2,3}$\thanks{E-mail: kyohei.kawaguchi@aei.mpg.de}
Nanae Domoto,$^{4}$
Sho Fujibayashi,$^{5,4,1}$
Hamid Hamidani,$^{4}$
Kota Hayashi,$^{1,3}$
\newauthor
Masaru Shibata,$^{1,3}$
Masaomi Tanaka,$^{4,6}$
and Shinya Wanajo$^{4,1}$
\\
$^{1}$Max Planck Institute for Gravitational Physics (Albert Einstein Institute), Am M\"{u}hlenberg 1, Potsdam-Golm, 14476, Germany\\
$^{2}$Institute for Cosmic Ray Research, The University of Tokyo, 5-1-5 Kashiwanoha, Kashiwa, Chiba 277-8582, Japan\\
$^{3}$Center for Gravitational Physics and Quantum Information, Yukawa Institute for Theoretical Physics, Kyoto University, Kyoto, 606-8502, Japan\\
$^{4}$Astronomical Institute, Tohoku University, Aoba, Sendai 980-8578, Japan\\
$^{5}$Frontier Research Institute for Interdisciplinary Sciences, Tohoku University, Sendai 980-8578, Japan\\
$^{6}$Division for the Establishment of Frontier Sciences, Organization for Advanced Studies, Tohoku University, Sendai 980-8577, Japan
}

\date{Accepted XXX. Received YYY; in original form ZZZ}

\pubyear{20XX}

\begin{document}
\label{firstpage}
\pagerange{\pageref{firstpage}--\pageref{lastpage}}
\maketitle

\begin{abstract}
We study long-term evolution of the matter ejected in a black-hole neutron-star (BH-NS) merger employing the results of a long-term numerical-relativity simulation and nucleosynthesis calculation, in which both dynamical and post-merger ejecta formation is consistently followed. In particular, we employ the results for the merger of a $1.35\,M_\odot$ NS and a $5.4\,M_\odot$ BH with the dimensionless spin of 0.75. We confirm the finding in the previous studies that thermal pressure induced by radioactive heating in the ejecta significantly modifies the morphology of the ejecta. We then compute the kilonova (KN) light curves employing the ejecta profile obtained by the long-term evolution. We find that our present BH-NS model results in a KN light curve that is fainter yet more enduring than that observed in AT2017gfo. This is due to the fact that the emission is primarily powered by the lanthanide-rich dynamical ejecta, in which a long photon diffusion time scale is realized by the large mass and high opacity. While the peak brightness of the KN emission in both the optical and near-infrared bands is fainter than or comparable to those of binary NS models, the time-scale maintaining the peak brightness is much longer in the near-infrared band for the BH-NS KN model. Our result indicates that a BH-NS merger with massive ejecta can observationally be identified by the long lasting ($>$two weeks) near-infrared emission. 
\end{abstract}




\section{Introduction}\label{sec:intro}
Neutron star (NS) mergers are known to be among the most promising targets of the ground-based gravitational-wave (GW) detectors (LIGO:~\citealt{TheLIGOScientific:2014jea}, Virgo:~\citealt{TheVirgo:2014hva}, KAGRA:~\citealt{Kuroda:2010zzb}) as well as one of the most important sources of high-energy astrophysical transients, such as gamma-ray bursts~\citep[GRB, ][]{1991AcA....41..257P,Nakar:2007yr,Berger:2013jza,LIGOScientific:2017zic}, kilonovae~\citep[KNe, ][]{Li:1998bw,Kulkarni:2005jw,Metzger:2010sy,Kasen:2013xka,Tanaka:2013ana}, jet heated cocoons~\citep{Nakar:2016cih,Hamidani:2022cyj,Hamidani:2022hvl}, and synchrotron flares~\citep{Nakar2011Natur,Hotokezaka:2015eja,Hotokezaka:2018gmo,Margalit2020MNRAS}. NS mergers are also considered to be important production sites of elements heavier than iron in the universe~\citep{Lattimer:1974slx,Eichler:1989ve,Freiburghaus1999a,Cowan:2019pkx}. The first detection of GWs from a binary neutron star (BNS) merger (GW170817;~\citealt{TheLIGOScientific:2017qsa}) and its multi-wavelength electromagnetic (EM) counterparts ~\citep{GBM:2017lvd} demonstrated that those simultaneous observations will provide a valuable opportunity to extend our knowledge of fundamental physics in the extreme (strongly self-gravitating, high-density, and high-temperature) environments.

Among NS mergers, the mergers of black-hole neutron-star (BH-NS) binaries can provide us with interesting insights that are different from BNS mergers. While the mass ratios of the compact stars in BNS binaries are expected to close to unity, BH-NS binaries can be more asymmetric in the mass ratio, and hence, will provide valuable opportunity to study higher-order GW multipole moments~\citep{LIGOScientific:2021qlt}. Also, if the NS is tidally disrupted before reaching the innermost circular orbit of the BH, an applicable amount of NS matter can remain outside the remnant BH and be ejected from the system. Such ejecta formed during the NS tidal disruption as well as the matter subsequently ejected during the evolution of the remnant BH-tours system will be the source of various EM counterparts to the GW event. In addition, since BH-NS mergers can potentially produce a large amount of very low ($\lesssim 0.1$) electron fraction ($Y_e$) ejecta, the nucleosynthetic abundances can be different to those in the case of BNS mergers. In fact, it has been pointed out that BH-NS mergers can provide an explanation to the observed elemental abundances of a subclass of $r$-process-enhanced stars, so-called "actinide-boosted" stars~\citep{Wanajo:2022jgw}.

To extract the physical information from the observation of EM counterparts, accurate modeling of the light curves and spectra consistent with the source properties are crucial. Since the detection of GW170817, light curve modeling of EM counterparts, particularly, for KNe has been significantly developed in this decade. In particular, the studies by employing numerical-simulation-based/motivated ejecta profiles and by performing radiative transfer (RT) simulations with realistic heating rates and/or detailed opacity tables enable us to directly connect the properties of the progenitor binary to the observables~\citep[e.g.,][]{Kasen:2013xka,Kasen:2014toa,Barnes:2016umi,Wollaeger:2017ahm,Tanaka:2017lxb,Wu:2018mvg,Kawaguchi:2018ptg,Hotokezaka:2019uwo,Kawaguchi:2019nju,Korobkin:2020spe,Bulla:2020jjr,Zhu:2020eyk,Barnes:2020nfi,Nativi:2020moj,Kawaguchi:2020vbf,Wu:2021ibi,Just:2021vzy,Just:2023wtj}. Previous studies showed that the complex ejecta profile in the presence of the multiple ejecta components of different mass ejection processes induces significant spatial dependences in radioactive heating as well as strong geometrical effects in radiative transfer, which have great impacts on the resulting light curves~\citep{Kasen:2014toa,Kawaguchi:2018ptg,Kawaguchi:2019nju,Bulla:2019muo,Zhu:2020inc,Darbha:2020lhz,Korobkin:2020spe,Almualla:2021znj,Kedia:2022onl,Collins:2022ocl,Shingles:2023kua,Collins:2023btn}. Hence, the employment of the realistic ejecta profile consistently taking multiple ejecta components into account is essential for the accurate prediction of KN light curves. 

One of the important missing links for the accurate prediction of KNe is the long-term hydrodynamics evolution of ejecta after the formation. While the ejecta formation takes place on a time scale of $\lesssim1$--10 s after the onset of a merger~\citep{Hayashi:2021oxy,Hayashi:2022cdq}, the KN emission peaks in a much longer time scale of 0.1--10\,d~\citep{Li:1998bw,Kulkarni:2005jw,Metzger:2010sy,Kasen:2013xka,Tanaka:2013ana}, at which the homologous expansion of ejecta has been achieved. Since ejected matter can be accelerated by the pressure gradient in it and interact with different ejecta components during these epochs, the ejecta profile at the time of the KN emission is non-trivial just from the ejecta properties at the time of formation. In fact, \cite{Rosswog:2013kqa} and \cite{Grossman:2013lqa} performed pseudo-Newtonian hydrodynamics simulations for BNS mergers, and studied the long-term evolution of the dynamical ejecta component until they reached the homologously expanding phase. They found that the thermal pressure induced by radioactive heating in ejecta significantly changes the ejecta morphology (see also~\citealt{Foucart:2021ikp}). ~\cite{Fernandez:2014bra} and \cite{Fernandez:2016sbf} performed long-term simulations for BH-NS mergers to investigate the effect of the interplay between the dynamical and post-merger components and found that the interaction of the multiple ejecta components can modify the ejecta profile. Thus, to accurately predict KN light curves, it is also important to follow the hydrodynamics evolution of the multiple ejecta components until the homologously expanding phase.

Recently, the development of numerical simulation techniques and the significant increase in the computational resources have enabled us to consistently follow the NS mergers from the onset of the merger up to the time that ejecta formation saturates ~\citep{Kiuchi:2022ubj,Fujibayashi:2022ftg,Fujibayashi:2020dvr,Shibata:2021xmo,Hayashi:2021oxy,Kiuchi:2022ubj,Fujibayashi:2022ftg,Hayashi:2022cdq,Kiuchi:2022nin,Just:2023wtj,Gottlieb:2023vuf,Kiuchi:2023obe}. In this paper, we study the KN emission associated with a BH-NS merger employing the results obtained by the numerical-relativity (NR) simulation and nucleosynthesis calculation consistently following the entire ejecta formation from the merger~\citep{Hayashi:2021oxy,Hayashi:2022cdq,Wanajo:2022jgw}. In particular, we focus on the KN emission from $\approx 1\,{\rm d}$ after the onset of the merger for the model of a large amount of dynamical ejecta with $\approx 0.04M_\odot$ in this paper.

This paper is organized as follows: In Section~\ref{sec:method}, we describe the method employed in this study. In Section~\ref{sec:model}, we describe the BH-NS model we study in this work. In Section~\ref{sec:result1}, we present the property of the ejecta obtained by the long-term hydrodynamics evolution. In Section~\ref{sec:result2}, we present the KN light curve obtained by RT simulations. Finally, we discuss the implication of this paper in Section~\ref{sec:discuss}. Throughout this paper, $c$ denotes the speed of light.

\section{Method}~\label{sec:method}

\subsection{hydrodynamics simulation}

In a BH-NS merger, matter ejected by various mechanisms is expected to experience hydrodynamics interactions between different ejecta components before eventually reaching a homologous expansion phase at $\sim 0.1$\,d~\citep{Kawaguchi:2020vbf}. In order to obtain the spatial profile of the rest-mass density, elemental abundances, and radioactive heating rate after 0.1\,d, which are necessary for accurate prediction of KN, we perform hydrodynamics simulations using the outflow data obtained by NR simulations as boundary conditions, as in our previous studies. To distinguish it from the NR simulation, the present hydrodynamics simulation is referred to as the HD simulation in this paper.

The simulation code for the HD simulation is a 3D extension of the code developed in our previous studies~\citep{Kawaguchi:2020vbf,Kawaguchi:2022bub,Kawaguchi:2023zln}. This code solves the relativistic Euler equations under a spherical coordinate system. In order to incorporate the effect of gravity, a fixed background metric for a non-rotating black hole expressed in isotropic coordinates is used (see Appendix~\ref{app:form} for the formulation of the basic equations). The effect of radioactive heating is incorporated 
by considering the neutrino loss\footnote{The experimentally evaluated neutrino-energy loss for each $\beta$-decay is adopted from the ENDF/B-VIII.0 library \citep{Brown:2018NDS}. For those with no experimental evaluation (relevant to the entire phase of $r$-processing), the fraction of neutrino-energy loss is assumed to be 0.4 according to \citet{Hotokezaka:2016}.}
in the same way as in the previous studies (see Appendix~\ref{app:form} and ~\citealt{Kawaguchi:2020vbf,Kawaguchi:2022bub,Kawaguchi:2023zln}; see also Appendix~\ref{app:pt} for the method of particle tracing used to employ the nucleosynthesis results in the HD simulation). We note that the equatorial symmetry is imposed for the HD simulation following the setup of the NR simulation.

For the equation of state (EOS), we consider both contributions from gas and radiation: the total pressure $P$ is given by $P=P_{\rm gas}+P_{\rm rad}$ with $P_{\rm gas}=n_{\rm B} k_{\rm B} T/\mu$ and $P_{\rm rad}=a_{\rm rad} T^4/3$, where $n_{\rm B}$, $T$, $ k_{\rm B}$, $\mu$, and $a_{\rm rad}$ are the baryon number density, temperature, Boltzmann constant, mean molecular weight, and radiation density constant, respectively. Here, we simplified the gas pressure assuming that atoms are fully ionized with $\mu=1$, and the gas pressure is dominated by the contribution from electrons (since the average atomic mass number is expected to be much larger than unity). We note that, although this simplification may overestimate the gas pressure component, the contribution of the gas pressure is found to be nevertheless subdominant. In fact, we confirm that the resulting ejecta profiles as well as the KN light curves are essentially unchanged even if we employ the ideal-gas EOS with the adiabatic index of $\Gamma=4/3$, which corresponds to the case that the radiation pressure dominates.

Note that the magnetic field effects are not taken into account in our present HD simulations. As a consequence, and due to the coarse grid resolution in the polar region, the collimated relativistic jet launched in the NR simulation is not well resolved in the present HD simulations. The previous study suggests that the presence of the jet may affect the ejecta profile and hence the KN light curves near the jet axis~\citep{Nativi:2020moj,Klion:2020efn}. Since resolving the propagation of the collimated relativistic jet in long-term three-dimensional simulations requires high computational costs, we leave the investigation of the effect of the jet for future work.

We employ the same time origin for the HD simulations as in the NR simulations. The uniform grid with $N_\theta$ and $N_\phi$ grid-cells along the polar angle $\theta$ and the longitudinal angle $\phi$, respectively, is prepared. For the radial direction, the following non-uniform grid structure is employed; for a given $j$-th radial grid-node
\begin{align}
	{\rm ln}\,r_j={\rm ln}\left(\frac{r_{\rm out}}{r_{\rm in}}\right)\frac{j-1}{N_r}+{\rm ln}\,r_{\rm in},\,j=1\cdots N_r+1,\label{eq:grid}
\end{align} 
where $r_{\rm in}$ and $r_{\rm out}$ denote the inner and outer radii of the computational domain, respectively, and $N_r$ denotes the total number of the grid-cells along the radial direction. In the present work, we employ $(N_r,N_\theta,N_\phi)=(1024,64,128)$, and $r_{\rm in}$ and $r_{\rm out}$ are initially set to be $3,000\,{\rm km}$ and $10^3\,r_{\rm in}$, respectively. We confirm that this grid resolution is sufficiently high enough for our purpose of the study by checking the results of the ejecta profile and KN light curves being semi-quantitatively unchanged for the HD simulation with $(N_r,N_\theta,N_\phi)=(512,32,64)$ (less than 10\% and 3\% difference in the total bolometric luminosity at 1\,d and 2\,d, respectively). 

The hydrodynamics properties of the outflow are extracted at $r=r_{\rm ext}$ in the NR simulations of~\citet{Hayashi:2021oxy,Hayashi:2022cdq}, and the time-sequential data are employed as the inner boundary condition of the present HD simulations. The outflow data obtained from the NR simulation run out at $t> 1$\,s, and after then, the HD simulation is continued by setting a very small floor value to the rest-mass density of the inner boundary. To follow the evolution of ejecta even after the high-velocity edge of the outflow reaches the outer boundary of our HD simulation, the radial grid points are added to the outside of the original outer boundary, while at the same time the innermost radial grid points are removed so as to keep the total number of the radial grid points. By this prescription, the value of $r_{\rm in}$ is increased in the late phase of the HD simulations. The outermost radial grids are added so that the location of the outer radial boundary, $r_{\rm out}$, is always $10^3 r_{\rm in}$. Note that the region of $r\gtrsim 10^{-3} ct$ is always covered with the computational domain up to $t=0.1\,{\rm d}$ in the HD simulations.

The so-called Courant–Friedrichs–Lewy (CFL) condition restricts the time steps in the HD simulation to ensure the numerical stability. For our setup, the time interval should be approximately less than the smallest value among $\Delta r_{\rm min}/c$, $r_{\rm in}\Delta \theta_{\rm min}/c$, and $r_{\rm in}{\rm sin}\theta_{\rm min} \Delta \phi_{\rm min}/c$ with $\theta_{\rm min}$, $\Delta r_{\rm min}$, $ \Delta \theta_{\rm min}$, and $\Delta \phi_{\rm min}$ being the minimum cell center value of the $\theta$ coordinate and the minimum cell sizes of $r$, $\theta$, and $\phi$ directions, respectively. For the present grid setup, the most strict constraint comes from the last condition of $r_{\rm in}{\rm sin}\theta_{\rm min} \Delta \phi_{\rm min}/c$, and this restricts the time interval to be so small that the computational costs becomes practically quite high. To relax this condition, we average over the conservative variables of hydrodynamics in the direction of $\phi$ for all the cells located in $\theta\leq\theta_{\rm c}$ for each sub-step of the evolution (see~\cite{Hirai:2022hbf} for the similar prescription). By this prescription, the HD simulation is kept numerically stable if the time interval is within $r_{\rm in}{\rm sin}\theta_{\rm c} \Delta \phi_{\rm min}/c$. For the present study, we choose $\theta_{\rm c}$ to be $\pi/24$, while we confirm that the resulting LCs are essentially unchanged even if we employ $\theta_{\rm c}=\pi/12$. 

\subsection{radiative-transfer simulation}

The light curves of KNe are calculated using a wavelength-dependent RT simulation code~\citep{Tanaka:2013ana,Tanaka:2017qxj,Tanaka:2017lxb,Kawaguchi:2019nju,Kawaguchi:2020vbf}. In this code, the photon transfer is simulated by a Monte Carlo method for given ejecta profiles composed of the density, velocity, and elemental abundance under the assumption of the homologous expansion. The time-dependent thermalization efficiency is taken into account following an analytic formula derived by~\citet{Barnes:2016umi}. 
In our RT code, the local gas temperature at each time step is calculated as $T_{\rm gas}=(u_{\rm rad}/a_{\rm rad})^{1/4}$, i.e., under the assumption of local thermodynamic equilibrium (LTE) using the Stefan-Boltzmann law with the radiation energy density, $u_\mathrm{rad}$, obtained by the RT simulation.
Then, the ionization and excitation states are determined from the gas density and temperature by using the Saha's ionization and Boltzmann excitation equations.

For the photon-matter interaction, bound-bound, bound-free, and free-free transitions, and electron scattering are taken into account for the transfer of optical and infrared photons~\citep{Tanaka:2013ana,Tanaka:2017qxj,Tanaka:2017lxb}. The formalism of the expansion opacity~\citep{1983ApJ...272..259F,1993ApJ...412..731E,Kasen:2006ce} and the new line list derived in~\citet{Domoto:2022cqp} are employed for the bound-bound transitions. In this line list, the atomic data of VALD~\citep{1995A&AS..112..525P,1999A&AS..138..119K,2015PhyS...90e4005R} or Kurucz's database~\citep{1995all..book.....K} is used for $Z=20$–$29$, while the results of atomic calculations from~\citet{Tanaka:2019iqp} are used for $Z=30$--$88$. For Sr II, Y I, Y II, Zr I, Zr II, Ba II, La III, and Ce III, which are the ions producing strong lines, the line data are replaced with those calibrated with the atomic data of VALD and NIST databases~\citep{NIST}. Note that, since our atomic data include only up to the triple ionization for all the ions, the early phase of the light curves ($t\le 0.5\,{\rm d}$) may not be very reliable due to high ejecta temperature (see~\citealt{Banerjee:2020myd,Banerjee:2022doa,Banerjee:2023gye} for the work taking the opacity contribution from higher ionization states into account).

The RT simulations are performed from $t=0.1\,{\rm d}$ to $30\,{\rm d}$ employing the density and internal energy profiles of the HD simulations at $t=0.1\,{\rm d}$ and assuming the homologous expansion for $t > 0.1$\,d. The spatial distributions of the heating rate and elemental abundances are determined by the table obtained by the nucleosynthesis calculations referring to the injected time and angle of the fluid elements. Note that, as an approximation, the elemental abundances at $t=1\,{\rm d}$ are used during the entire time evolution in the RT simulations to reduce the computational cost, but this simplified prescription gives an only minor systematic error on the resultant light curves as illustrated in~\citet{Kawaguchi:2020vbf}. 

A three-dimensional cylindrical grid is applied for storing the local elemental abundances and radioactive heating rate as well as for solving the temperature and opacity. The 50, 50, and 32 cells are set to the cylindrical radius, vertical, and longitudinal directions, which cover the domain with the coordinate ranges of $(0,0.6\,ct)$, $(0,0.6\,ct)$, and $(0,2\pi)$, respectively. We confirm that the resulting light curves are unchanged by changing each cell numbers from 50, 50, and 32 cells to $40$, $40$, and $28$ cells or changing the maximum cylindrical radius and vertical coordinate ranges from $0.6\,ct$ to $0.75\,ct$.

\section{The BH-NS model}~\label{sec:model}
In this work, we employ the NR outflow profiles and nucleosynthetic data obtained in~\citet{Hayashi:2021oxy,Hayashi:2022cdq} and~\citet{Wanajo:2022jgw} as the input for the HD simulations. In particular, we employ the outflow data of model Q4B5H in~\citet{Hayashi:2021oxy}. For this model, a BH-NS binary of which the NS mass, BH mass and dimensionless spin are initially $1.35\,M_\odot$, $5.4\,M_\odot$ (thus 4 times larger than the NS mass), and 0.75, respectively, is considered with the DD2 EOS~\citep{Banik:2014qja}. The poloidal magnetic field with the maximum strength of $5\times 10^{16}\,{\rm G}$ is initially set in the NS; with this setting, the magnetorotational instability~\citep{Balbus:1998ja} in the remnant disk is resolved and the $\alpha$-$\Omega$ dynamo subsequently in action can be captured. We set $6.6\,M_\odot$ as the BH mass of the metric employed in the HD simulations, which approximately agrees with the summation of the remnant BH mass and matter outside the BH measured at $t=0.1\,{\rm s}$.

We note that, although BHs may be born with low spins~\citep[e.g.,][]{Fuller:2019sxi,Gottlieb:2023cgm}, 
we assume a relatively large spin (dimensionless BH spin of 0.75), 
which tends to give a larger ejecta mass and brighter kilonova ~\citep[e.g.,][]{Colombo:2023une}. We also note that the large initial magnetic field is artificially set for the numerical prescription, while~\citet{Hayashi:2022cdq} showed that the resulting ejecta profile may not be very sensitive to the initial magnetic-field strength and configuration. However it is important to keep in mind systematic errors associated with these uncertainty.

For model Q4B5H, the NS experiences tidal disruption before it reaches the inner-most stable circular orbit of the binary ($t\approx 10\,{\rm ms}$). This leads to the formation of massive ejecta and disk around the remnant BH. Ejecta formed at the time of the NS tidal disruption, which often referred to as the dynamical ejecta, are concentrated in the vicinity of equatorial plane and exhibit significant non-axisymmetric geometry. The dynamical ejecta typically have low electron fraction (0.03--0.07) because those are driven primarily by gravitational torque and do not appreciably go through weak processes in the merger.

Subsequently, the magnetic field is amplified in the remnant disk, and the effective viscosity is induced by the magnetohydrodynamical turbulence, driven by the magnetorotational instability~\citep{Balbus:1998ja} and $\alpha$-$\Omega$ dynamo. Initially, viscous heating in the disk is balanced with neutrino cooling. As the disk rest-mass density and temperature drop due to the expansion driven by angular momentum transport, neutrino cooling becomes inefficient, and viscosity-driven mass ejection sets in at $t\approx 0.2$--$0.3\,{\rm s}$ \citep{Fernandez:2013tya,Just:2014fka,Fujibayashi:2020qda,Just:2021vzy}. In parallel, magneto-centrifugal force in the central region might play a role for enhancing mass ejection. Mass ejection in this stage, which is referred to as the post-merger mass ejection, lasts for $\sim 1$--$10\,{\rm s}$. In contrast to the dynamical ejecta, since thermal and weak processes play important roles during the post-merger stages, the electron fraction of ejecta has a broad distribution in the range of 0.1--0.4 with its peak being 0.24.

For model Q4B5H, the masses of the dynamical and post-merger ejecta are found to be $0.045\,M_\odot$ and $0.028\,M_\odot$, respectively, at the end time of the NR simulation. It is worth being remarked that the combination of dynamical and post-merger ejecta approximately reproduces a solar-like $r$-process pattern~\citep{Wanajo:2022jgw}. 

In this paper, we study the ejecta and KN property for one case of a BH-NS merger among available NR results as the first step for the end-to-end KN simulation. However, we should note that the disk and ejecta masses formed in BH-NS mergers can have large variety depending on the binary parameters, such as the BH and NS masses, BH spin, and NS radius~\citep{Rosswog:2005su,Shibata:2007zm,Etienne:2008re,Lovelace:2013vma,Kyutoku:2015gda,Foucart:2018rjc}, as well as the adopted EOS~\citep{Hayashi:2022cdq}. For example, the smaller amount of disk and ejecta would be formed for the case that the NS radius is smaller due to softer EOS, such as the SFHo EOS~\citep{Steiner:2012rk,Hayashi:2022cdq}. Hence, the resulting property of the KN light curves can also have a large diversity. Therefore, we emphasize that the ejecta and KN property found for a particular model (Q4B5H) with the DD2 EOS may not be universal property for every case of BH-NS mergers, and we leave the investigation of the binary parameter and EOS dependences for future work. 

\section{Results: Hydrodynamics simulation}~\label{sec:result1}

\subsection{Ejecta mass and energy evolution}
\begin{figure}
 	 \includegraphics[width=\linewidth]{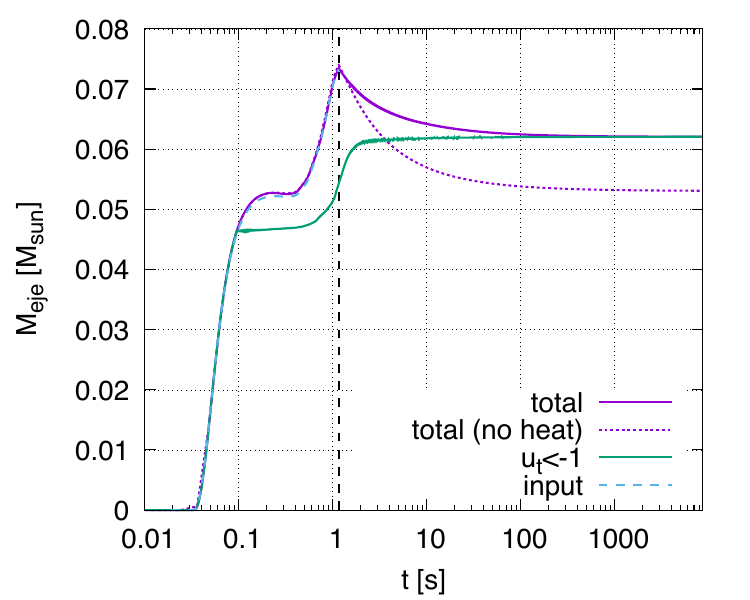}
 	 \caption{Time evolution of the total rest mass in the computational domain of the HD simulation (the purple curves). The solid and dotted purple curves denote the results for the HD simulations in which radioactive heating are turned on and off, respectively. The green curve denotes the same as for the purple curve but only for the matter which satisfies the geodesic criterion ($u_t<-1$ where $u_t$ is the lower time component of the four velocity). The blue dashed curve with the label ``input" denotes the rest mass obtained by integrating the mass flux of the NR outflow data which is employed as the inner boundary condition of the HD simulation. The black dashed line denotes the time at which the NR outflow data run out.}
	 \label{fig:t_mass}
\end{figure}

Figure~\ref{fig:t_mass} shows the total rest mass in the computational domain as a function of time. We can consider that the ejecta have reached the homologously expanding phase at $t=0.1\,{\rm d}$, because the total internal energy of ejecta
is smaller by 4 order of magnitudes than the total kinetic energy. In general, two distinct ejecta components are seen in Figure~\ref{fig:t_mass}. One saturated in $t\sim\,0.2$ s corresponds to the dynamical ejecta, and the subsequent increase found for $t\gtrsim\,0.5$ s corresponds to the post-merger ejecta. 

After the NR outflow data run out at $t\approx 1\,{\rm s}$, we impose a floor rest-mass density value to the inner boundary. It is clearly seen in Figure~\ref{fig:t_mass} that the total mass in the computational domain decreases after that time, indicating that the matter is artificially falling back and escaping through the inner boundary. This happens because the pressure support from the inner boundary vanishes after the outflow data run out. However, the total mass of the matter with gravitationally unbound orbits remains increasing even after the time when the NR outflow data run out, as a  consequence of the acceleration of the matter in the presence of the thermal pressure gradient. In fact, the mass of the fall back matter is larger due to lower pressure for the case in which radioactive heating is turned off (see the purple dotted curve in Figure~\ref{fig:t_mass}).

After $t\approx100\,{\rm s}$, approximately all the ejecta matter remaining in the computational domain becomes gravitationally unbound, and the value of the total mass in the computational domain converges to $0.063\,M_\odot$. This value is smaller than the ejecta mass estimated in ~\citet{Hayashi:2022cdq} by $\approx0.01\,M_\odot$. We interpret this discrepancy as a consequence of the mismatch in the employed EOS between the NR and HD simulations and the inconsistency of the matter flux at the inner boundary. In fact, ~\cite{Fernandez:2014bra} show similar results: they performed BH-disk simulations to follow the formation of the post-merger wind ejecta and used the extracted ejecta property as the inner boundary condition of the subsequent simulation for long-term ejecta evolution in the same manner as our present work. They found that the mass of the post-merger wind ejecta which becomes gravitationally unbound in the subsequent simulation decreases by a factor of $\approx2$ from the values estimated in the BH-disk simulations. They interpreted this difference as a consequence of the discrepancy between the stresses at the inner boundary and those that would be obtained in a self-consistent simulation. 

Nevertheless, by performing the HD simulation with artificially modified inner boundary conditions, we confirmed that our main results are essentially the same and the modification to the resulting KN light curves is only minor: we perform a HD simulation in which the ejecta injection is sustained with the final value of the mass flux at $1\,{\rm s}$ after the NR outflow data run out. By this prescription, the total ejecta mass in the HD simulation at the homologously expanding phase increases by $0.01\,M_\odot$, but the bolometric luminosity increases only at most $\approx 10\%$ since the unbound matter increased by this prescription has the velocity only less than $\lesssim0.05\,c$ and hence has a long diffusion time scale, which gives a minor contribution to the brightness of the emission.

\begin{figure}
 	 \includegraphics[width=\linewidth]{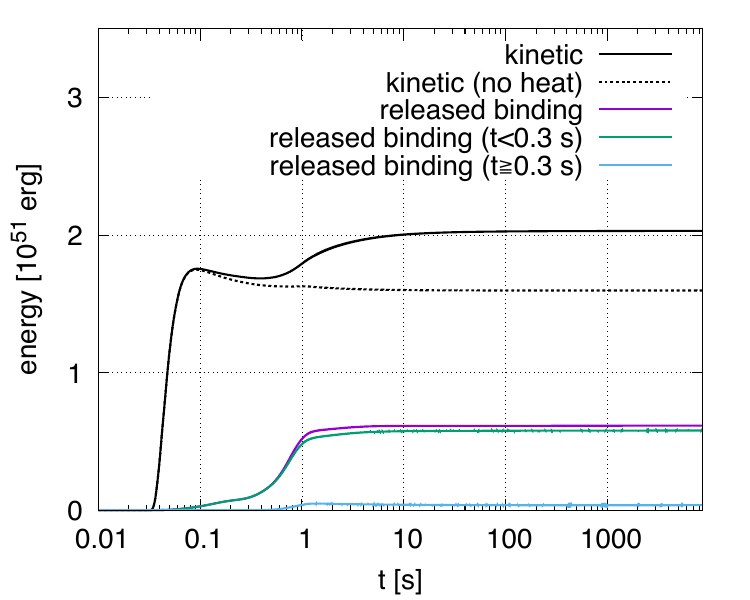}\\
          \includegraphics[width=\linewidth]{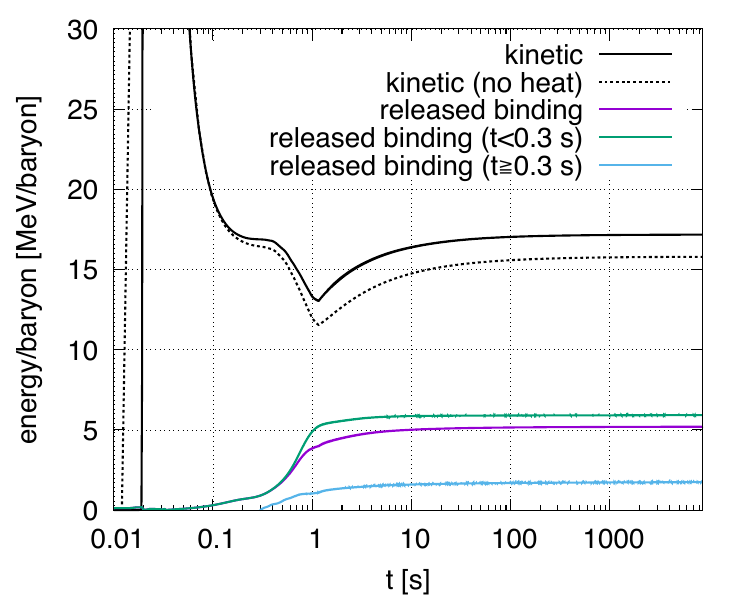}
 	 \caption{Time evolution of the ejecta kinetic energy and total released nuclear binding energy. The top and bottom panels show the total value and mass-averaged value per baryon, respectively. The black solid and dotted curves denote the kinetic energy for the HD simulations in which radioactive heating are turned on and off, respectively. The purple curve denotes the total nuclear binding energy released during the radioactive decay. The green and blue curves denote the same as the purple curve but only taking into account the ejecta components which pass the inner boundary at $t<0.3\,{\rm s}$ (the dynamical ejecta) and $t\geq0.3\,{\rm s}$ (the post-merger ejecta), respectively.}
	 \label{fig:heat_check}
\end{figure}

Figure~\ref{fig:heat_check} shows the time evolution of the ejecta kinetic energy and total released nuclear binding energy. The top panel of Figure~\ref{fig:heat_check} shows that the ejecta kinetic energy increases around $t\sim 0.04\,{\rm s}$ and $t\sim 0.5\,{\rm s}$. The former increase is due to the entry of the dynamical ejecta into the computational domain. On the other hand, the increase at $t\sim 0.5\,{\rm s}$ is due to the acceleration induced by the radioactive heating. In fact, the increase is not found for the case in which the radioactive heating is turned off. The increase in the kinetic energy by the radioactive heating agrees with $\approx 70\%$ of the released nuclear binding energy; the 30\% reduction is owing to the neutrino loss. The slight decrease in the kinetic energy after $\sim 0.1\,{\rm s}$ is due to the fall back of the matter to the outside of the computational domain. After $t\sim 1000\,{\rm s}$ the kinetic energy becomes mostly a constant, which indicates that the homologous expansion of the system is achieved. The radioactive heating is dominated by the component which passes the inner boundary at $t<0.3\,{\rm s}$, that is, the dynamical ejecta.

The bottom panel of Figure~\ref{fig:heat_check} shows the mass-averaged values of the ejecta kinetic energy and total released nuclear binding energy per baryon. The released nuclear binding energy saturates with $\approx 5\,{\rm MeV}/{\rm baryon}$. That for the dynamical ejecta saturates with a larger value ($\approx 6\,{\rm MeV}/{\rm baryon}$) than the post-merger ejecta ($\approx 2\,{\rm MeV}/{\rm baryon}$). This indicates that the radioactive heating in the dynamical ejecta is more efficient because of the stronger $r$-process with fission recycling. Note that the difference in the mass-averaged kinetic energy per baryon between the HD simulations in which radioactive heating are turned on and off is less pronounced compared to that in the top panel of Figure~\ref{fig:heat_check}. This is because the mass of the fall-back matter is larger for the case without radioactive heating (see Figure~\ref{fig:t_mass}), and hence, the mass-averaged value increases.

\subsection{Ejecta profiles at homologously expanding phase}

\begin{figure*}
 	 \includegraphics[width=.48\linewidth]{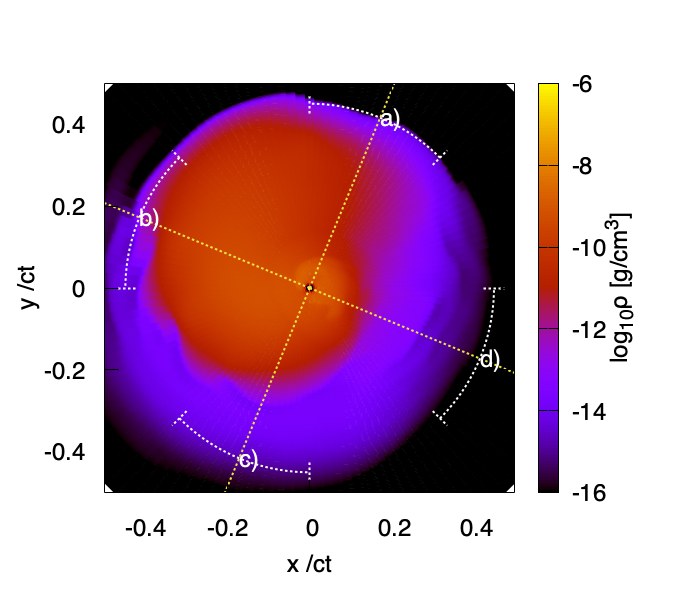}
 	 \includegraphics[width=.48\linewidth]{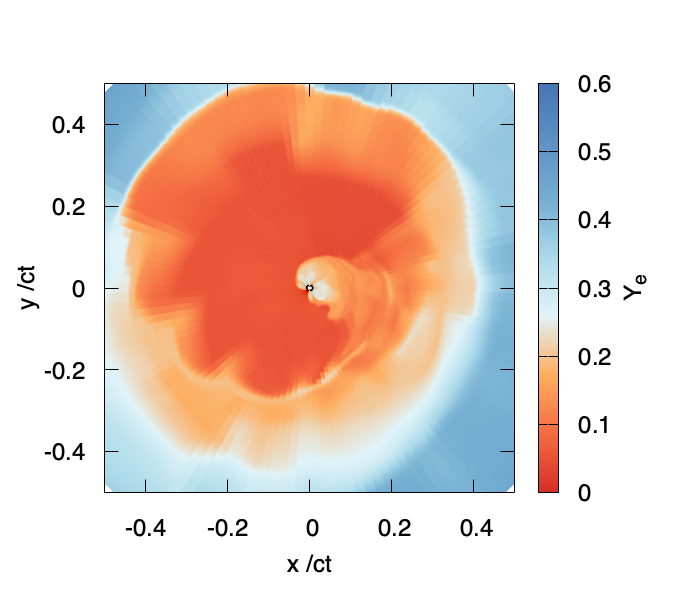}
 	 \caption{Rest-mass density and electron fraction ($Y_e$) profiles on the equatorial plane at $t=0.1\,{\rm d}$. The yellow dotted lines denote the angles for which the meridional ejecta profiles are shown in Figures~\ref{fig:dens_prof} and \ref{fig:ye_prof}. The white dotted curves denote the longitudinal angle ranges in which the KN light curves shown in Figures~\ref{fig:lbol_angle} and~\ref{fig:mag_angle} are obtained.}  
	 \label{fig:xy_prof}
\end{figure*}

\begin{figure*}
 	 \includegraphics[width=.48\linewidth]{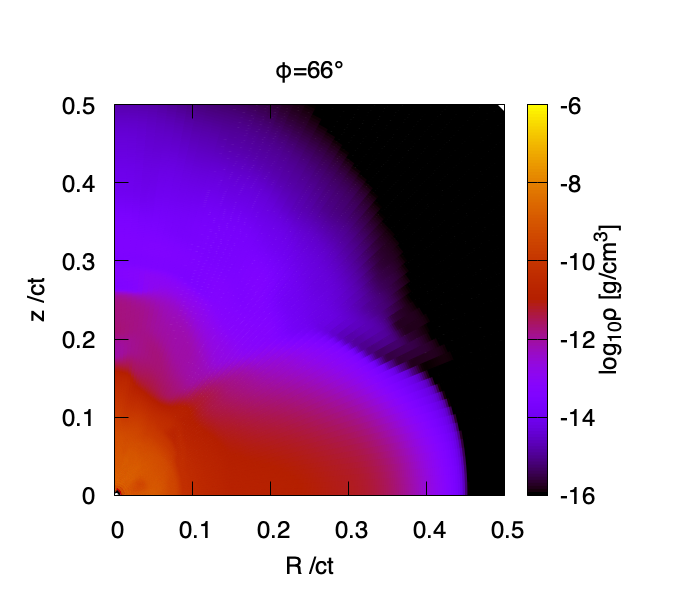}
 	 \includegraphics[width=.48\linewidth]{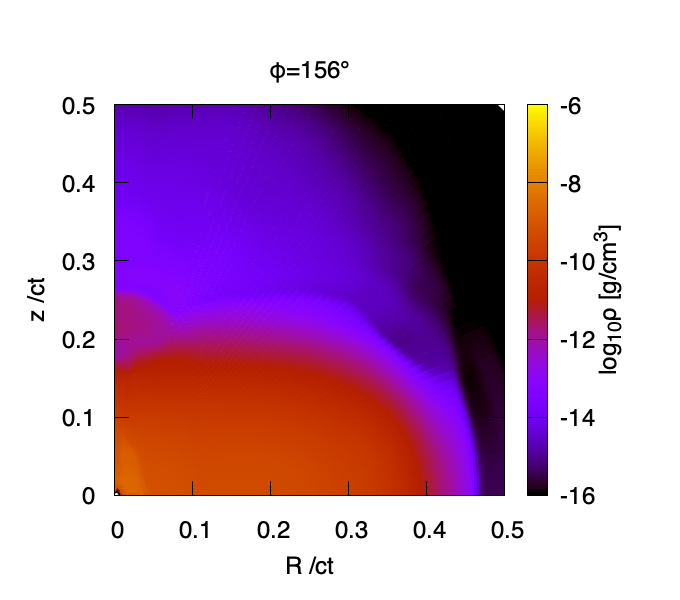}\\
 	 \includegraphics[width=.48\linewidth]{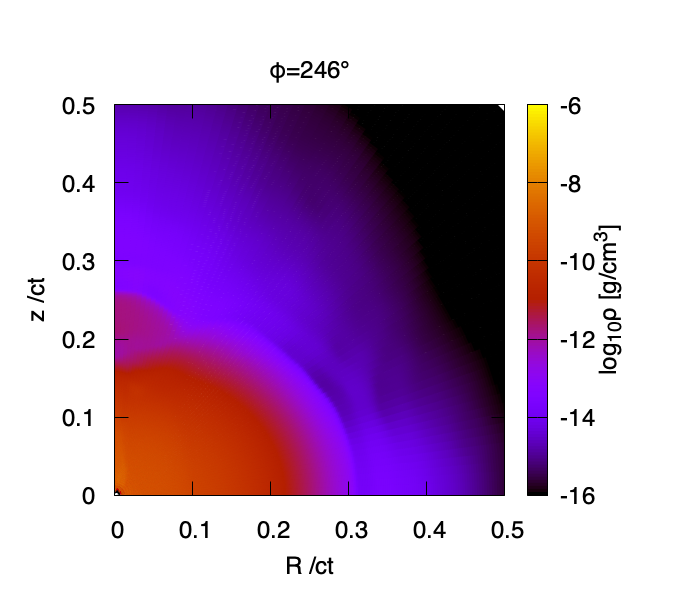}
 	 \includegraphics[width=.48\linewidth]{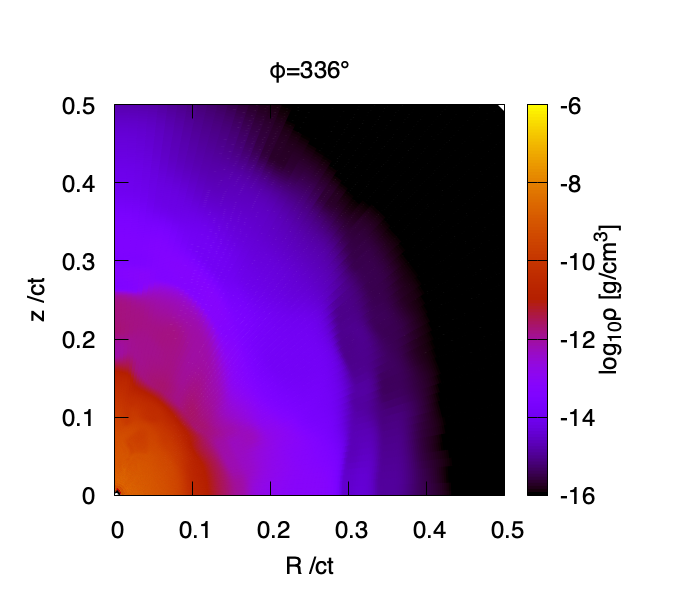}
 	 \caption{Rest-mass density profiles on the meridional planes at $t=0.1\,{\rm d}$. The top left, top right, bottom left, and bottom right panels denote the profiles on the $\phi\approx66^\circ$, $156^\circ$, $246^\circ$, $336^\circ$ planes, respectively (see also the left panel of Figure~\ref{fig:xy_prof} for the location of each plane). $R$ denotes the cylindrical radius.}
	 \label{fig:dens_prof}
\end{figure*}

\begin{figure*}
 	 \includegraphics[width=.48\linewidth]{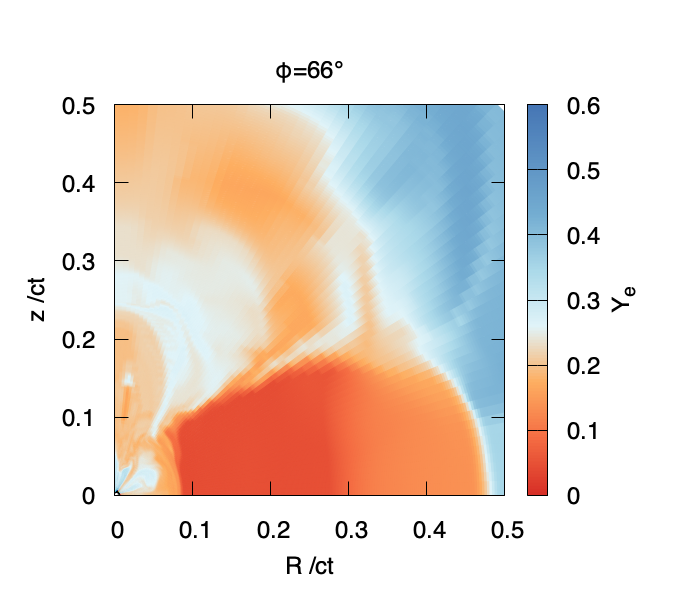}
 	 \includegraphics[width=.48\linewidth]{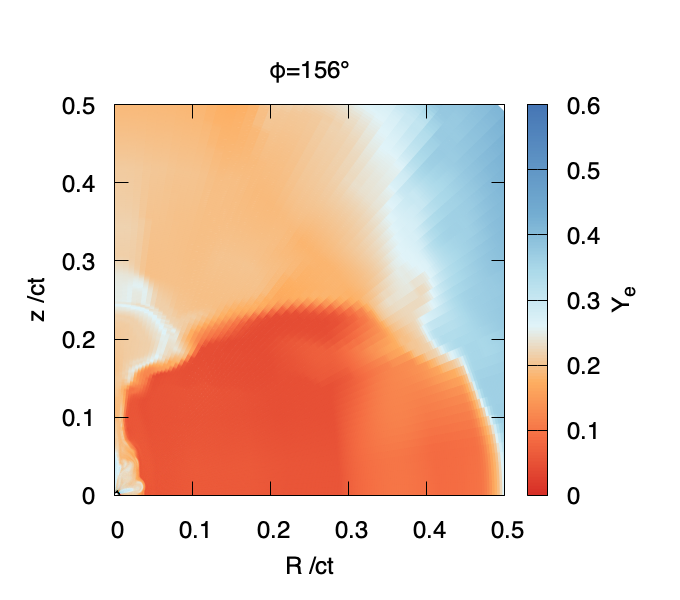}\\
 	 \includegraphics[width=.48\linewidth]{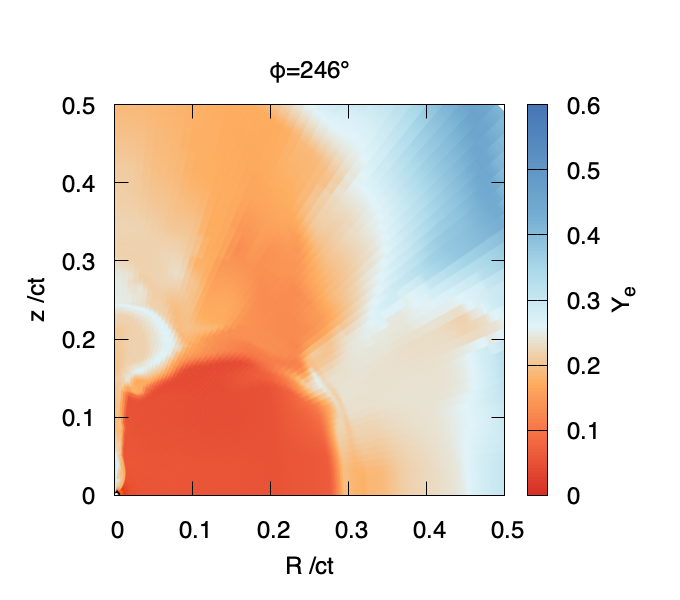}
 	 \includegraphics[width=.48\linewidth]{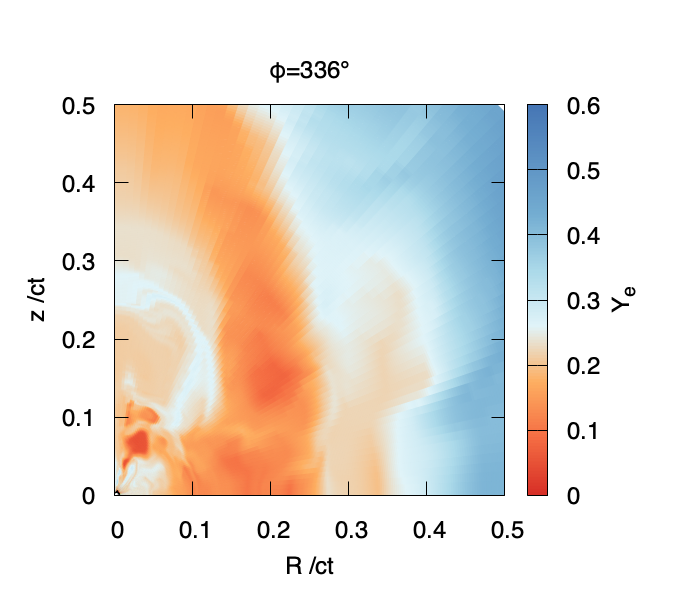}
 	 \caption{Same as Figure~\ref{fig:dens_prof} but for electron fraction $Y_e$ (see also the left panel of Figure~\ref{fig:xy_prof} for the location of each plane).}
	 \label{fig:ye_prof}
\end{figure*}

Figures~\ref{fig:xy_prof}, \ref{fig:dens_prof}, and \ref{fig:ye_prof} show the rest-mass density and electron fraction ($Y_e$) profiles of ejecta with two-dimensional various cross sections at $t=0.1\,{\rm d}$ obtained by the HD simulation. We note that the $Y_e$ profile shown here is employed as the initial condition of the nucleosynthesis calculation (see Appendix~\ref{app:pt} and~\cite{Wanajo:2022jgw} for the detail). The center of mass for the matter with $Y_e<0.1$ is located in the direction of $\phi\approx 144^\circ$ with $\phi$ being the longitudinal angle measured from the $+x$ axis. The longitudinal angles of the meridional planes shown in Figures~\ref{fig:dens_prof} and \ref{fig:ye_prof} are selected to show the profiles in which the dynamical ejecta are approximately mostly (`{\bf b)}': $\phi\approx156^\circ$), moderately (`{\bf a)}': $\phi\approx66^\circ$ and '{\bf c)}': $\phi\approx246^\circ$), and least (`{\bf d)}': $\phi\approx336^\circ$) present. As we mentioned above, the entire ejecta have reached the homologously expanding phase at this epoch. Broadly speaking, the dynamical and post-merger ejecta are present around the regions where the cylindrical radius is larger and smaller than $\approx 0.05$--$0.1 ct$, respectively. Those two components are clearly distinguishable with the value of $Y_e$. The value of $Y_e$ for the dynamical ejecta is typically below $0.1$, which primarily reflects the original $Y_e$ values of the disrupted NS. On the other hand, the post-merger ejecta have a wider range of $Y_e$ values from $0.1$ to $0.4$.

The rest-mass density profile of the dynamical ejecta exhibits clear non-axisymmetric geometry, with its mass mostly distributed in the fan-like shape in $70^\circ\lesssim \phi\lesssim 250^\circ$. The dynamical ejecta are extended up to $\approx0.5 ct$ in the cylindrical radius direction, while their vertical extent is $\approx 0.2 ct$. The aspect ratio of the cylindrical and vertical extents for the dynamical ejecta is close to unity. This is in contrast to the fact that the dynamical ejecta are launched initially confined around the equatorial plane within the latitudinal opening angle of $\sim 10^\circ$~\citep{Kyutoku:2013wxa,Foucart:2014nda}. As we show below, this ejecta expansion is due to thermal pressure enhanced by radioactive heating. 

On the other hand, the post-merger ejecta exhibit a more axisymmetric shape. It has two distinct components with one having approximately a spherical shape and the other having the cone-like shape. The former is concentrated in the region within $\approx 0.05 ct$ while the latter is more extended in the vertical direction with the polar opening angle of $\approx10^\circ$ and the vertical extent reaches $\approx 0.25 ct$. As we show below, this complex geometry of the post-merger ejecta is realized by the interaction with the dynamical ejecta which significantly expand due to thermal pressure enhanced by radioactive heating.

\begin{figure*}
 	 \includegraphics[width=.48\linewidth]{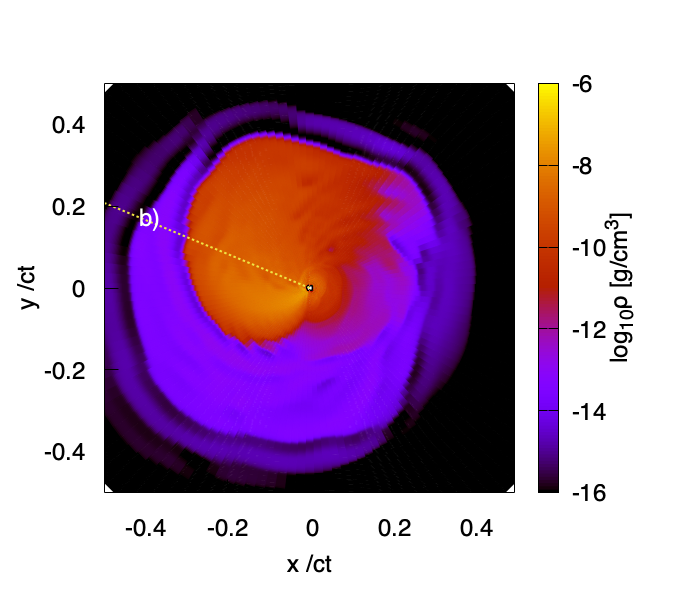}
 	 \includegraphics[width=.48\linewidth]{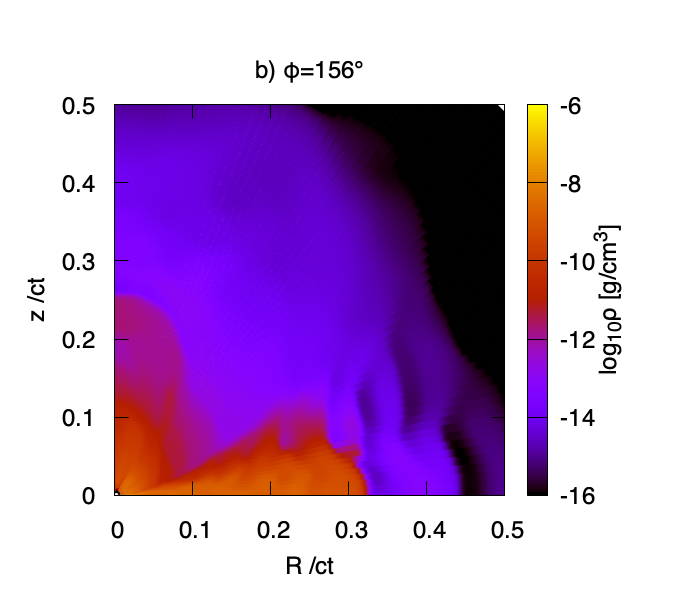}\\
 	 \includegraphics[width=.48\linewidth]{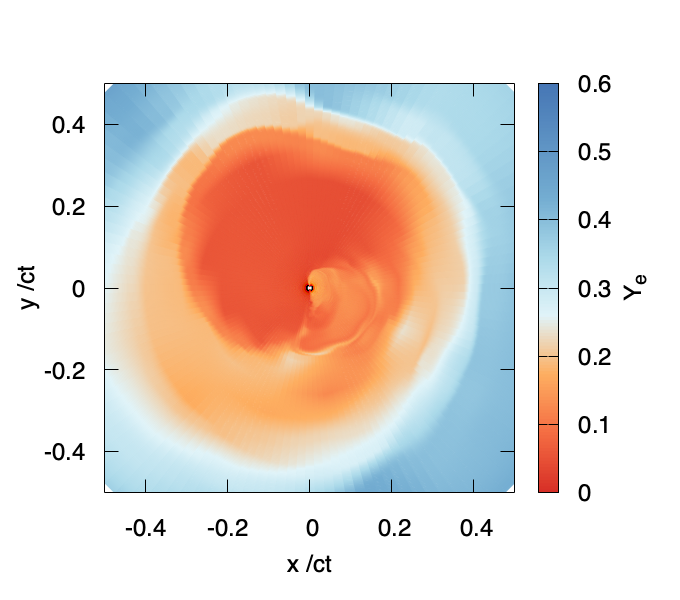}
 	 \includegraphics[width=.48\linewidth]{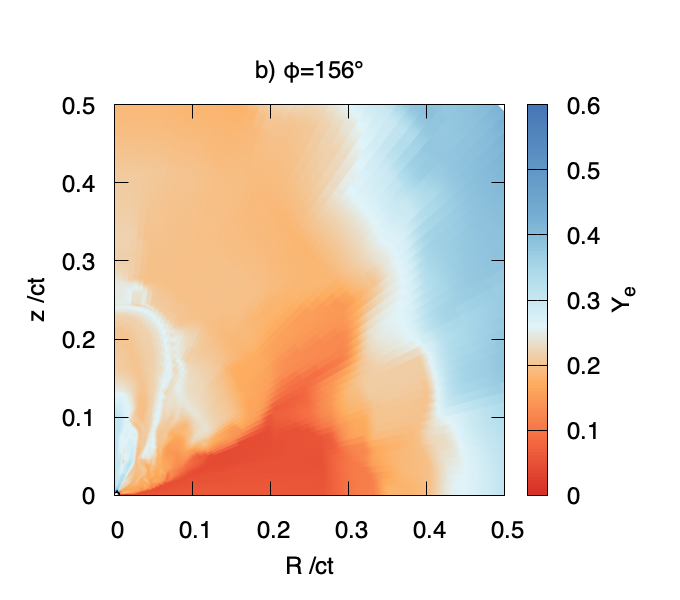}
 	 \caption{Rest-mass density and electron fraction ($Y_e$) profiles on the equatorial plane at $t=0.1\,{\rm d}$ for the HD simulation in which radioactive heating is turned off.}
	 \label{fig:noheat_prof}
\end{figure*}

\begin{figure}
 	 \includegraphics[width=\linewidth]{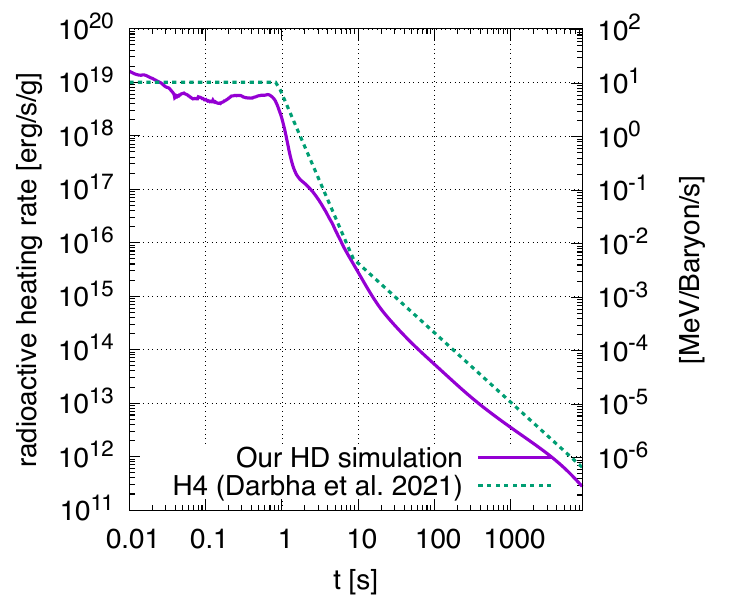}
 	 \caption{Mass weighted average of the total specific radioactive heating rate in our HD simulation. The specific heating rate of model H4 in~\citet{Darbha:2021rqj} is also shown.}
	 \label{fig:qtot_av}
\end{figure}

Figure~\ref{fig:noheat_prof} shows the rest-mass density and electron fraction profiles of ejecta on the equatorial and meridional planes at $t=0.1\,{\rm d}$ obtained by the HD simulation but switching off radioactive heating. Under the presence of radioactive heating, the dynamical ejecta expand significantly due to the increase
in thermal pressure and the inhomogeneities in the rest-mass density are also smoothed out, as clearly seen in Figures~\ref{fig:xy_prof} and~\ref{fig:dens_prof}. These results are consistent with the finding of~\cite{Rosswog:2013kqa,Grossman:2013lqa} in the context of BNSs, and of~\cite{Fernandez:2014bra,Darbha:2021rqj} in the context of BH-NSs. In fact, the resulting aspect ratio of the dynamical ejecta is found to be close to unity as the model H4 in~\cite{Darbha:2021rqj}. The radioactive heating rate as a function of time for the dynamical ejecta in our model is similar to that in~\cite{Darbha:2021rqj} (see Figure~\ref{fig:qtot_av}) , although the latter appears to be a factor of a few larger than ours\footnote{We note that, although it is claimed to be realistic heating rates for $Y_e\approx 0.1$~(according to \citealt{Metzger:2010sy,Lippuner:2015gwa}), the heating rate model H4 in~\cite{Darbha:2021rqj} is unphysical such that the total amount of energy released in this model is $\approx 13$ MeV/baryon (larger than the highest binding energy among radioactively stable nuclei, 8.8 MeV/baryon for $^{62}$Ni). On the other hand, the total amount of energy released by radioactive heating in our model is $\approx 5$ MeV/baryon (see Figure~\ref{fig:heat_check}), which is consistent with the typical value $\approx 8$~MeV/baryon for stable $r$-process nuclei by considering the neutrino-energy loss.}.

The comparison between Figures~\ref{fig:ye_prof} and~\ref{fig:noheat_prof} shows that the profile of the post-merger ejecta is modified by the dynamical ejecta. Figure~\ref{fig:noheat_prof} shows that, in the absence of radioactive heating, the post-merger ejecta exhibit a prolate shape with the extension of $0.1 ct$ and $0.25 ct$ in the equatorial and vertical directions, respectively. On the other hand, the radioactive heating significantly expands the dynamical ejecta, which compress the post-merger ejecta in $0.05 ct\le z\le 0.15 ct$ and confine the ejecta in the region of $\lesssim 0.05 ct$ as found in Figure~\ref{fig:ye_prof}. This is due to the higher typical electron fraction of the post-merger ejecta. The higher electron fraction results in the relatively small radioactive heating rate and hence small enhancement of the pressure of the ejecta compared to the dynamical component (see Figure~\ref{fig:heat_check}).

We note that, while the dynamical ejecta apparently expand significantly, the total increase in the ejecta total kinetic energy is not as significant as that one might expect from the change in the ejecta edge ($\approx0.1\,c\rightarrow\approx0.2\,c$ in the $z$-direction and $\approx0.3\,c\rightarrow\approx0.4\,c$ in the $R$-direction). This is because only a small fraction of the ejecta matter is accelerated to such high velocity. We stress that the increase in the ejecta kinetic energy is consistent with the energy deposited by radioactive heating (see~Figure~\ref{fig:heat_check}).

Significant expansion of the dynamical ejecta and enforced confinement of the post-merger ejecta in the presence of radioactive heating are not found in the BNS models in our previous studies~\citep{Kawaguchi:2020vbf,Kawaguchi:2022bub,Kawaguchi:2023zln}. This is because the dynamical ejecta of the present BH-NS model, as well as of the BNS models studied in~\cite{Rosswog:2013kqa,Grossman:2013lqa}, are more massive than the post-merger ejecta, and in addition, much more confined around the equatorial plane than in the BNS models. As a result, higher internal energy density and high thermal pressure are realized. This illustrates that the importance of the radioactive heating will depend on the density and isotopic-abundance profiles of ejecta, which can have a variety even among BH-NS mergers depending on the binary parameters or the adopted EOS.

\section{Results: KN light curves}~\label{sec:result2}

\subsection{bolometric light curves}
\begin{figure}
 	 \includegraphics[width=\linewidth]{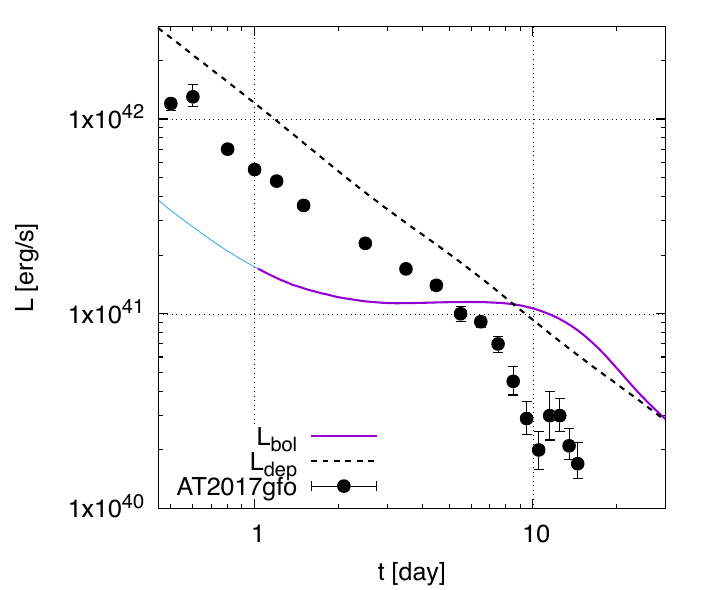}
 	 \caption{Total bolometric luminosity (solid curve) and total energy deposition rate (dashed curve). The isotropically equivalent bolometric luminosity observed in AT2017gfo with the distance of 40 Mpc is shown by the filled circles adopting the data in~\citet{Waxman:2017sqv}, which assume a black-body fit to the photo-metric observations. Note that the bolometric light curve before 1\,d is hidden since it is not reliable due to the lack of opacity data in the high temperature regime ($\gtrsim20,000\,{\rm K}$).}
	 \label{fig:lbol}
\end{figure}

Figure~\ref{fig:lbol} shows the bolometric luminosity calculated by the RT simulation employing the profiles of the ejecta rest-mass density, elemental abundance, and radioactive heating rate obtained by the combination of the results of the HD simulation and nucleosynthesis calculation~\citep{Wanajo:2022jgw}. The total energy deposition rate taking the thermalization efficiency into account is also plotted in Figure~\ref{fig:lbol}. As we mentioned in Section~\ref{sec:method}, our atomic data include only up to the triple ionization for all the ions, and the opacity of ejecta in the early phase ($t\leq 0.5\,{\rm d}$) may be underestimated due to high temperature ($\gtrsim20,000\,{\rm K}$). Hence, hereafter, we only focus on the light curves after 1\,d.

For 1--10\,d, the bolometric luminosity is approximately constant with the value of 1--2$\times 10^{\rm 41}\,{\rm erg/s}$. It decreases only slowly and the change is only by a factor of 2 during this epoch. However, after 10\,d, the bolometric luminosity starts decreasing more rapidly, and it decreases by more than a factor of 3 during 10--30 d. This faint and long-lasting emission is caused by the fact that it is primarily powered by the lanthanide-rich dynamical ejecta, in which a long photon diffusion time scale is realized by the large mass and high opacity. This behaviour of the bolometric luminosity is qualitatively the same as that found in the models with massive dynamical ejecta studied in a pioneering study (MS1Q3a75 and H4Q3a75 in~\citealt{Tanaka:2013ixa}). The bolometric luminosity converges to the total deposition rate after 20\,d, which suggests that the entire thermal photons created in the ejecta immediately diffuse out from the ejecta after this epoch. As we show below, however, the viewing-angle dependence of the emission due to the aspherical profile of the ejecta opacity still plays a role up to 30\,d.

Our present BH-NS KN model shows significantly distinct light curves from the observation of AT2017gfo. Specifically, our BH-NS model is fainter by a factor of 3 around $\sim1\,{\rm d}$ than AT2017gfo. However, due to the slow decrease in the bolometric luminosity, compared to AT2017gfo, our BH-NS KN model becomes comparably bright at 5\,d, and brighter by a factor of 5 at 10\,d. This result clearly shows that a BH-NS binary which we study particularly in this work is not likely to be the progenitor of AT2017gfo.

\begin{figure*}
\centering
 	 \includegraphics[width=.48\linewidth]{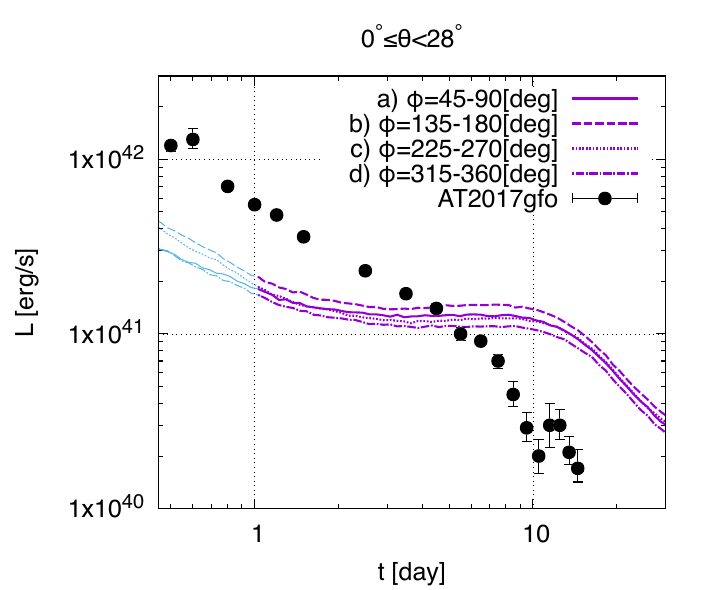}
 	 \includegraphics[width=.48\linewidth]{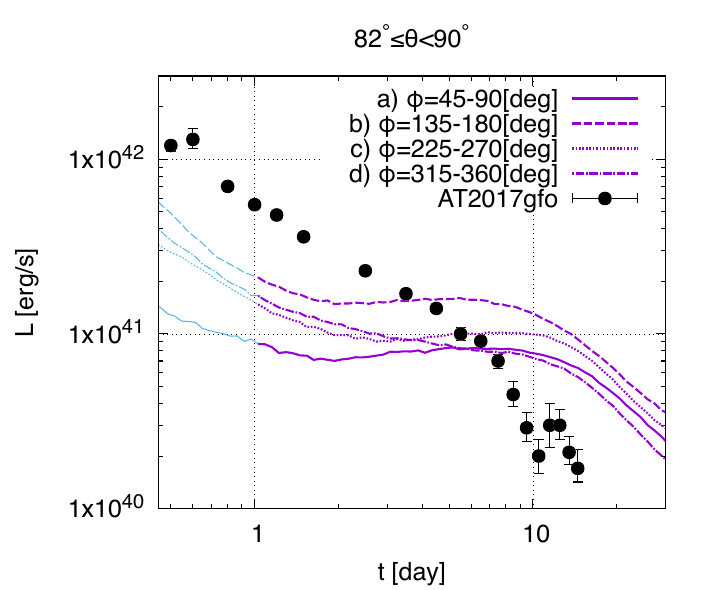}\\
 	 \includegraphics[width=.48\linewidth]{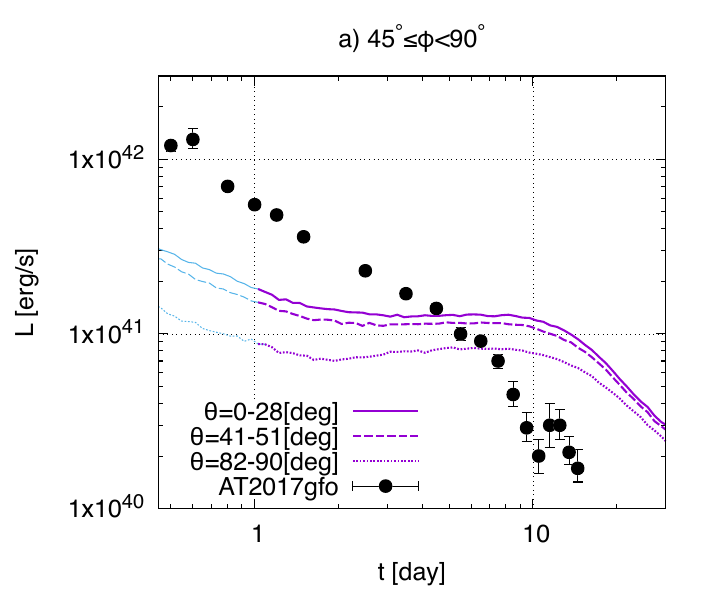}
 	 \includegraphics[width=.48\linewidth]{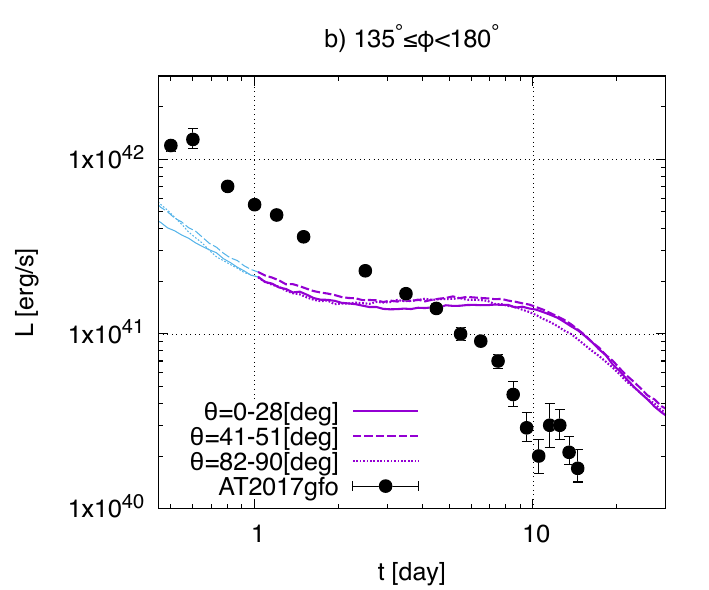}\\
 	 \includegraphics[width=.48\linewidth]{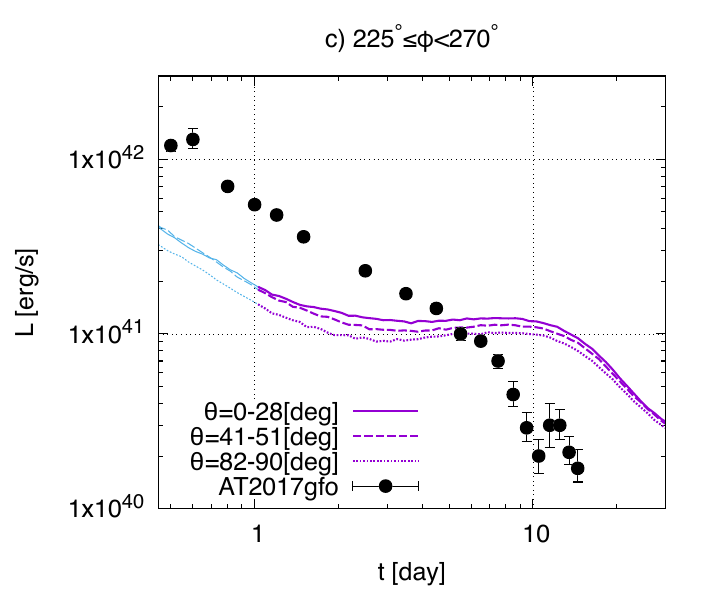}
 	 \includegraphics[width=.48\linewidth]{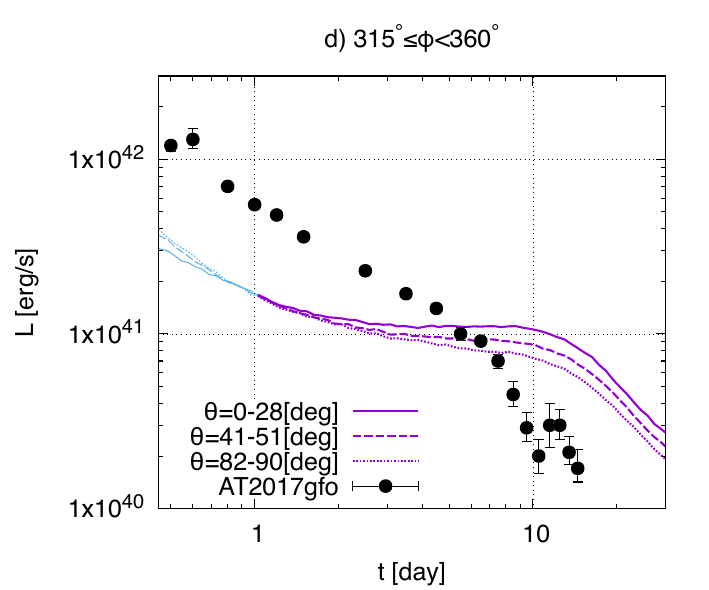}
 	 \caption{Isotropically equivalent bolometric luminosities observed from various viewing angles. The top panels compare the results among different longitudinal directions, while the middle and bottom panels compare the results among different latitudinal directions. The isotropically equivalent bolometric luminosity observed in AT2017gfo with the distance of 40 Mpc is also shown by the filled circles adopting the data in~\citet{Waxman:2017sqv}, which assume a black-body fit to the photo-metric observations. Note that the light curves before 1\,d are hidden since they are not reliable due to the lack of opacity data in the high temperature regime ($\gtrsim20,000\,{\rm K}$).}
	 \label{fig:lbol_angle}
\end{figure*}

Figure~\ref{fig:lbol_angle} shows the results of the isotropically equivalent bolometric light curves observed from various viewing angles for the present model. Focusing on the observer in the polar direction with $\theta\leq28^\circ$, the KN emission is brightest in {\bf b)}: $135^\circ\leq \phi < 180^\circ$. This direction approximately matches to the longitudinal direction in which the dynamical ejecta have most of their mass (see Figures~\ref{fig:xy_prof} and~\ref{fig:dens_prof}). On the other hand, the faintest emission is observed from the direction in which the dynamical ejecta least present (`{\bf d)}': $315^\circ\leq \phi < 360^\circ$). Nevertheless, the variation in the bolometric luminosity is not so large and it is always within 40\%. This is reasonable because the observers with different longitudinal angles are in similar directions for the polar view $\theta\leq28^\circ$.

On the other hand, the longitudinal variation is larger for the equatorial view ($82^\circ\le\theta\le90^\circ$). For this case, the KN emission is also brightest in {\bf b)}: $135^\circ\leq \phi < 180^\circ$ in which direction the dynamical ejecta are most abundant. By contrast, the bolometric luminosity is faintest in {\bf a)}: $45^\circ\leq \phi < 90^\circ$ for $t\leq5\,{\rm d}$. This is because, in this direction, the relatively thin part of the dynamical ejecta present in $R\gtrsim 0.3 ct$ around the equatorial plane (see Figure~\ref{fig:dens_prof}) suppresses radiation from the ejecta center\footnote{Note that such suppression is not significant from the direction of {\bf c)}: $225^\circ\leq \phi < 270^\circ$ due to the absence of the low density ejecta above $R\approx 0.3 ct$}. Meanwhile, the emission from the post-merger ejecta enhances the luminosity in this view ($82^\circ\le\theta\le90^\circ$) in which the dynamical ejecta are least present (`{\bf d)}': $315^\circ\leq \phi < 360^\circ$). The variation in the bolometric luminosity is larger for the equatorial view than that for the polar view, and it is larger than a factor of 2 for 1--7\,d.

The latitudinal viewing-angle dependence of the bolometric luminosity is not significant and the variation in the bolometric luminosity is always less than a factor of 2 in our present model except for the equatorial view in {\bf a)}: $45^\circ\leq \phi < 90^\circ$. The dependence of the KN brightness on the latitudinal direction is weak in particular from the viewing angle of {\bf b)}: $135^\circ\leq \phi < 180^\circ$, in which direction the KN emission is brightest. As we show below, the latitudinal viewing-angle dependence is much weaker than that for BNS KNe. This is due to the fact that, for the present BH-NS KN model, the emission is dominated by the dynamical ejecta of which the aspect ratio is close to unity. In fact, compared with a previous study~\citep{Tanaka:2013ixa}, our present BH-NS KN model shows a less significant viewing-angle dependence on the latitudinal direction. This is because the ejecta in this study have larger aspect ratio than those in the models of their previous study. This comparison indicates that the modification of the ejecta morphology due to radioactive heating has a great impact on the viewing-angle dependence of the KN emission. Hence, this work demonstrates the importance of modeling KN light curves taking the ejecta long-term evolution into account.

Interestingly, the latitudinal viewing-angle dependence is still present in {\bf d)}: $315^\circ\leq \phi < 360^\circ$ even after 20\,d at which the total bolometric luminosity converges to the total deposition rate (see Figure~\ref{fig:lbol}). This can be understood by the fact that the post-merger ejecta are present in between the high-density part of the dynamical ejecta and the observer in the direction of {\bf d)}: $315^\circ\leq \phi < 360^\circ$ (see Figures~\ref{fig:xy_prof} and ~\ref{fig:dens_prof}). While the post-merger ejecta have a minor contribution to the luminosity in the late epoch due to their relatively low heating rate, they still contribute as an opacity source (because of the relatively high density) to prevent photons emitted in the high-density part of the dynamical ejecta diffusing out to the direction of {\bf d)}: $315^\circ\leq \phi < 360^\circ$. Although it is not quantitatively significant, this long-lasting viewing-angle dependence due to the non-axisymmetric geometry of the ejecta might have an impact on estimating the total deposition rate in the ejecta from the late-time observation.

\subsection{broad-band magnitudes}
\begin{figure*}
 	 \includegraphics[width=.47\linewidth]{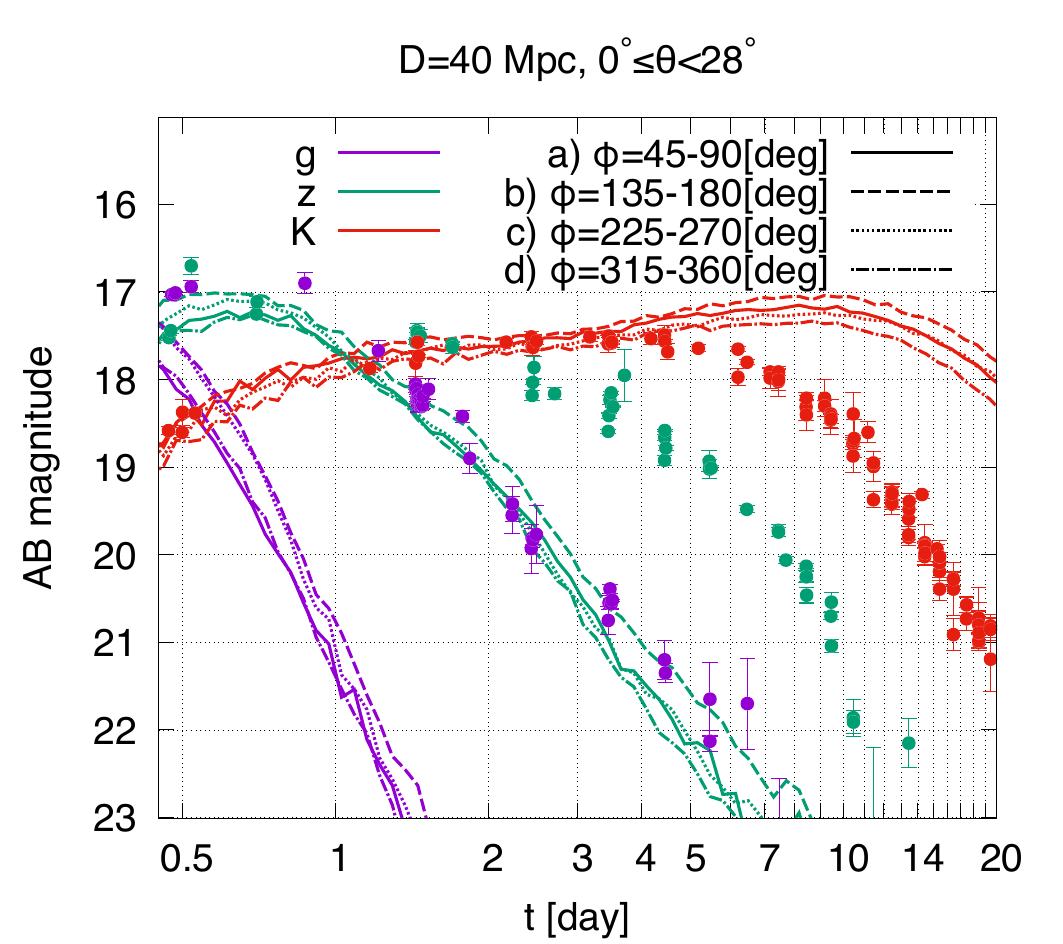}
 	 \includegraphics[width=.47\linewidth]{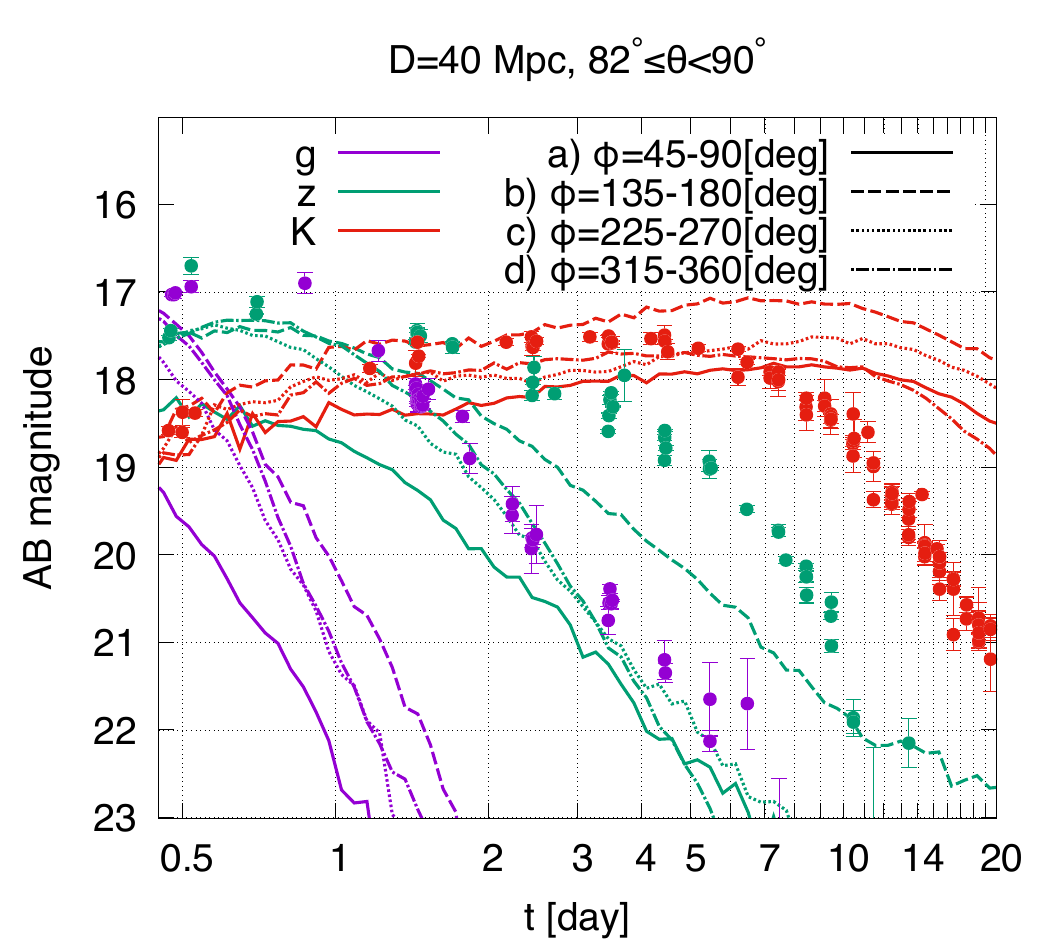}\\
 	 \includegraphics[width=.47\linewidth]{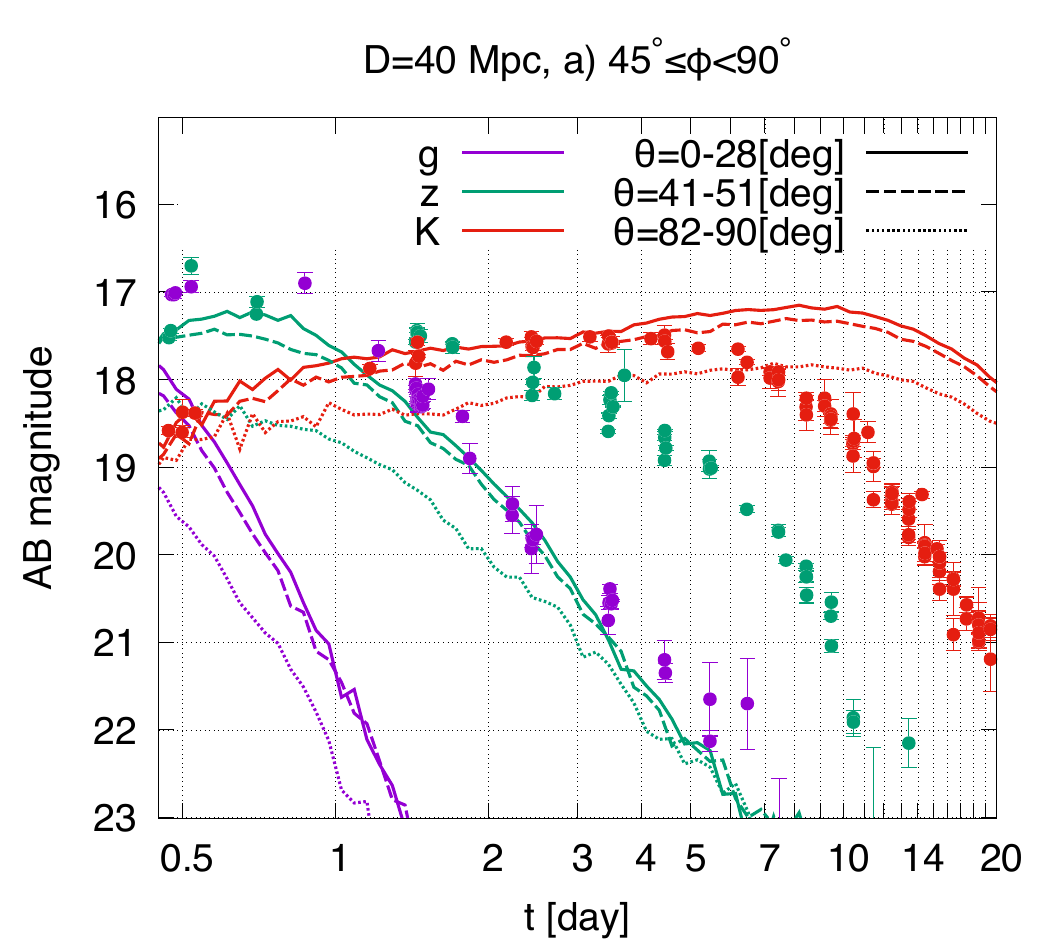}
 	 \includegraphics[width=.47\linewidth]{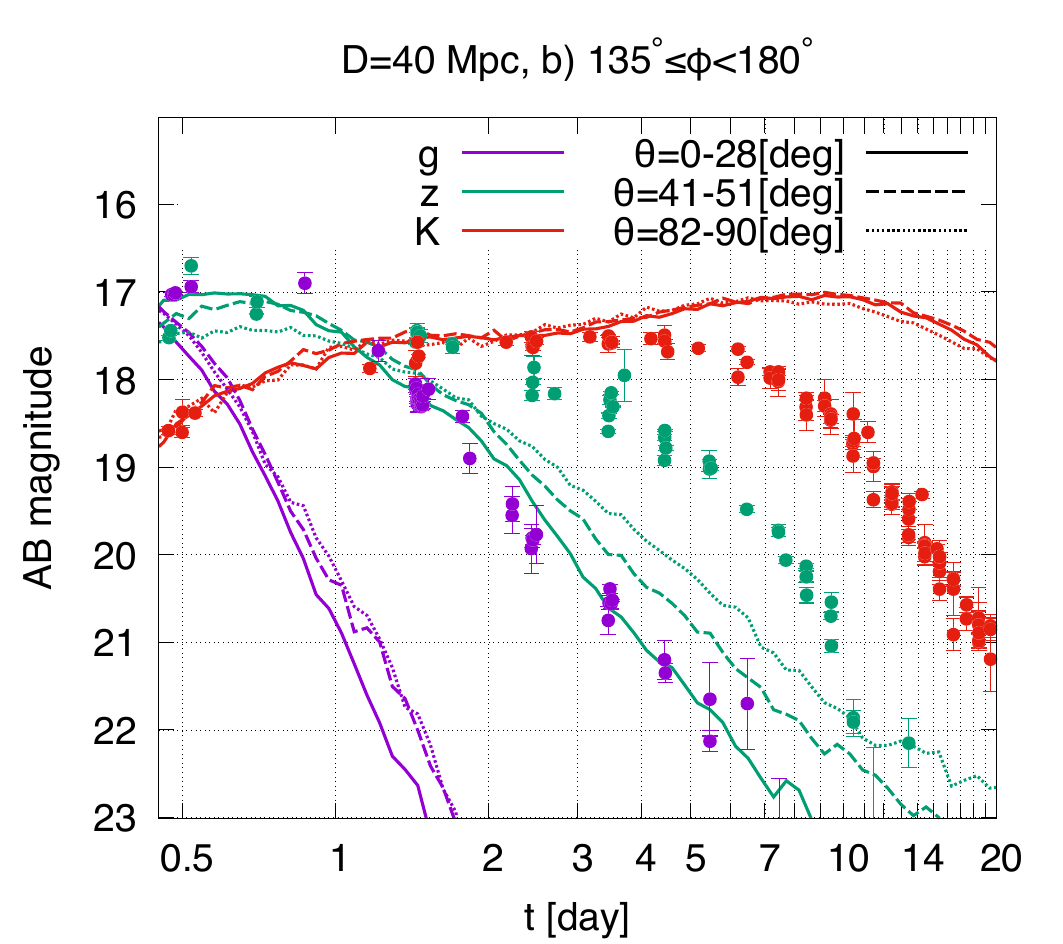}\\
 	 \includegraphics[width=.47\linewidth]{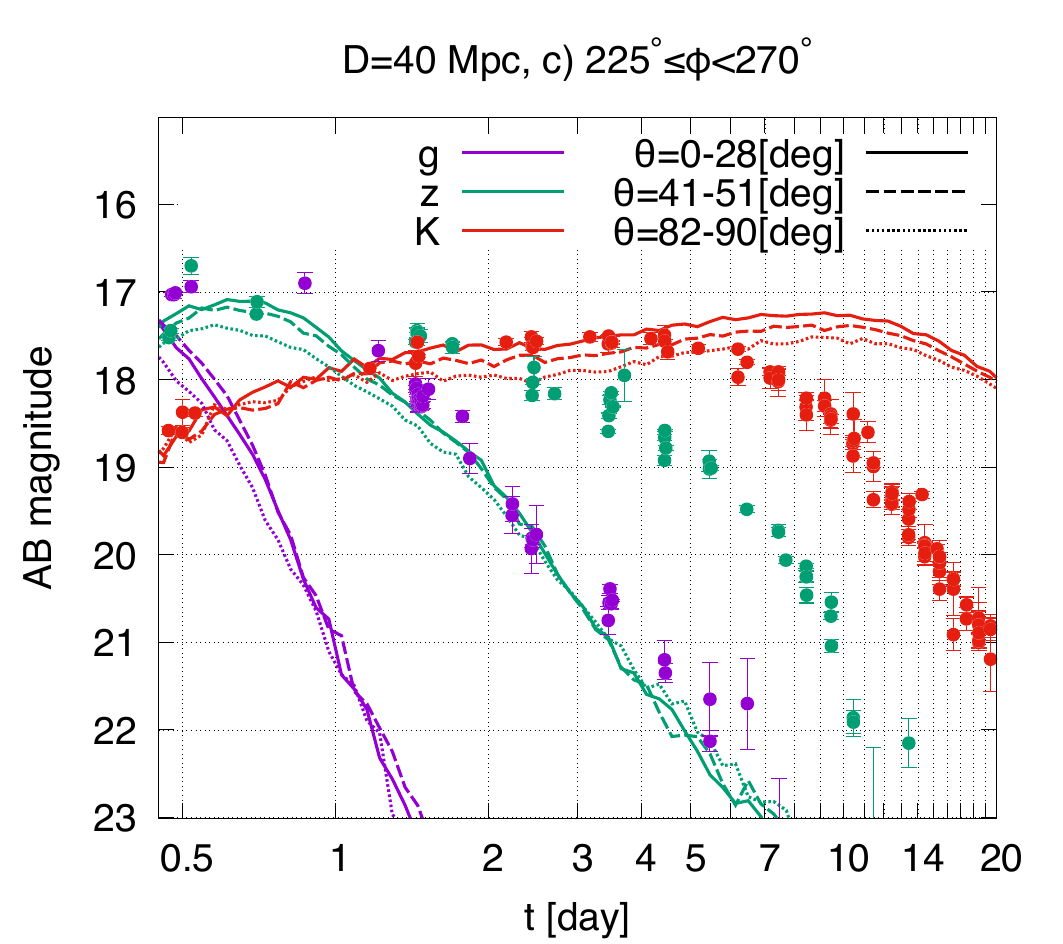}
 	 \includegraphics[width=.47\linewidth]{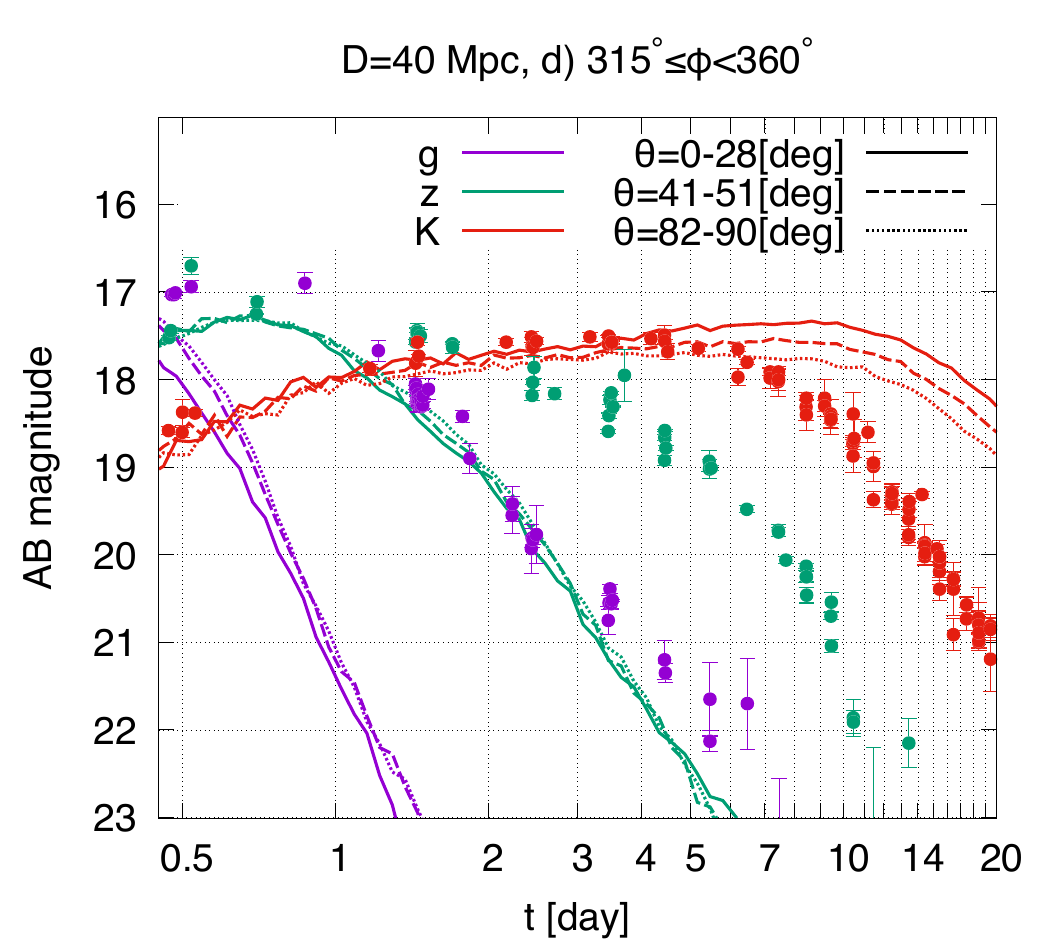}
 	 \caption{{\it gzK}-band light curves for observed from various viewing angles with the distance of 40 Mpc. The top panels compare the results among different longitudinal directions, while the middle and bottom panels compare the results among different latitudinal directions. The data points denote the AB magnitudes of AT2017gfo taken from~\citet{Villar:2017wcc}.}
	 \label{fig:mag_angle}
\end{figure*}

Figure~\ref{fig:mag_angle} shows the optical (the {\it g} and {\it z} bands) and near-infrared (the {\it K}-band) light curves observed from various viewing angles. As is the case for the bolometric luminosity, for the polar view ($\theta\le28^\circ$), the {\it gzK}-bands are brightest and faintest in {\bf b)}: $135^\circ\leq \phi < 180^\circ$ and in {\bf d)}: $315^\circ\leq \phi \leq 360^\circ$, in which the dynamical ejecta are mostly and least present, respectively. The viewing-angle dependence of the emission is weak from the polar view, and the variation is always within 0.5 mag around the peak magnitudes.

For the equatorial view ($82^\circ\leq\theta\leq90^\circ$), the {\it gzK}-band emission is brightest in {\bf b)}: $135^\circ\leq \phi < 180^\circ$, while the emission in {\bf a)}: $45^\circ\leq \phi \leq 90^\circ$ becomes faintest. The longitudinal viewing-angle dependence of the emission is significant in the {\it gz}-bands, and it is always larger than 1 mag among different longitudinal directions. The variation in the {\it K}-band magnitude among different longitudinal directions is relatively small compared to that in the {\it gz}-bands, and it is always approximately within 1 mag among the all viewing angles.

The {\it g}-band emission is fainter and declines faster than AT2017gfo even if it is observed from the brightest direction (`{\bf b)}': $135^\circ\leq \phi < 180^\circ$). The peak brightness in the {\it z}-band is comparable to the observation of AT2017gfo, but it declines much faster. In contrast, the {\it K}-band emission is comparable to that of AT2017gfo in a few days, and then, it becomes brighter after 4\,d. The {\it K}-band magnitude finally reaches its peak at $\approx10\,{\rm d}$ after the onset of the merger with its emission brighter than AT2017gfo by more than 1 mag.

Interestingly, the {\it gz}-band emission observed from {\bf b)}: $135^\circ\leq \phi < 180^\circ$ becomes slightly brighter in the equatorial direction than in the polar direction after 1\,d for the present BH-NS model. This brightness dependence on the latitudinal direction is opposite compared to the BNS KN models, for which the emission becomes brighter in the polar direction. 
The same latitudinal-angle dependence is also found for the emission observed in the direction of the ejecta bulk motion for the model in~\cite{Darbha:2021rqj} in which the radioactive heating rate broadly agrees with our model (model H4). The brighter emission in the equatorial plane is explained by the enhancement of the radiation energy flux due to the Doppler effect induced by the bulk motion of the ejecta. It is also important to have the aspect ratio of the dynamical ejecta close to unity to realize the present latitudinal-angle dependence of the emission brightness: otherwise the Doppler effect can be obscured by the suppression of the emission due to the decrease in the projected area toward the observer for the case that the dynamical ejecta have a more oblate shape (see ~\citealt{Darbha:2021rqj} for the discussion). 

\subsection{radiative-transfer effect of non-axisymmetric ejecta geometry }

\begin{figure*}
 	 \includegraphics[width=.48\linewidth]{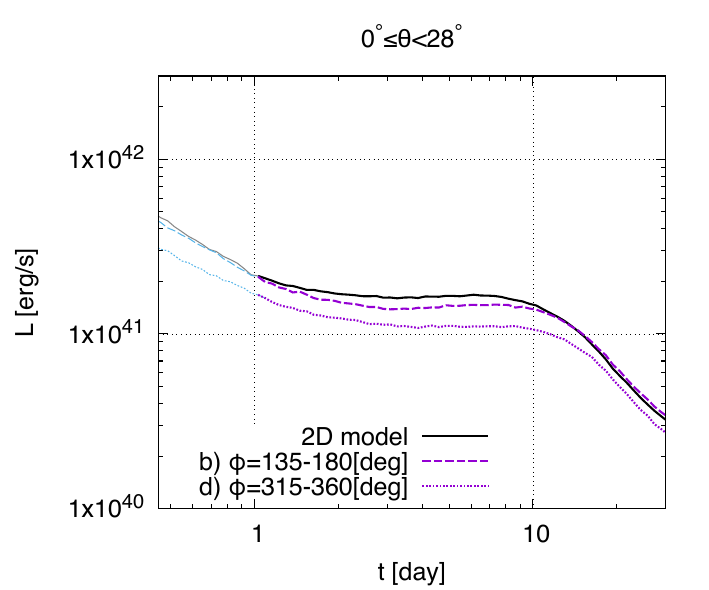}
 	 \includegraphics[width=.48\linewidth]{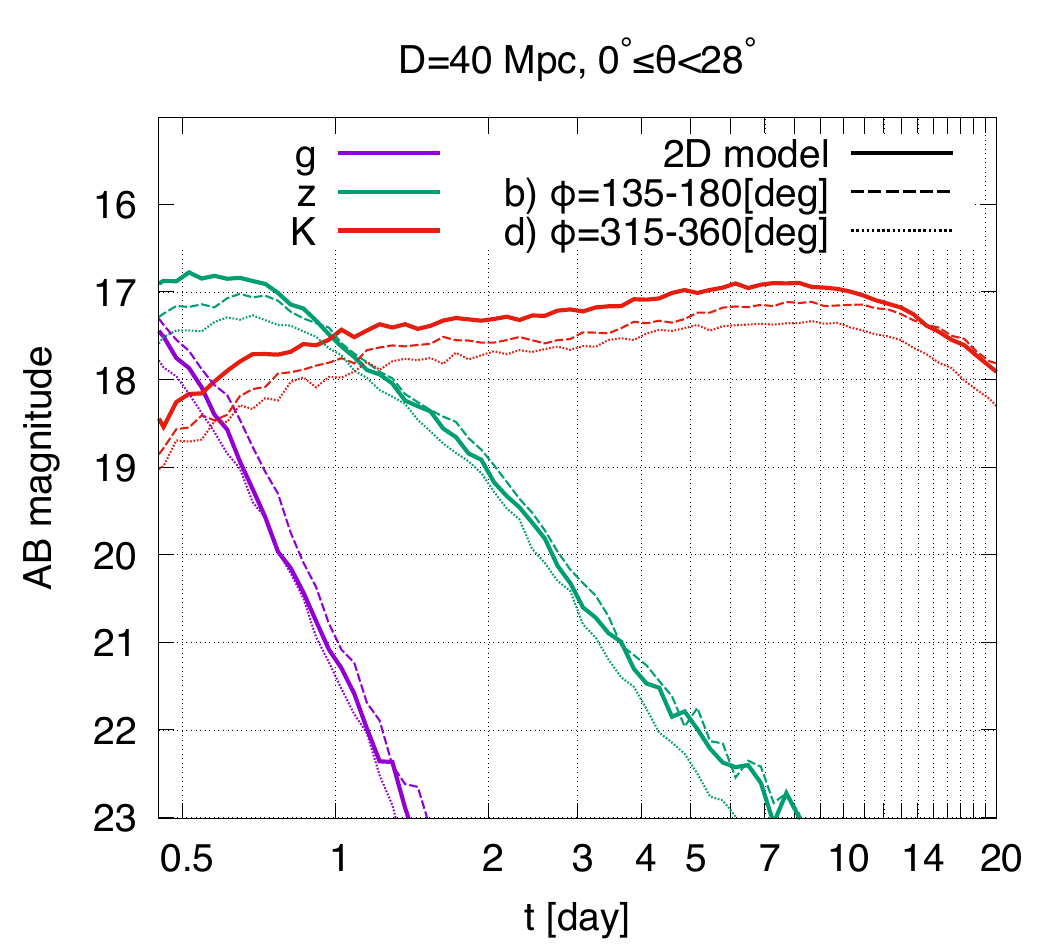}\\
 	 \includegraphics[width=.48\linewidth]{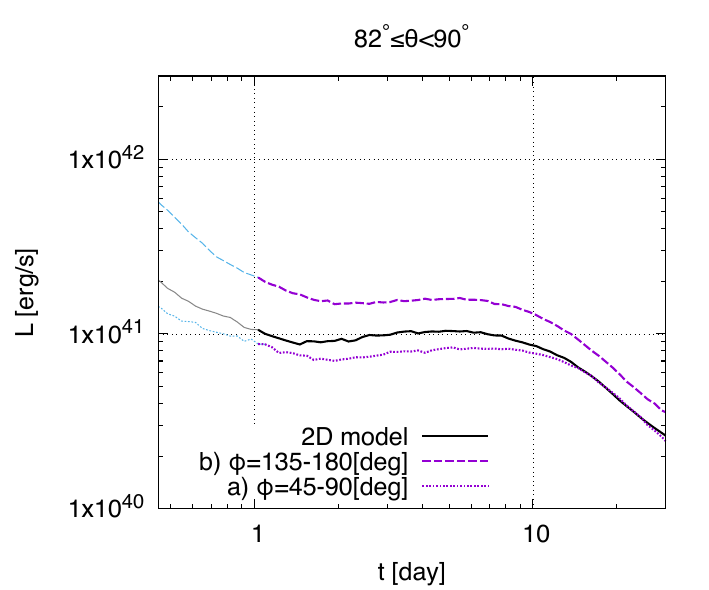}
 	 \includegraphics[width=.48\linewidth]{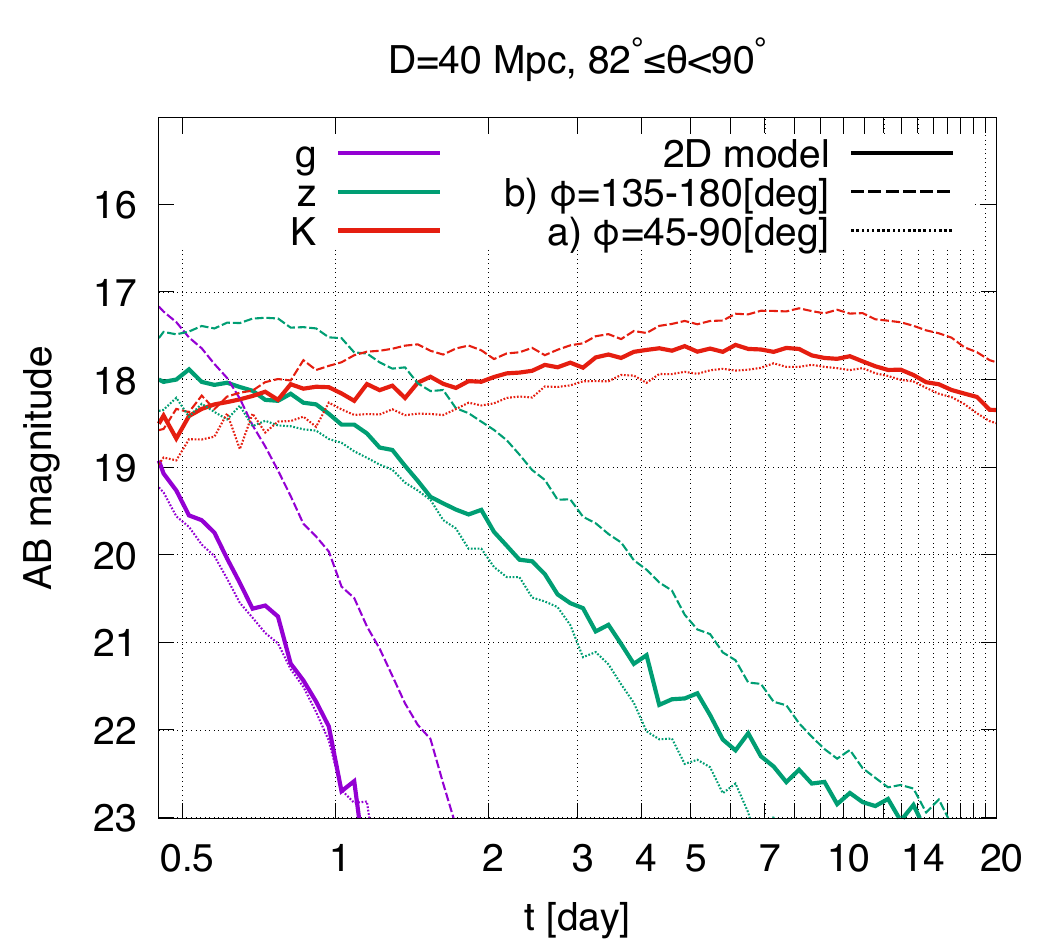}
 	 \caption{Comparison of the isotropically equivalent bolometric luminosities (left) and {\it gzK}-band light curves (right) between the fiducial (the same as in Figures~\ref{fig:lbol_angle} and~\ref{fig:mag_angle}) and axisymmetrized (labeled as ``2D model'') models. The top and bottom panels denote the light curves observed from $0^\circ\le\theta\le28^\circ$ and $82^\circ\le\theta\le90^\circ$, respectively. The solid and dashed curves denote the light curves of the axisymmetrized model and those for the fiducial model observed from {\bf b)}: $135^\circ\leq\phi<180^\circ$, respectively. The dotted curves in the upper and bottom panels denote the light curves of the fiducial model observed from {\bf d)}: $315^\circ\leq\phi<360^\circ$ and {\bf a)}: $45^\circ\leq \phi <90^\circ$, respectively.}
	 \label{fig:comp_2d}
\end{figure*}

To clarify the RT effect of the non-axisymmetric ejecta geometry, we perform a RT simulation for an axisymmetrized ejecta profile. The axisymmetrized ejecta profile is generated by averaging over the rest-mass density, specific internal energy, elemental abundances, and radioactive heating rate profiles obtained by the HD simulation at $t=0.1\,{\rm d}$ with respect to the longitudinal direction. Note that the volume and mass in each grid cell are used as the weights of the average for the rest-mass density and the latter three quantities, respectively. Figure~\ref{fig:comp_2d} compares the isotropically equivalent bolometric luminosities and {\it gzK}-band light curves observed from the polar ($0^\circ\le\theta\le28^\circ$) and equatorial ($82^\circ\le\theta\le90^\circ$) directions between the axisymmetrized and fiducial models (the same as in Figures~\ref{fig:lbol_angle} and~\ref{fig:mag_angle}). For the fiducial model, the light curves observed from {\bf b)}: $135^\circ\leq \phi < 180^\circ$, {\bf d)}: $315^\circ\leq \phi < 360^\circ$ (for the polar view), and {\bf a)}: $45^\circ\leq \phi < 90^\circ$ (for the equatorial view) are shown. 

The upper panels of Figure~\ref{fig:comp_2d} show that the KN emission observed from the polar direction becomes slightly brighter for the axisymmetrized model than for the fiducial model except for the {\it g}-band emission. This reflects the fact that the area of the dynamical ejecta projected toward the observer increases for the axisymmetrized model due to the longitudinal average. The bolometric light curves declines slightly earlier than the original fiducial model because the optical depth decreases due to the decrease in the rest-mass density of the dynamical ejecta for the axisymmetrized model. Nevertheless, the effect of the longitudinal average is found to be minor for the polar view, particularly, in the {\it gzK}-band magnitudes, for which the differences between the axisymmetrized model and fiducial model are always smaller than $0.5$ mag for $t\ge0.5\,{\rm d}$.

The difference in the brightness between the axisymmetrized model and fiducial model is more pronounced for the emission observed from the equatorial direction. For the equatorial view, the bolometric luminosity observed from the longitudinal direction of the brightest region (`{\bf b)}': $135^\circ\leq \phi < 180^\circ$) is brighter approximately by a factor of 1.5--2 for the fiducial model than that for the axisymmetrized model. The {\it gz}-band magnitudes observed from the same direction are also brighter than those for the axisymmetrized model by $\sim 1\,{\rm mag}$. On the other hand, the KN brightness observed from the longitudinal direction of the faintest region (`{\bf a)}': $45^\circ\leq \phi < 90^\circ$) is comparable to or slightly fainter for the fiducial model than that for the axisymmetrized model. This discrepancy in the equatorial brightness between the fiducial and axisymmetrized models is due to the difference in the ejecta aspect ratio: as a consequence of the longitudinal average, the polar projected area of the ejecta is larger than that in the equatorial direction for the axisymmetrized model, which makes photons to preferentially diffuse in the polar direction and thus the equatorial brightness to be fainter.

\subsection{Comparison with various BNS KN models}


\begin{figure*}
 	 \includegraphics[width=0.48\linewidth]{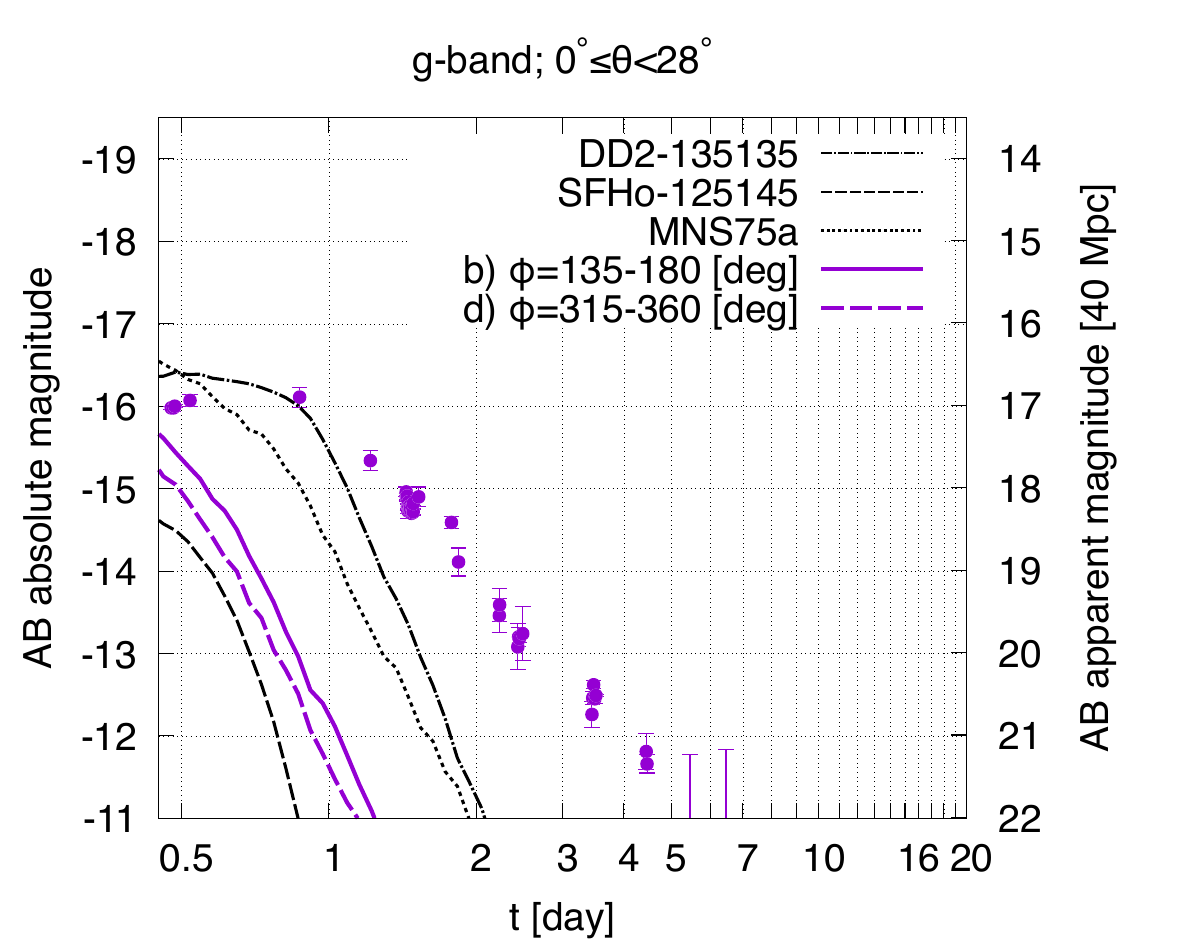}
 	 \includegraphics[width=0.48\linewidth]{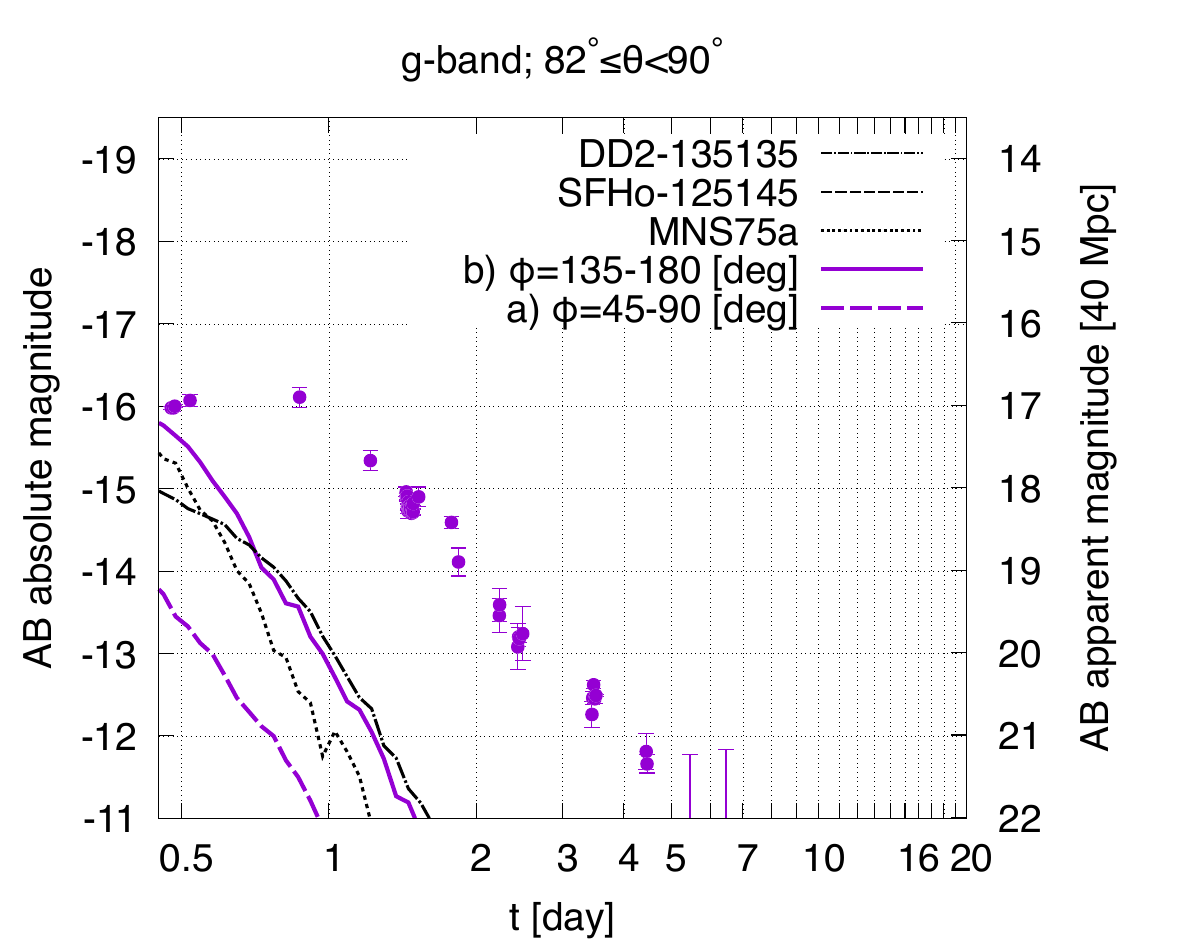}\\
 	 \includegraphics[width=0.48\linewidth]{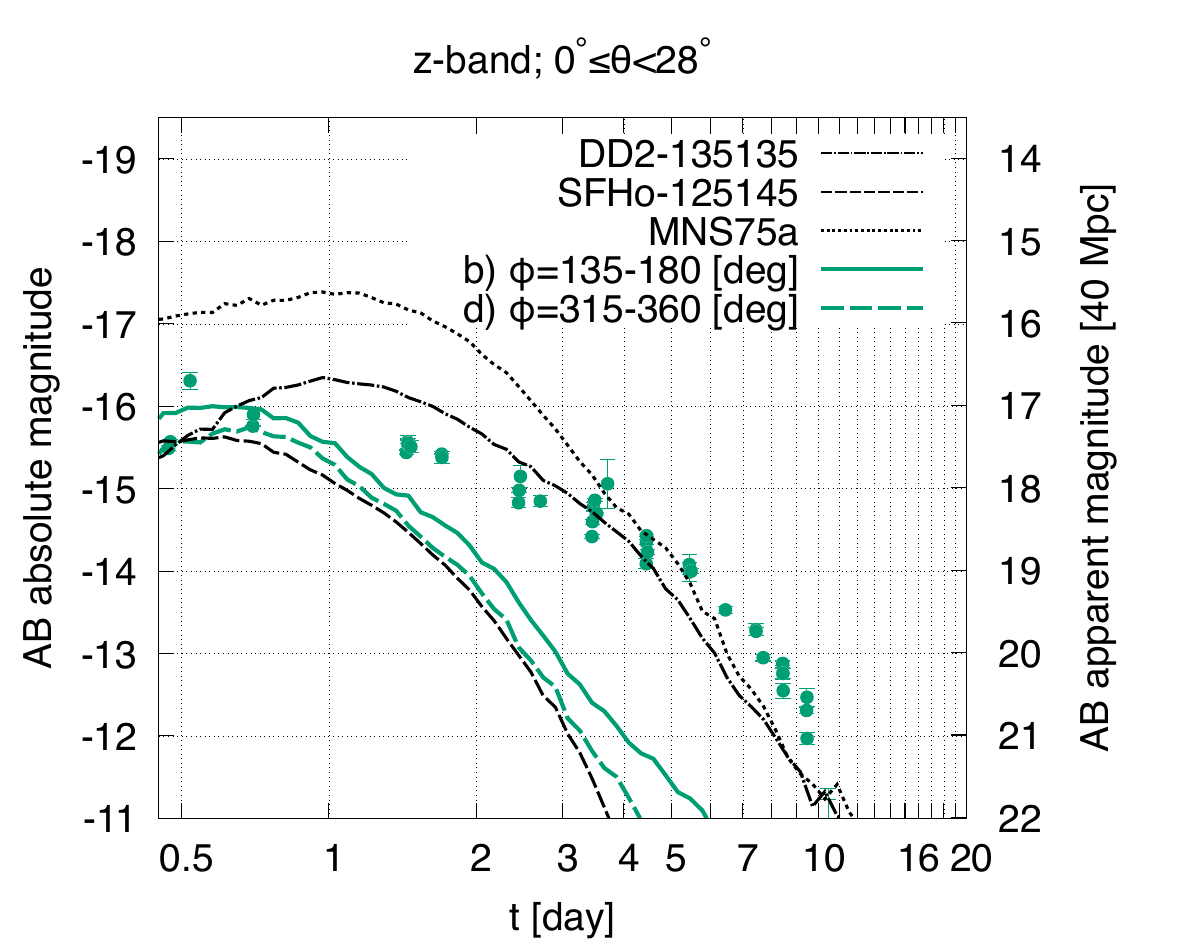}
 	 \includegraphics[width=0.48\linewidth]{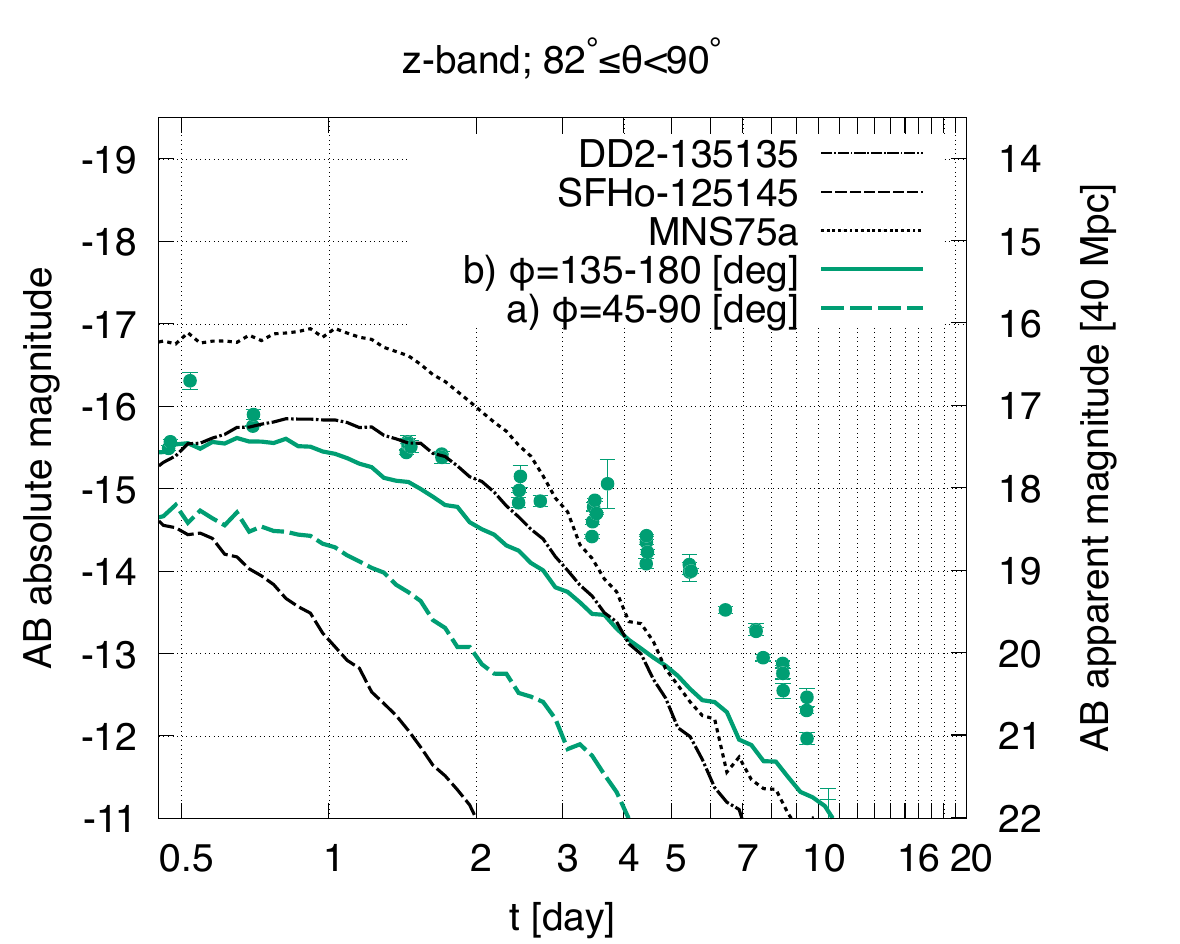}\\
 	 \includegraphics[width=0.48\linewidth]{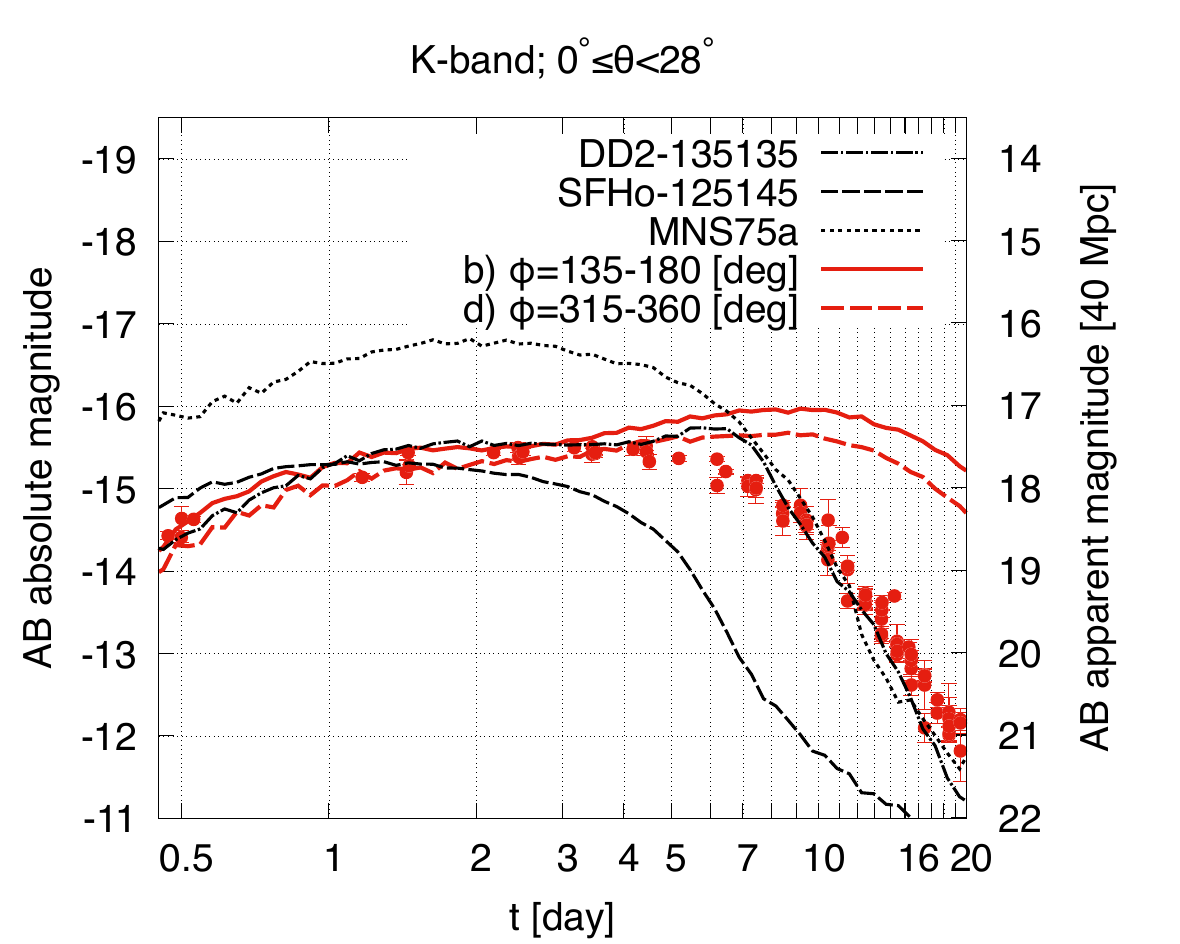}
 	 \includegraphics[width=0.48\linewidth]{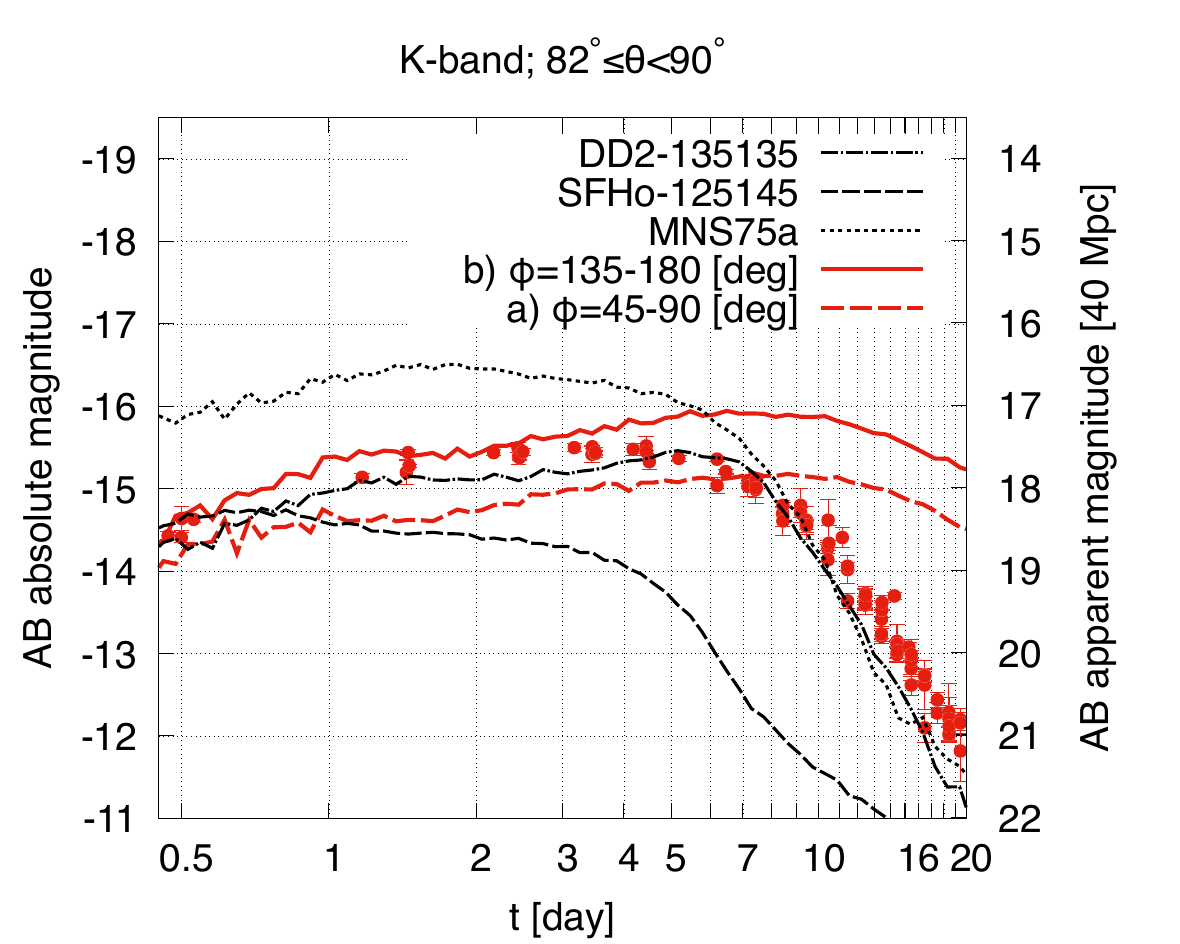}
 	 \caption{Comparison of the bolometric and {\it gzK}-band light curves among the present BH-NS KN model and various BNS KN models. For BNS KN models, three cases are shown: a case in which the remnant MNS survives for a short time (the dashed curves; SFHo-125145,~\citealt{Kiuchi:2022ubj,Fujibayashi:2022ftg,Kawaguchi:2023zln}), a case in which the remnant MNS survives for a long time (the dash-dotted curves; DD2-135135, ~\citealt{Fujibayashi:2020dvr,Kawaguchi:2022bub}), and a case in which large-scale magnetic field significantly plays a role in the long-surviving remnant MNS (the dotted curves; MNS75a, ~\citealt{Shibata:2021xmo,Kawaguchi:2022bub}). Note that the light curve for model SFHo-125145 in the top-right panel is below the plot range. Note also that due to different angle binning setups, the light curves observed from $0^\circ\le\theta\le20^\circ$ and $86^\circ\le\theta\le90^\circ$ are used for BNS KN models to compare with the BH-NS KN light curves observed from $0^\circ\le\theta\le28^\circ$ and $82^\circ\le\theta\le90^\circ$, respectively. The data points denote the AB magnitudes of AT2017gfo taken from~\citet{Villar:2017wcc}.}
	 \label{fig:mag_comp}
\end{figure*}

Figure~\ref{fig:mag_comp} compares the {\it gzK}-band light curves among the present BH-NS KN model and various BNS KN models obtained in our previous studies~\citep{Kawaguchi:2020vbf,Kawaguchi:2022bub,Kawaguchi:2023zln}. For BNS KN models, three cases are shown as representative: a case in which the remnant massive NS (MNS) survives for a short time (the dashed curves; SFHo-125145,~\citealt{Kiuchi:2022ubj,Fujibayashi:2022ftg,Kawaguchi:2023zln}), a case in which the remnant MNS survives for a long time (the dash-dot curves; DD2-135135, ~\citealt{Fujibayashi:2020dvr,Kawaguchi:2022bub}), and a case in which large-scale magnetic field significantly plays a role in the long-surviving remnant MNS (the dotted curves; MNS75a, ~\citealt{Shibata:2021xmo,Kawaguchi:2022bub}). 

For model SFHo-125145, the ejecta mass and average velocity are $0.012\,M_\odot$ and $0.17\,c$, respectively. The dynamical and post-merger ejecta have approximately the same mass and exhibit approximately spherical and mildly prolate shapes, respectively. While the $Y_e$ value of the post-merger ejecta is $\gtrsim 0.3$ and shows a weak spatial dependence, the dynamical ejecta have a significant $Y_e$ dependence on the latitudinal angle with the value being higher and lower than 0.3 for $\theta\lesssim 45^\circ$ and $\gtrsim 45^\circ$, respectively. For model DD2-135135, the ejecta mass and average velocity are $0.065\,M_\odot$ and $0.094\,c$, respectively. The morphology and $Y_e$ profiles of the dynamical and post-merger ejecta are similar to those in model SFHo-125145, but the dynamical ejecta mass is $1.5\times10^{-3}\,M_\odot$, and the post-merger ejecta dominate the total ejecta mass. For model MNS75a, the same BNS configuration as in model DD2-135135 is considered, but the post-merger ejecta are significantly accelerated by taking into account the magneto-centrifugal effect. As the consequence, the ejecta mass and average velocity are increased compared to those in model DD2-135135 and are $0.084\,M_\odot$ and $0.42\,c$, respectively. 

We note that the BNS KN models are obtained by imposing axisymmetry in all the post-merger NR simulations, subsequent HD simulations, and RT simulations. For the BH-NS model, we show the light curves observed from {\bf b)}: $135^\circ\leq \phi < 180^\circ$, {\bf d)}: $315^\circ\leq \phi < 360^\circ$ (for the polar view), and {\bf a)}: $45^\circ\leq \phi < 90^\circ$ (for the equatorial view), which represent the longitudinal directions for the brightest and faintest emission, respectively.

The {\it gz}-band emission observed from the polar direction ($0^\circ\le \theta \le 28^\circ$) for the present BH-NS KN model is by 0.5--1 mag brighter than that for the BNS models in which the remnant MNS survives only for a short time ($<10\,{\rm ms}$, SFHo-125145), but is by more than $\approx$ 1 mag fainter than that for the BNS models in which the remnant MNS survives for a long time ($>1\,{\rm s}$, DD2-135135 and MNS75a). On the other hand, the {\it gz}-band emission observed from the equatorial direction ($82^\circ\le \theta \le 90^\circ$) is comparably bright or brighter than those for the BNS models in which remnant MNSs survive for a long time except for the {\it z}-band emission of model MNS75a. This is due to the fact that the BNS KN models show stronger latitudinal viewing-angle dependence than the BH-NS KN model and become significantly faint in the equatorial view. The difference in the latitudinal viewing-angle dependence reflects the fact that the dynamical ejecta are the primary source of the emission in the optical wavelength for the BH-NS model, while for the BNS models, the post-merger ejecta are main source of the emission and the dynamical ejecta are mostly acting as the opacity source rather than the emission source (lanthanide curtain effect;~\citealt{Kasen:2014toa,Kawaguchi:2018ptg,Kawaguchi:2019nju,Bulla:2019muo,Zhu:2020inc,Darbha:2020lhz,Korobkin:2020spe}). We note that, in the BNS cases, enhancement of the brightness due to the Doppler effect is obscured by the latitudinal-angle dependence of the emission induced by the angle-dependent opacity of the dynamical ejecta. 

The {\it K}-band emission for the present BH-NS KN model has comparable peak brightness to that for the BNS models without significant large-scale magnetic field effect in the remnant NS (SFHo-125145 and DD2-135135). However, it is only the BH-NS model that maintains the {\it K}-band brightness within 1 mag of its peak for a two-week period. The BNS models in which the remnant MNSs survive for short and long periods of time become fainter than the BH-NS model after $1$--$2\,{\rm d}$ and $5$--$7\,{\rm d}$, respectively. The BNS model in which large-scale magnetic field significantly plays a role in the remnant NS shows bright {\it K}-band emission for a week, but the brightness declines much faster than that for the BH-NS model. Hence, the observation of a KN with long-lasting near-infrared emission which is bright for more than two weeks will indicate that the progenitor of a KN is a BH-NS merger with massive ejecta (in particular dynamical ejecta) formation.

\section{Summary and Discussions}\label{sec:discuss}

In this paper, we studied the long-term evolution of the matter ejected in a BH-NS merger by employing the results of the NR simulation and nucleosynthesis calculation, in which both dynamical and post-merger ejecta are followed consistently. In particular, we employed the results for the merger of a $1.35\,M_\odot$ NS and $5.4\,M_\odot$ BH with the dimensionless spin of 0.75. We confirmed the finding in the previous studies that, thermal pressure induced by radioactive heating in the ejecta could significantly modify the morphology of the ejecta. In our studied case for the BH-NS binary, the dynamical ejecta expand significantly and the aspect ratio becomes close to unity with the fine structure being smeared out in the presence of radioactive heating. On the other hand, the post-merger ejecta were compressed and confined in the region with the radial velocity $\lesssim 0.05\,c$ due to the significant expansion of the dynamical component.

We then computed the KN light curves employing the ejecta profile obtained by the HD simulation of the ejecta matter. We found that our present BH-NS model results in KN light curves that are fainter but longer lasting than those observed in AT2017gfo, reflecting the fact that the emission is primarily powered by the lanthanide-rich massive dynamical ejecta. The optical-band emission is comparable to or fainter than those for the various BNS models obtained in our previous studies. While the peak brightness of the near-infrared emission is also comparable to the BNS models, the time-scale maintaining the brightness is much longer, and the emission comparable to the peak brightness within 1\,mag is sustained for more than two weeks for the BH-NS model. The wide-field infrared observations with the ground-based telescopes, such as VISTA~\citep{Ackley:2020qkz}, WINTER~\citep{Frostig:2021vkt} and PRIME~\citep{2023AJ....165..254K}, can detect such bright infrared KN emission up to $\approx 14$\,d if the distance to the event is within 150\,Mpc since the {\it K}-band emission will be apparently brighter than 21\,mag for all the viewing angles. However, the field of view of infrared telescopes is typically not as large as those for the optical telescopes for a given sensitivity~\citep{Nissanke:2012dj}. Therefore, a tight constraint of the localization area by the GW data analysis or the follow-up observation within $\approx 1$\,d in the optical bands is crucial to detect the KN emission unless the event occurs as close as in the case of AT2017gfo. 

We found that the non-axisymmetric geometry of the ejecta induces various interesting radiative-transfer effects in the viewing-angle dependence of the KN emission. In particular, we found the Doppler effect induced by the bulk velocity of the ejecta to the emission, which is pointed out by~\cite{Fernandez:2016sbf} and~\cite{Darbha:2021rqj}, is in fact appreciable. Due to this effect, the optical light curves observed from the direction of the bulk ejecta motion show a slightly inverted latitudinal angle dependence to those found in the BNS models: The optical-band emission observed from {\bf b)}: $135^\circ\leq \phi < 180^\circ$ becomes slightly brighter in the equatorial direction than in the polar direction for the present BH-NS model. Since the KNe emission becomes fainter in the equatorial direction than in the polar direction for BNS mergers, our results suggest that, for the edge-on view, the KN emission for BH-NS mergers can be brighter in the optical-band than that of BNS mergers.

Our results indicate that the long-lasting near-infrared emission is the key to distinguish the types of progenitors by the KN observation. If the {\it K}-band emission of which brightness comparable to its peak is maintained for more than two weeks, it may indicate that the progenitor is a BH-NS merger with massive ejecta formation. This is consistent with our finding in the previous study~\citep{Kawaguchi:2019nju}. On the other hand, the BH-NS KN light curves in the optical band is similar to those associated with BNS mergers, and hence, it may be difficult to infer the information of the progenitor only from the optical emission. We should note that the ejecta mass and hence the brightness of the KN of BH-NS mergers can have large variety depending on the binary parameters, such as the BH and NS masses, BH spin, and NS radius~\citep{Rosswog:2005su,Shibata:2007zm,Etienne:2008re,Lovelace:2013vma,Kyutoku:2015gda,Foucart:2018rjc}, as well as on the adopted EOS~\citep{Hayashi:2022cdq}. We also note that the assumption of LTE employed in our RT simulation will not be valid in the region where the rest-mass density has been significantly dropped. It is implied in~\cite{Hotokezaka:2021ofe,Pognan2022MNRAS} that the matter temperature can be higher than that estimated under the assumption of LTE if non-LTE effects take place. In such a case, the infrared emission can be dimmer with the combination of the suppression of the neutral and poorly ionized atoms~\citep{Kawaguchi:2020vbf,Kawaguchi:2022bub,Kawaguchi:2023zln}. We should keep in mind that the absence of the long-lasting bright near-infrared emission does not necessarily rule out the possibility that the progenitor of the observed KN is a BH-NS merger.

So far, four candidates have been reported for BH-NS GW events: GW190814~\citep{LIGOScientific:2020zkf}, GW200105/GW200115~\citep{LIGOScientific:2021qlt}, and GW230529~\citep{LIGOScientific:2024elc}. Among them, according to the inferred masses and spins of the binary, the latest GW event, GW230529, was most likely to be accompanied by EM counterparts. Unfortunately, the EM counterpart was not found in GW230529 due to the poorly constrained sky localization, although the luminosity distance to the event was relatively close ($201^{+102}_{-96}\,{\rm Mpc}$ with the error bar being the 90\% credible intervals). Nevertheless, the discovery of this system increases the expected rate of the GW detection of a BH-NS merger with EM counterparts in the future. For this event, (under the assumption that this event was a BH-NS merger) the ratio of the BH mass to the NS mass and the dimensionless BH spin were less than $\approx 4$ and larger than $\approx0$, respectively. For such a case of a BH-NS merger, the ratio of the post-merger ejecta mass to the dynamical one can be larger compared to the BH-NS model which we studied in this paper~\citep{Hayashi:2020zmn}. This indicates that the resulting KN may become bluer than the present result, while it is not always trivial since the long-term hydrodynamics evolution of ejecta may also differ. Hence, the systematic study on the KNe for various configurations of BH-NS binaries would be crucial to quantitatively interpret the EM observational data in the future.

\begin{figure*}
 	 \includegraphics[width=.45\linewidth]{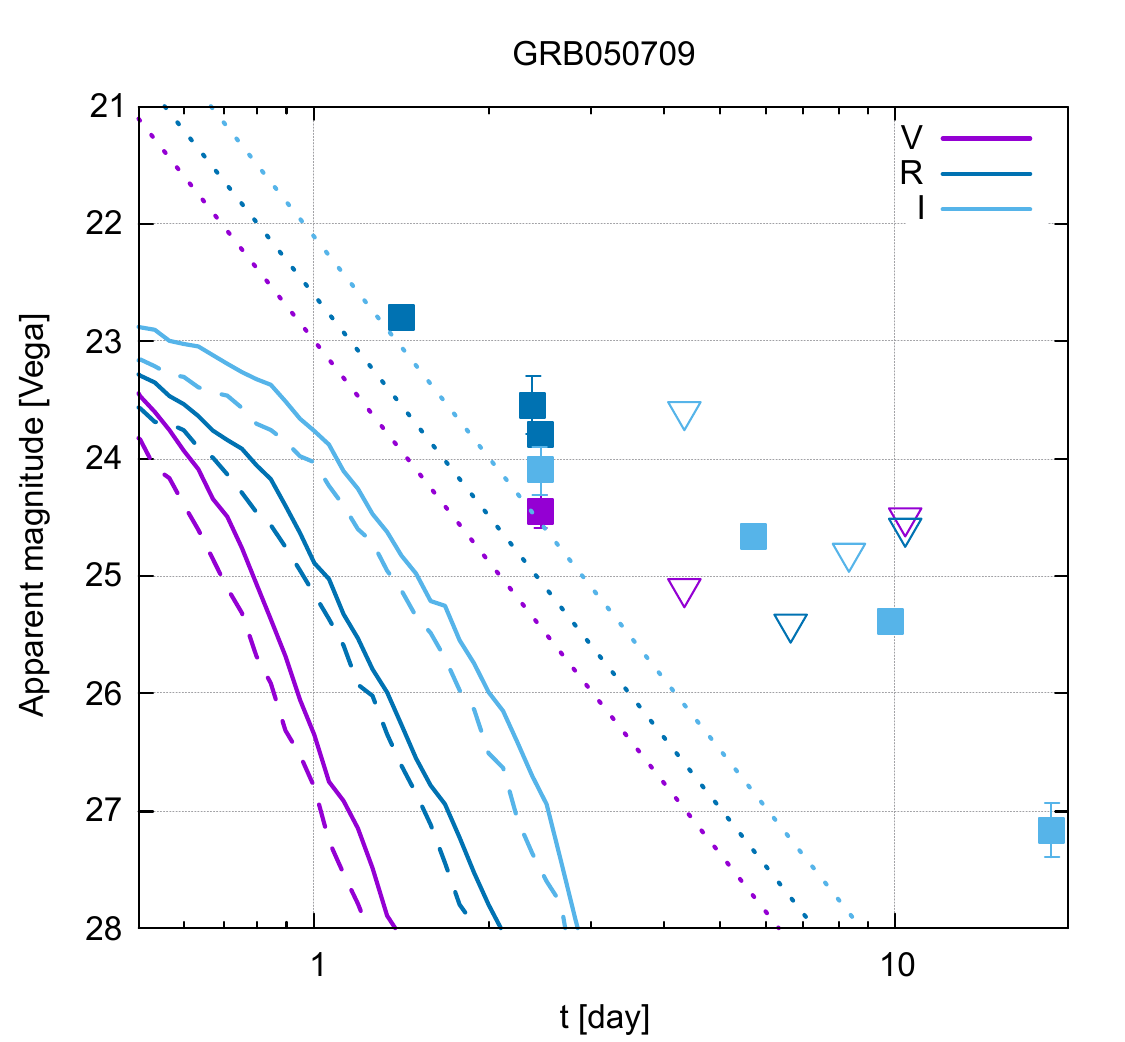}
 	 \includegraphics[width=.45\linewidth]{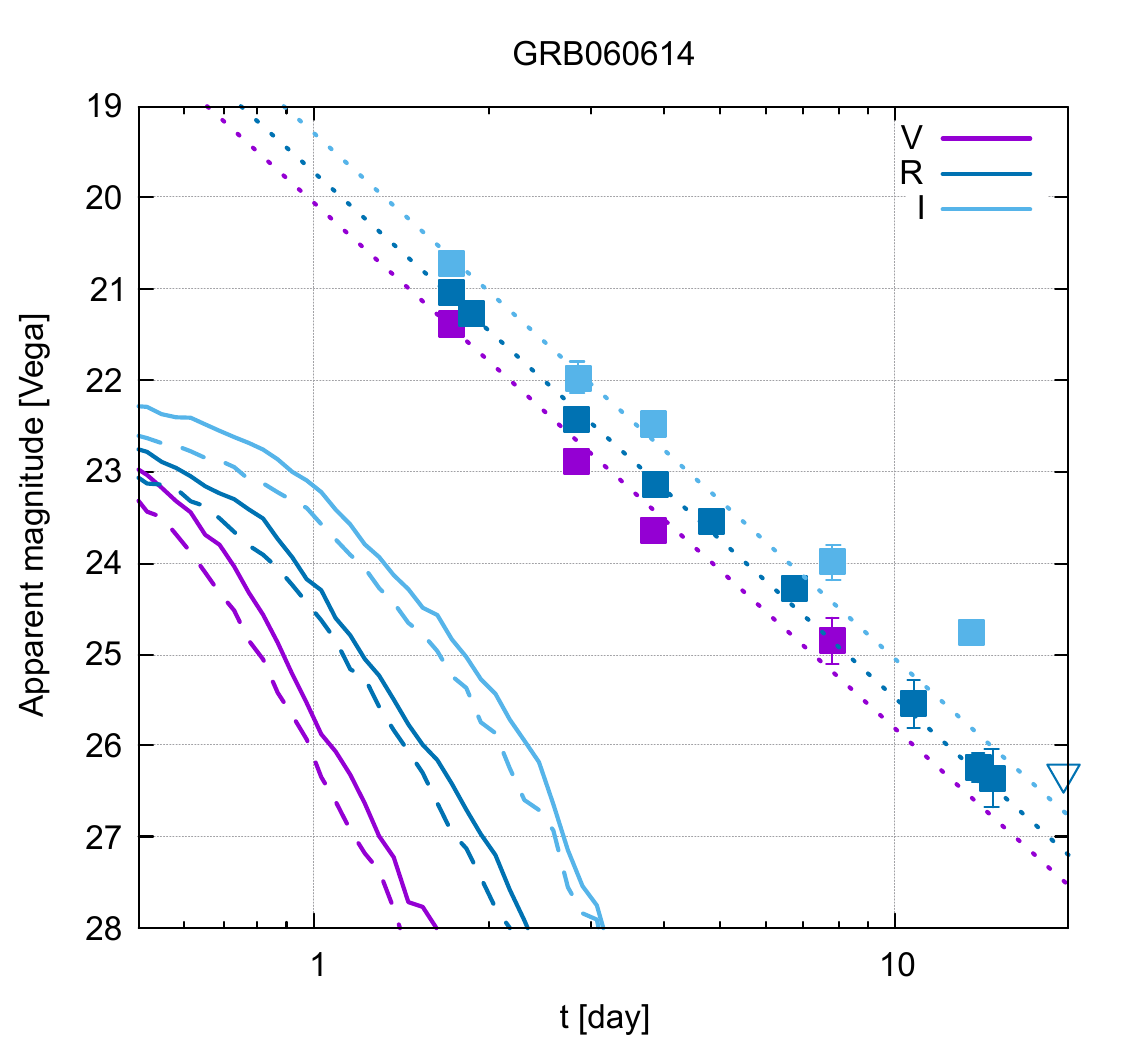}\\
 	 \includegraphics[width=.45\linewidth]{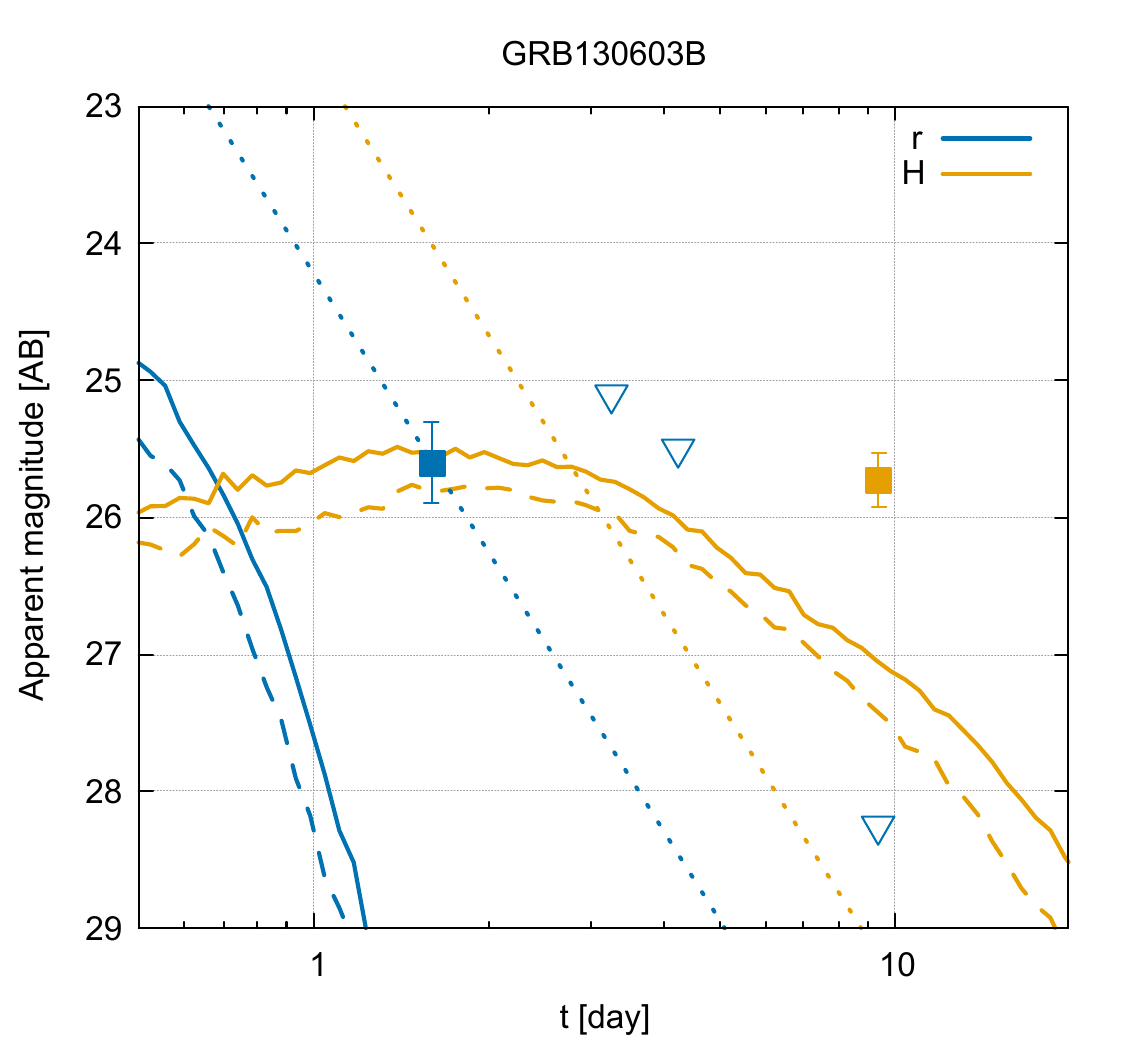}
 	 \includegraphics[width=.45\linewidth]{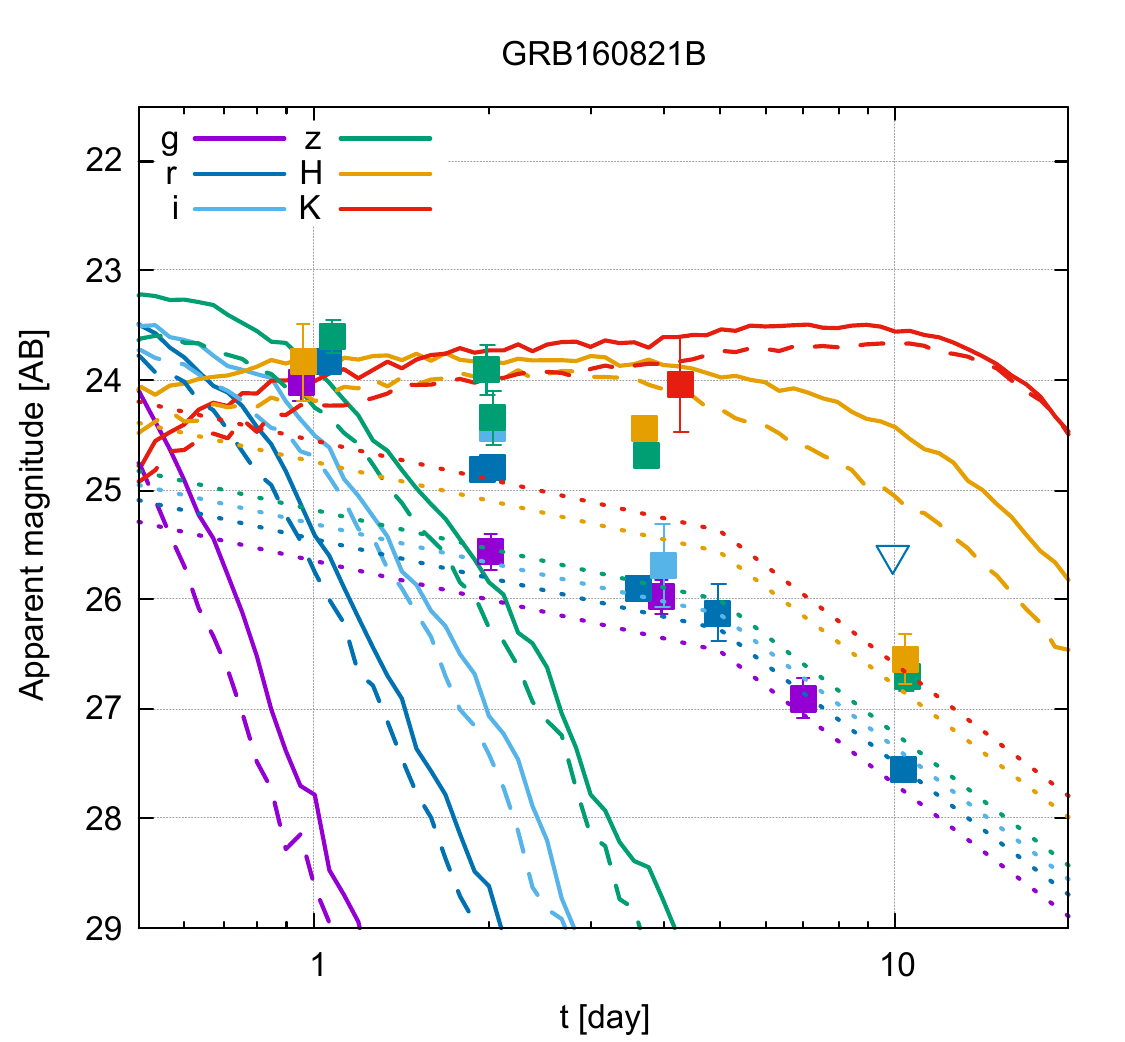}\\
 	 \includegraphics[width=.45\linewidth]{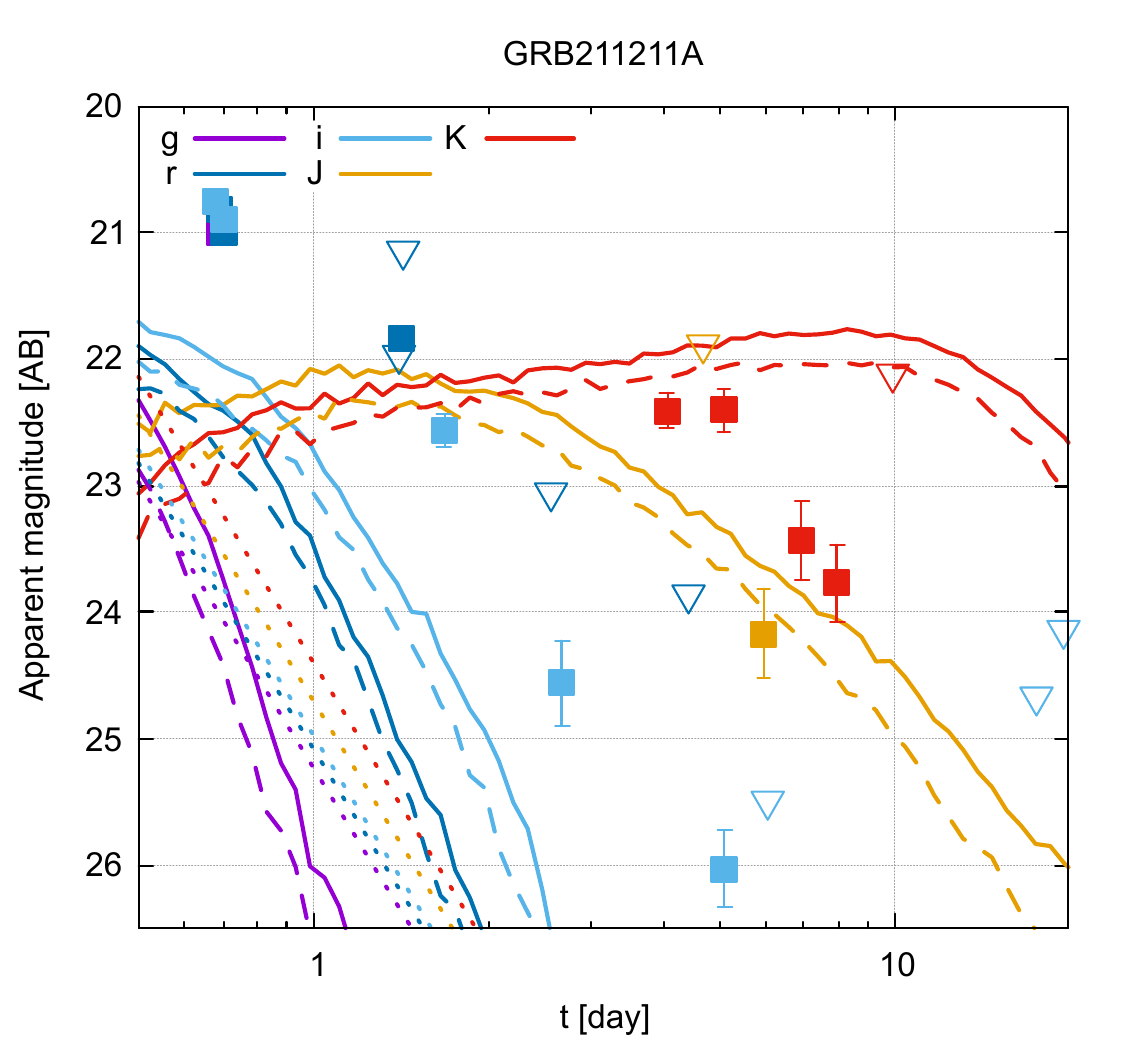}
 	 \includegraphics[width=.45\linewidth]{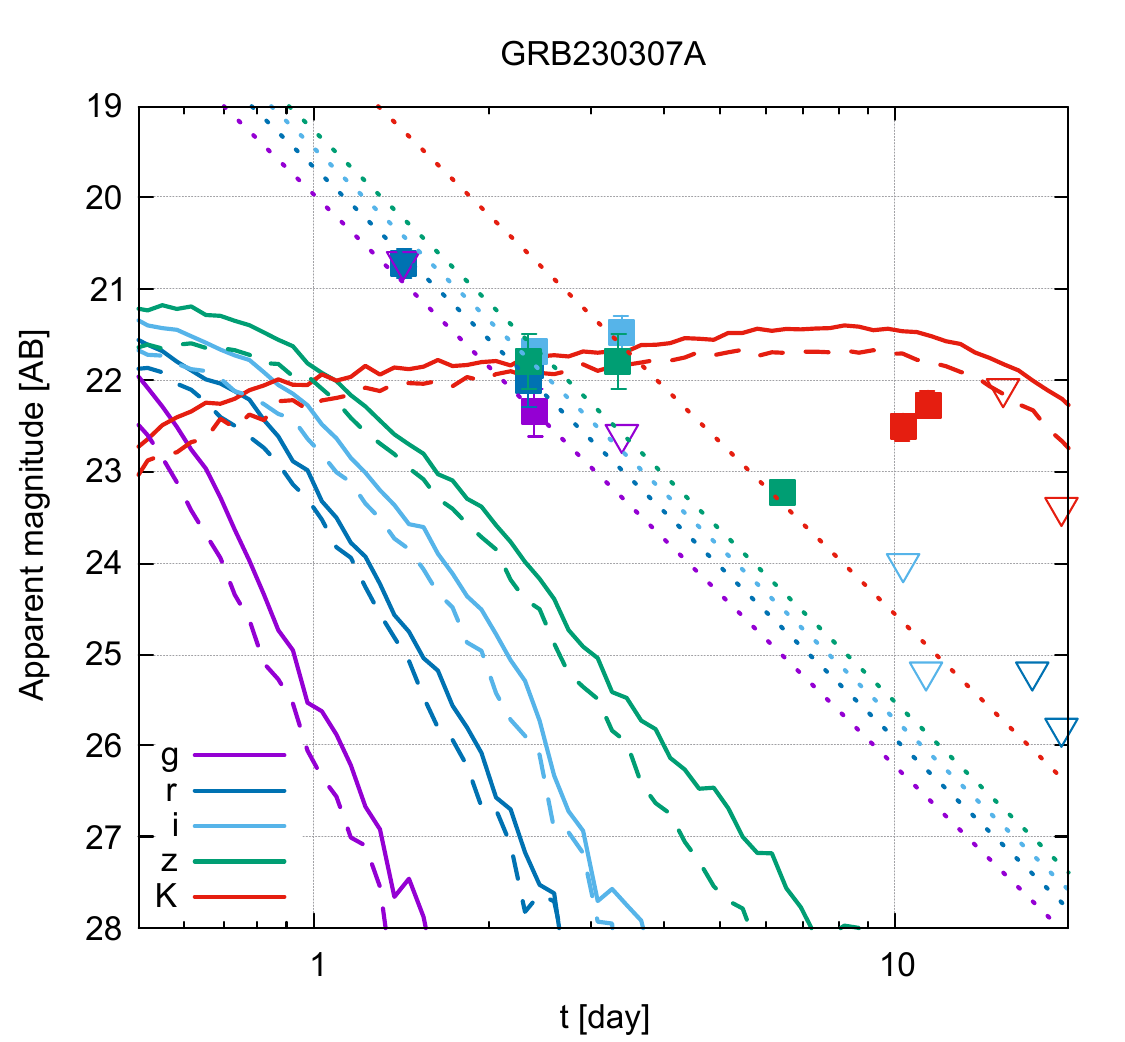}
 	 \caption{Comparison between the present BH-NS KN model and GRB KN candidates. The solid and dashed curves denote the polar light curves in the observer frame ($0^\circ\le\theta\le28^\circ$ ) for the present BH-NS KN model observed from {\bf b)}: $135^\circ\leq\phi<180^\circ$ and {\bf d)}: $315^\circ\leq\phi<360^\circ$, respectively. The square and triangle symbols denote, respectively, the observed magnitudes and upper-limits of the optical and near-infrared counterparts of GRBs taken from~\citet{GRB050709,GRB060614J,GRB060614Y,GRB130603BB,GRB130603BT,GRB160821BG,GRB160821BT,GRB211211AR,GRB211211AT,GRB211211AG,GRB230307AL}. The afterglow models which broadly reproduce the models in the literature are also plotted in the dotted curves.}
	 \label{fig:comp_GRB}
\end{figure*}

There are a number of KN candidates reported which are associated with the observation of GRBs: GRB050709~\citep{GRB050709}, GRB060614~\citep{GRB060614J,GRB060614Y}, GRB130603B~\citep{GRB130603BB,GRB130603BT}, GRB160821B~\citep{GRB160821BG,GRB160821BT}, GRB211211A~\citep{GRB211211AR,GRB211211AT,GRB211211AG}, and GRB230307A~\citep{GRB230307AL}. In Figure~\ref{fig:comp_GRB}, we compare our present BH-NS KN model with these observational data. The optical and near-infrared brightness of KN candidates found in GRB211211A and GRB230311 are comparable to that of AT2017gfo. We find that our present BH-NS KN model is too faint to explain the optical brightness of these KN candidates at a few days, while the {\it K}-band emission of our present BH-NS KN model after 4\,d is too bright to be consistent with the later time observations and upper limits. Our present BH-NS KN model is also too faint in the optical bands to explain the KN candidates found in GRB050709, GRB060614, and GRB160821B after $\approx2$\,d. The {\it K}-band brightness of GRB160821B at 4.3\,d is comparable to that of our present BH-NS KN model, while the BNS model in which the remnant MNS survives for a long time ($>1\,{\rm s}$, DD2-135135 in Figure~\ref{fig:mag_comp}) also has comparable {\it K}-band brightness at that epoch. Interestingly, despite the bright and long-lasting {\it K}-band emission, our present BH-NS model has fainter {\it H}-band emission at $\sim10\,{\rm d}$ than that observed in GRB130603B. This is due to the fact that the KN of our present BH-NS model is very red and the peak of the red-shifted spectrum is located in the wavelength longer than the {\it H}-band at that epoch. In summary, currently we do not find a KN candidate that can be only explained by our present BH-NS model with significant dynamical ejecta formation. However, as we mentioned above, we cannot rule out the possibility that some of those KN candidates are KNe associated with BH-NS mergers, since BH-NS KNe can have a large diversity reflecting the variety of binary parameters as well as the adopted EOS.

We found more than a factor of 2 variation in the KN brightness depending on the viewing angle in our present BH-NS model. Such a variation in the brightness can induce the same degree of the systematic error in conducting the ejecta parameter estimation for the ejecta properties, such as the mass, velocity, and effective ejecta opacity, from the observational data. We should also note that we only focus on one single case of a BH-NS merger with the DD2 EOS, and it is not clear whether the property of KN is always the same for other setups of BH-NS mergers. For example, if the longitudinal opening angle of the ejecta is close to $2\pi$, the BH-NS KN can have the viewing-angle dependence in the brightness comparably strong to those of BNS mergers as we indeed see in the results of the axisymmetrized model (see Fig.~\ref{fig:comp_2d}). We further should note that uncomprehended systematic errors in the opacity and heating rate can induce large systematic errors in the ejecta parameter inference. In particular, the latter can be severe for KNe from BH-NS mergers since the uncertainty is more significant for the ejecta with low values of $Y_e$ ($<0.24$, see ~\citealt{Barnes:2020nfi,Zhu:2020eyk}). Hence, it is essential to consider that these systematic errors can significantly alter the results of the ejecta parameter inference, and those estimated values should be used with a great caution.

\section*{Acknowledgements}
We thank the anonymous referee for the careful reading of our manuscript and valuable comments. Numerical computation was performed on Yukawa21 at Yukawa Institute for Theoretical Physics, Kyoto University and the Yamazaki, Sakura, Momiji, Cobra, and Raven clusters at Max Planck Computing and Data Facility, and the Cray XC50 at CfCA of the National Astronomical Observatory of Japan. ND acknowledges support from Graduate Program on Physics for the Universe (GP-PU) at Tohoku University. This work was supported by Grant-in-Aid for Scientific Research of JSPS/MEXT (20H00158, 21H04997, 22KJ0317, 23H00127, 23H04894, 23H04900, 23H05432, and 23H01172) and JST FOREST Program (JPMJFR212Y).

\section*{Data availability}
Data and results underlying this article will be shared on reasonable
request to the corresponding author.



\bibliographystyle{mnras}

\begin{thebibliography}{}
\makeatletter
\relax
\def\mn@urlcharsother{\let\do\@makeother \do\$\do\&\do\#\do\^\do\_\do\%\do\~}
\def\mn@doi{\begingroup\mn@urlcharsother \@ifnextchar [ {\mn@doi@}
  {\mn@doi@[]}}
\def\mn@doi@[#1]#2{\def\@tempa{#1}\ifx\@tempa\@empty \href
  {http://dx.doi.org/#2} {doi:#2}\else \href {http://dx.doi.org/#2} {#1}\fi
  \endgroup}
\def\mn@eprint#1#2{\mn@eprint@#1:#2::\@nil}
\def\mn@eprint@arXiv#1{\href {http://arxiv.org/abs/#1} {{\tt arXiv:#1}}}
\def\mn@eprint@dblp#1{\href {http://dblp.uni-trier.de/rec/bibtex/#1.xml}
  {dblp:#1}}
\def\mn@eprint@#1:#2:#3:#4\@nil{\def\@tempa {#1}\def\@tempb {#2}\def\@tempc
  {#3}\ifx \@tempc \@empty \let \@tempc \@tempb \let \@tempb \@tempa \fi \ifx
  \@tempb \@empty \def\@tempb {arXiv}\fi \@ifundefined
  {mn@eprint@\@tempb}{\@tempb:\@tempc}{\expandafter \expandafter \csname
  mn@eprint@\@tempb\endcsname \expandafter{\@tempc}}}

\bibitem[\protect\citeauthoryear{Aasi et~al.}{Aasi
  et~al.}{2015}]{TheLIGOScientific:2014jea}
Aasi J.,  et~al., 2015, \mn@doi [Class. Quant. Grav.]
  {10.1088/0264-9381/32/7/074001}, 32, 074001

\bibitem[\protect\citeauthoryear{Abbott et~al.}{Abbott
  et~al.}{2017a}]{TheLIGOScientific:2017qsa}
Abbott B.,  et~al., 2017a, \mn@doi [Phys. Rev. Lett.]
  {10.1103/PhysRevLett.119.161101}, 119, 161101

\bibitem[\protect\citeauthoryear{Abbott et~al.}{Abbott
  et~al.}{2017b}]{GBM:2017lvd}
Abbott B.~P.,  et~al., 2017b, \mn@doi [Astrophys. J.]
  {10.3847/2041-8213/aa91c9}, 848, L12

\bibitem[\protect\citeauthoryear{Abbott et~al.}{Abbott
  et~al.}{2017c}]{LIGOScientific:2017zic}
Abbott B.~P.,  et~al., 2017c, \mn@doi [Astrophys. J. Lett.]
  {10.3847/2041-8213/aa920c}, 848, L13

\bibitem[\protect\citeauthoryear{Abbott et~al.}{Abbott
  et~al.}{2020}]{LIGOScientific:2020zkf}
Abbott R.,  et~al., 2020, \mn@doi [Astrophys. J. Lett.]
  {10.3847/2041-8213/ab960f}, 896, L44

\bibitem[\protect\citeauthoryear{Abbott et~al.}{Abbott
  et~al.}{2021}]{LIGOScientific:2021qlt}
Abbott R.,  et~al., 2021, \mn@doi [Astrophys. J. Lett.]
  {10.3847/2041-8213/ac082e}, 915, L5

\bibitem[\protect\citeauthoryear{Acernese et~al.}{Acernese
  et~al.}{2015}]{TheVirgo:2014hva}
Acernese F.,  et~al., 2015, \mn@doi [Class. Quant. Grav.]
  {10.1088/0264-9381/32/2/024001}, 32, 024001

\bibitem[\protect\citeauthoryear{Ackley et~al.}{Ackley
  et~al.}{2020}]{Ackley:2020qkz}
Ackley K.,  et~al., 2020, \mn@doi [Astron. Astrophys.]
  {10.1051/0004-6361/202037669}, 643, A113

\bibitem[\protect\citeauthoryear{{Almualla}, {Ning}, {Salehi}, {Bulla},
  {Dietrich}, {Coughlin}  \& {Guessoum}}{{Almualla}
  et~al.}{2021}]{Almualla:2021znj}
{Almualla} M.,  {Ning} Y.,  {Salehi} P.,  {Bulla} M.,  {Dietrich} T.,
  {Coughlin} M.~W.,   {Guessoum} N.,  2021, \mn@doi [arXiv e-prints]
  {10.48550/arXiv.2112.15470}, \href
  {https://ui.adsabs.harvard.edu/abs/2021arXiv211215470A} {p. arXiv:2112.15470}

\bibitem[\protect\citeauthoryear{Balbus \& Hawley}{Balbus \&
  Hawley}{1998}]{Balbus:1998ja}
Balbus S.~A.,  Hawley J.~F.,  1998, \mn@doi [Rev. Mod. Phys.]
  {10.1103/RevModPhys.70.1}, 70, 1

\bibitem[\protect\citeauthoryear{Banerjee, Tanaka, Kawaguchi, Kato  \&
  Gaigalas}{Banerjee et~al.}{2020}]{Banerjee:2020myd}
Banerjee S.,  Tanaka M.,  Kawaguchi K.,  Kato D.,   Gaigalas G.,  2020, \mn@doi
  [Astrophys. J.] {10.3847/1538-4357/abae61}, 901, 29

\bibitem[\protect\citeauthoryear{Banerjee, Tanaka, Kato, Gaigalas, Kawaguchi
  \& Domoto}{Banerjee et~al.}{2022}]{Banerjee:2022doa}
Banerjee S.,  Tanaka M.,  Kato D.,  Gaigalas G.,  Kawaguchi K.,   Domoto N.,
  2022, \mn@doi [Astrophys. J.] {10.3847/1538-4357/ac7565}, 934, 117

\bibitem[\protect\citeauthoryear{Banerjee, Tanaka, Kato  \& Gaigalas}{Banerjee
  et~al.}{2024}]{Banerjee:2023gye}
Banerjee S.,  Tanaka M.,  Kato D.,   Gaigalas G.,  2024, \mn@doi [Astrophys.
  J.] {10.3847/1538-4357/ad4029}, 968, 64

\bibitem[\protect\citeauthoryear{Banik, Hempel  \& Bandyopadhyay}{Banik
  et~al.}{2014}]{Banik:2014qja}
Banik S.,  Hempel M.,   Bandyopadhyay D.,  2014, \mn@doi [Astrophys. J. Suppl.]
  {10.1088/0067-0049/214/2/22}, 214, 22

\bibitem[\protect\citeauthoryear{Barnes, Kasen, Wu  \&
  Mart{\'i}nez-Pinedo}{Barnes et~al.}{2016}]{Barnes:2016umi}
Barnes J.,  Kasen D.,  Wu M.-R.,   Mart{\'i}nez-Pinedo G.,  2016, \mn@doi
  [Astrophys. J.] {10.3847/0004-637X/829/2/110}, 829, 110

\bibitem[\protect\citeauthoryear{Barnes, Zhu, Lund, Sprouse, Vassh, McLaughlin,
  Mumpower  \& Surman}{Barnes et~al.}{2021}]{Barnes:2020nfi}
Barnes J.,  Zhu Y.~L.,  Lund K.~A.,  Sprouse T.~M.,  Vassh N.,  McLaughlin
  G.~C.,  Mumpower M.~R.,   Surman R.,  2021, \mn@doi [Astrophys. J.]
  {10.3847/1538-4357/ac0aec}, 918, 44

\bibitem[\protect\citeauthoryear{Berger}{Berger}{2014}]{Berger:2013jza}
Berger E.,  2014, \mn@doi [Ann. Rev. Astron. Astrophys.]
  {10.1146/annurev-astro-081913-035926}, 52, 43

\bibitem[\protect\citeauthoryear{Berger, Fong  \& Chornock}{Berger
  et~al.}{2013}]{GRB130603BB}
Berger E.,  Fong W.,   Chornock R.,  2013, \mn@doi [Astrophys. J. Lett.]
  {10.1088/2041-8205/774/2/L23}, 774, L23

\bibitem[\protect\citeauthoryear{{Brown} et~al.,}{{Brown}
  et~al.}{2018}]{Brown:2018NDS}
{Brown} D.~A.,  et~al., 2018, \mn@doi [Nuclear Data Sheets]
  {10.1016/j.nds.2018.02.001}, \href
  {https://ui.adsabs.harvard.edu/abs/2018NDS...148....1B} {148, 1}

\bibitem[\protect\citeauthoryear{Bulla}{Bulla}{2019}]{Bulla:2019muo}
Bulla M.,  2019, \mn@doi [Mon. Not. Roy. Astron. Soc.] {10.1093/mnras/stz2495},
  489, 5037

\bibitem[\protect\citeauthoryear{Bulla et~al.,}{Bulla
  et~al.}{2021}]{Bulla:2020jjr}
Bulla M.,  et~al., 2021, \mn@doi [Mon. Not. Roy. Astron. Soc.]
  {10.1093/mnras/staa3796}, 501, 1891

\bibitem[\protect\citeauthoryear{Collins, Bauswein, Sim, Vijayan,
  Mart\'\i{}nez-Pinedo, Just, Shingles  \& Kromer}{Collins
  et~al.}{2023}]{Collins:2022ocl}
Collins C.~E.,  Bauswein A.,  Sim S.~A.,  Vijayan V.,  Mart\'\i{}nez-Pinedo G.,
   Just O.,  Shingles L.~J.,   Kromer M.,  2023, \mn@doi [Mon. Not. Roy.
  Astron. Soc.] {10.1093/mnras/stad606}, 521, 1858

\bibitem[\protect\citeauthoryear{Collins et~al.}{Collins
  et~al.}{2024}]{Collins:2023btn}
Collins C.~E.,  et~al., 2024, \mn@doi [Mon. Not. Roy. Astron. Soc.]
  {10.1093/mnras/stae571}, 529, 1333

\bibitem[\protect\citeauthoryear{Colombo et~al.}{Colombo
  et~al.}{2024}]{Colombo:2023une}
Colombo A.,  et~al., 2024, \mn@doi [Astron. Astrophys.]
  {10.1051/0004-6361/202348384}, 686, A265

\bibitem[\protect\citeauthoryear{Cowan, Sneden, Lawler, Aprahamian, Wiescher,
  Langanke, Mart\'\i{}nez-Pinedo  \& Thielemann}{Cowan
  et~al.}{2021}]{Cowan:2019pkx}
Cowan J.~J.,  Sneden C.,  Lawler J.~E.,  Aprahamian A.,  Wiescher M.,  Langanke
  K.,  Mart\'\i{}nez-Pinedo G.,   Thielemann F.-K.,  2021, \mn@doi [Rev. Mod.
  Phys.] {10.1103/RevModPhys.93.015002}, 93, 15002

\bibitem[\protect\citeauthoryear{{Darbha} \& {Kasen}}{{Darbha} \&
  {Kasen}}{2020}]{Darbha:2020lhz}
{Darbha} S.,  {Kasen} D.,  2020, \mn@doi [\apj] {10.3847/1538-4357/ab9a34},
  \href {https://ui.adsabs.harvard.edu/abs/2020ApJ...897..150D} {897, 150}

\bibitem[\protect\citeauthoryear{Darbha, Kasen, Foucart  \& Price}{Darbha
  et~al.}{2021}]{Darbha:2021rqj}
Darbha S.,  Kasen D.,  Foucart F.,   Price D.~J.,  2021, \mn@doi [Astrophys.
  J.] {10.3847/1538-4357/abff5d}

\bibitem[\protect\citeauthoryear{Domoto, Tanaka, Kato, Kawaguchi, Hotokezaka
  \& Wanajo}{Domoto et~al.}{2022}]{Domoto:2022cqp}
Domoto N.,  Tanaka M.,  Kato D.,  Kawaguchi K.,  Hotokezaka K.,   Wanajo S.,
  2022, \mn@doi [Astrophys. J.] {10.3847/1538-4357/ac8c36}, 939, 8

\bibitem[\protect\citeauthoryear{{Eastman} \& {Pinto}}{{Eastman} \&
  {Pinto}}{1993}]{1993ApJ...412..731E}
{Eastman} R.~G.,  {Pinto} P.~A.,  1993, \mn@doi [\apj] {10.1086/172957}, \href
  {http://adsabs.harvard.edu/abs/1993ApJ...412..731E} {412, 731}

\bibitem[\protect\citeauthoryear{Eichler, Livio, Piran  \& Schramm}{Eichler
  et~al.}{1989}]{Eichler:1989ve}
Eichler D.,  Livio M.,  Piran T.,   Schramm D.~N.,  1989, \mn@doi [Nature]
  {10.1038/340126a0}, 340, 126

\bibitem[\protect\citeauthoryear{Etienne, Liu, Shapiro  \& Baumgarte}{Etienne
  et~al.}{2009}]{Etienne:2008re}
Etienne Z.~B.,  Liu Y.~T.,  Shapiro S.~L.,   Baumgarte T.~W.,  2009, \mn@doi
  [Phys. Rev.] {10.1103/PhysRevD.79.044024}, D79, 044024

\bibitem[\protect\citeauthoryear{Fern\'andez, Quataert, Schwab, Kasen  \&
  Rosswog}{Fern\'andez et~al.}{2015}]{Fernandez:2014bra}
Fern\'andez R.,  Quataert E.,  Schwab J.,  Kasen D.,   Rosswog S.,  2015,
  \mn@doi [Mon. Not. Roy. Astron. Soc.] {10.1093/mnras/stv238}, 449, 390

\bibitem[\protect\citeauthoryear{Fern\'andez, Foucart, Kasen, Lippuner, Desai
  \& Roberts}{Fern\'andez et~al.}{2017}]{Fernandez:2016sbf}
Fern\'andez R.,  Foucart F.,  Kasen D.,  Lippuner J.,  Desai D.,   Roberts
  L.~F.,  2017, \mn@doi [Class. Quant. Grav.] {10.1088/1361-6382/aa7a77}, 34,
  154001

\bibitem[\protect\citeauthoryear{Fernández \& Metzger}{Fernández \&
  Metzger}{2013}]{Fernandez:2013tya}
Fernández R.,  Metzger B.~D.,  2013, \mn@doi [Mon. Not. Roy. Astron. Soc.]
  {10.1093/mnras/stt1312}, 435, 502

\bibitem[\protect\citeauthoryear{Foucart et~al.,}{Foucart
  et~al.}{2014}]{Foucart:2014nda}
Foucart F.,  et~al., 2014, \mn@doi [Phys. Rev.] {10.1103/PhysRevD.90.024026},
  D90, 024026

\bibitem[\protect\citeauthoryear{Foucart, Hinderer  \& Nissanke}{Foucart
  et~al.}{2018}]{Foucart:2018rjc}
Foucart F.,  Hinderer T.,   Nissanke S.,  2018, \mn@doi [Phys. Rev.]
  {10.1103/PhysRevD.98.081501}, D98, 081501

\bibitem[\protect\citeauthoryear{Foucart, Moesta, Ramirez, Wright, Darbha  \&
  Kasen}{Foucart et~al.}{2021}]{Foucart:2021ikp}
Foucart F.,  Moesta P.,  Ramirez T.,  Wright A.~J.,  Darbha S.,   Kasen D.,
  2021, \mn@doi [Phys. Rev. D] {10.1103/PhysRevD.104.123010}, 104, 123010

\bibitem[\protect\citeauthoryear{{Freiburghaus}, {Rosswog}  \&
  {Thielemann}}{{Freiburghaus} et~al.}{1999}]{Freiburghaus1999a}
{Freiburghaus} C.,  {Rosswog} S.,   {Thielemann} F.~K.,  1999, \mn@doi [\apjl]
  {10.1086/312343}, \href
  {https://ui.adsabs.harvard.edu/abs/1999ApJ...525L.121F} {525, L121}

\bibitem[\protect\citeauthoryear{{Friend} \& {Castor}}{{Friend} \&
  {Castor}}{1983}]{1983ApJ...272..259F}
{Friend} D.~B.,  {Castor} J.~I.,  1983, \mn@doi [\apj] {10.1086/161289}, \href
  {https://ui.adsabs.harvard.edu/abs/1983ApJ...272..259F} {272, 259}

\bibitem[\protect\citeauthoryear{Frostig et~al.}{Frostig
  et~al.}{2022}]{Frostig:2021vkt}
Frostig D.,  et~al., 2022, \mn@doi [Astrophys. J.] {10.3847/1538-4357/ac4508},
  926, 152

\bibitem[\protect\citeauthoryear{Fujibayashi, Shibata, Wanajo, Kiuchi, Kyutoku
  \& Sekiguchi}{Fujibayashi et~al.}{2020a}]{Fujibayashi:2020qda}
Fujibayashi S.,  Shibata M.,  Wanajo S.,  Kiuchi K.,  Kyutoku K.,   Sekiguchi
  Y.,  2020a, \mn@doi [Phys. Rev. D] {10.1103/PhysRevD.101.083029}, 101, 083029

\bibitem[\protect\citeauthoryear{Fujibayashi, Wanajo, Kiuchi, Kyutoku,
  Sekiguchi  \& Shibata}{Fujibayashi et~al.}{2020b}]{Fujibayashi:2020dvr}
Fujibayashi S.,  Wanajo S.,  Kiuchi K.,  Kyutoku K.,  Sekiguchi Y.,   Shibata
  M.,  2020b, \mn@doi [Astrophys. J.] {10.3847/1538-4357/abafc2}, 901, 122

\bibitem[\protect\citeauthoryear{Fujibayashi, Kiuchi, Wanajo, Kyutoku,
  Sekiguchi  \& Shibata}{Fujibayashi et~al.}{2023}]{Fujibayashi:2022ftg}
Fujibayashi S.,  Kiuchi K.,  Wanajo S.,  Kyutoku K.,  Sekiguchi Y.,   Shibata
  M.,  2023, \mn@doi [Astrophys. J.] {10.3847/1538-4357/ac9ce0}, 942, 39

\bibitem[\protect\citeauthoryear{Fuller \& Ma}{Fuller \&
  Ma}{2019}]{Fuller:2019sxi}
Fuller J.,  Ma L.,  2019, \mn@doi [Astrophys. J. Lett.]
  {10.3847/2041-8213/ab339b}, 881, L1

\bibitem[\protect\citeauthoryear{Gompertz et~al.}{Gompertz
  et~al.}{2023}]{GRB211211AG}
Gompertz B.~P.,  et~al., 2023, \mn@doi [Nature Astron.]
  {10.1038/s41550-022-01819-4}, 7, 67

\bibitem[\protect\citeauthoryear{Gottlieb, Jacquemin-Ide, Lowell, Tchekhovskoy
  \& Ramirez-Ruiz}{Gottlieb et~al.}{2023a}]{Gottlieb:2023cgm}
Gottlieb O.,  Jacquemin-Ide J.,  Lowell B.,  Tchekhovskoy A.,   Ramirez-Ruiz
  E.,  2023a, \mn@doi [Astrophys. J. Lett.] {10.3847/2041-8213/ace779}, 952,
  L32

\bibitem[\protect\citeauthoryear{Gottlieb et~al.,}{Gottlieb
  et~al.}{2023b}]{Gottlieb:2023vuf}
Gottlieb O.,  et~al., 2023b, \mn@doi [Astrophys. J. Lett.]
  {10.3847/2041-8213/acec4a}, 953, L11

\bibitem[\protect\citeauthoryear{Grossman, Korobkin, Rosswog  \&
  Piran}{Grossman et~al.}{2014}]{Grossman:2013lqa}
Grossman D.,  Korobkin O.,  Rosswog S.,   Piran T.,  2014, \mn@doi [Mon. Not.
  Roy. Astron. Soc.] {10.1093/mnras/stt2503}, 439, 757

\bibitem[\protect\citeauthoryear{Hamidani \& Ioka}{Hamidani \&
  Ioka}{2023a}]{Hamidani:2022cyj}
Hamidani H.,  Ioka K.,  2023a, \mn@doi [Mon. Not. Roy. Astron. Soc.]
  {10.1093/mnras/stad041}, 520, 1111

\bibitem[\protect\citeauthoryear{Hamidani \& Ioka}{Hamidani \&
  Ioka}{2023b}]{Hamidani:2022hvl}
Hamidani H.,  Ioka K.,  2023b, \mn@doi [Mon. Not. Roy. Astron. Soc.]
  {10.1093/mnras/stad1933}, 524, 4841

\bibitem[\protect\citeauthoryear{Hayashi, Kawaguchi, Kiuchi, Kyutoku  \&
  Shibata}{Hayashi et~al.}{2021}]{Hayashi:2020zmn}
Hayashi K.,  Kawaguchi K.,  Kiuchi K.,  Kyutoku K.,   Shibata M.,  2021,
  \mn@doi [Phys. Rev. D] {10.1103/PhysRevD.103.043007}, 103, 043007

\bibitem[\protect\citeauthoryear{Hayashi, Fujibayashi, Kiuchi, Kyutoku,
  Sekiguchi  \& Shibata}{Hayashi et~al.}{2022}]{Hayashi:2021oxy}
Hayashi K.,  Fujibayashi S.,  Kiuchi K.,  Kyutoku K.,  Sekiguchi Y.,   Shibata
  M.,  2022, \mn@doi [Phys. Rev. D] {10.1103/PhysRevD.106.023008}, 106, 023008

\bibitem[\protect\citeauthoryear{Hayashi, Kiuchi, Kyutoku, Sekiguchi  \&
  Shibata}{Hayashi et~al.}{2023}]{Hayashi:2022cdq}
Hayashi K.,  Kiuchi K.,  Kyutoku K.,  Sekiguchi Y.,   Shibata M.,  2023,
  \mn@doi [Phys. Rev. D] {10.1103/PhysRevD.107.123001}, 107, 123001

\bibitem[\protect\citeauthoryear{Hirai \& Podsiadlowski}{Hirai \&
  Podsiadlowski}{2022}]{Hirai:2022hbf}
Hirai R.,  Podsiadlowski P.,  2022, \mn@doi [Mon. Not. Roy. Astron. Soc.]
  {10.1093/mnras/stac3007}, 517, 4544

\bibitem[\protect\citeauthoryear{{Hotokezaka} \& {Nakar}}{{Hotokezaka} \&
  {Nakar}}{2020}]{Hotokezaka:2019uwo}
{Hotokezaka} K.,  {Nakar} E.,  2020, \mn@doi [\apj] {10.3847/1538-4357/ab6a98},
  \href {https://ui.adsabs.harvard.edu/abs/2020ApJ...891..152H} {891, 152}

\bibitem[\protect\citeauthoryear{Hotokezaka \& Piran}{Hotokezaka \&
  Piran}{2015}]{Hotokezaka:2015eja}
Hotokezaka K.,  Piran T.,  2015, \mn@doi [Mon. Not. Roy. Astron. Soc.]
  {10.1093/mnras/stv620}, 450, 1430

\bibitem[\protect\citeauthoryear{{Hotokezaka}, {Wanajo}, {Tanaka}, {Bamba},
  {Terada}  \& {Piran}}{{Hotokezaka} et~al.}{2016}]{Hotokezaka:2016}
{Hotokezaka} K.,  {Wanajo} S.,  {Tanaka} M.,  {Bamba} A.,  {Terada} Y.,
  {Piran} T.,  2016, \mn@doi [\mnras] {10.1093/mnras/stw404}, \href
  {https://ui.adsabs.harvard.edu/abs/2016MNRAS.459...35H} {459, 35}

\bibitem[\protect\citeauthoryear{Hotokezaka, Kiuchi, Shibata, Nakar  \&
  Piran}{Hotokezaka et~al.}{2018}]{Hotokezaka:2018gmo}
Hotokezaka K.,  Kiuchi K.,  Shibata M.,  Nakar E.,   Piran T.,  2018, \mn@doi
  [Astrophys. J.] {10.3847/1538-4357/aadf92}, 867, 95

\bibitem[\protect\citeauthoryear{{Hotokezaka}, {Tanaka}, {Kato}  \&
  {Gaigalas}}{{Hotokezaka} et~al.}{2021}]{Hotokezaka:2021ofe}
{Hotokezaka} K.,  {Tanaka} M.,  {Kato} D.,   {Gaigalas} G.,  2021, \mn@doi
  [\mnras] {10.1093/mnras/stab1975}, \href
  {https://ui.adsabs.harvard.edu/abs/2021MNRAS.506.5863H} {506, 5863}

\bibitem[\protect\citeauthoryear{Jin, Li, Cano, Covino, Fan  \& Wei}{Jin
  et~al.}{2015}]{GRB060614J}
Jin Z.-P.,  Li X.,  Cano Z.,  Covino S.,  Fan Y.-Z.,   Wei D.-M.,  2015,
  \mn@doi [Astrophys. J. Lett.] {10.1088/2041-8205/811/2/L22}, 811, L22

\bibitem[\protect\citeauthoryear{Jin et~al.,}{Jin et~al.}{2016}]{GRB050709}
Jin Z.-P.,  et~al., 2016, \mn@doi [Nature Commun.] {10.1038/ncomms12898}, 7,
  12898

\bibitem[\protect\citeauthoryear{Just, Bauswein, Pulpillo, Goriely  \&
  Janka}{Just et~al.}{2015}]{Just:2014fka}
Just O.,  Bauswein A.,  Pulpillo R.~A.,  Goriely S.,   Janka H.~T.,  2015,
  \mn@doi [Mon. Not. Roy. Astron. Soc.] {10.1093/mnras/stv009}, 448, 541

\bibitem[\protect\citeauthoryear{{Just}, {Kullmann}, {Goriely}, {Bauswein},
  {Janka}  \& {Collins}}{{Just} et~al.}{2022}]{Just:2021vzy}
{Just} O.,  {Kullmann} I.,  {Goriely} S.,  {Bauswein} A.,  {Janka} H.~T.,
  {Collins} C.~E.,  2022, \mn@doi [\mnras] {10.1093/mnras/stab3327}, \href
  {https://ui.adsabs.harvard.edu/abs/2022MNRAS.510.2820J} {510, 2820}

\bibitem[\protect\citeauthoryear{Just et~al.,}{Just
  et~al.}{2023}]{Just:2023wtj}
Just O.,  et~al., 2023, \mn@doi [Astrophys. J. Lett.]
  {10.3847/2041-8213/acdad2}, 951, L12

\bibitem[\protect\citeauthoryear{Kasen, Thomas  \& Nugent}{Kasen
  et~al.}{2006}]{Kasen:2006ce}
Kasen D.,  Thomas R.~C.,   Nugent P.,  2006, \mn@doi [Astrophys. J.]
  {10.1086/506190}, 651, 366

\bibitem[\protect\citeauthoryear{Kasen, Badnell  \& Barnes}{Kasen
  et~al.}{2013}]{Kasen:2013xka}
Kasen D.,  Badnell N.~R.,   Barnes J.,  2013, \mn@doi [Astrophys. J.]
  {10.1088/0004-637X/774/1/25}, 774, 25

\bibitem[\protect\citeauthoryear{Kasen, Fernandez  \& Metzger}{Kasen
  et~al.}{2015}]{Kasen:2014toa}
Kasen D.,  Fernandez R.,   Metzger B.,  2015, \mn@doi [Mon. Not. Roy. Astron.
  Soc.] {10.1093/mnras/stv721}, 450, 1777

\bibitem[\protect\citeauthoryear{Kawaguchi, Shibata  \& Tanaka}{Kawaguchi
  et~al.}{2018}]{Kawaguchi:2018ptg}
Kawaguchi K.,  Shibata M.,   Tanaka M.,  2018, \mn@doi [Astrophys. J.]
  {10.3847/2041-8213/aade02}, 865, L21

\bibitem[\protect\citeauthoryear{{Kawaguchi}, {Shibata}  \&
  {Tanaka}}{{Kawaguchi} et~al.}{2020}]{Kawaguchi:2019nju}
{Kawaguchi} K.,  {Shibata} M.,   {Tanaka} M.,  2020, \mn@doi [\apj]
  {10.3847/1538-4357/ab61f6}, \href
  {https://ui.adsabs.harvard.edu/abs/2020ApJ...889..171K} {889, 171}

\bibitem[\protect\citeauthoryear{Kawaguchi, Fujibayashi, Shibata, Tanaka  \&
  Wanajo}{Kawaguchi et~al.}{2021}]{Kawaguchi:2020vbf}
Kawaguchi K.,  Fujibayashi S.,  Shibata M.,  Tanaka M.,   Wanajo S.,  2021,
  \mn@doi [Astrophys. J.] {10.3847/1538-4357/abf3bc}, 913, 100

\bibitem[\protect\citeauthoryear{Kawaguchi, Fujibayashi, Hotokezaka, Shibata
  \& Wanajo}{Kawaguchi et~al.}{2022}]{Kawaguchi:2022bub}
Kawaguchi K.,  Fujibayashi S.,  Hotokezaka K.,  Shibata M.,   Wanajo S.,  2022,
  \mn@doi [Astrophys. J.] {10.3847/1538-4357/ac6ef7}, 933, 22

\bibitem[\protect\citeauthoryear{Kawaguchi, Fujibayashi, Domoto, Kiuchi,
  Shibata  \& Wanajo}{Kawaguchi et~al.}{2023}]{Kawaguchi:2023zln}
Kawaguchi K.,  Fujibayashi S.,  Domoto N.,  Kiuchi K.,  Shibata M.,   Wanajo
  S.,  2023, Mon. Not. Roy. Astron. Soc.

\bibitem[\protect\citeauthoryear{Kedia et~al.,}{Kedia
  et~al.}{2023}]{Kedia:2022onl}
Kedia A.,  et~al., 2023, \mn@doi [Phys. Rev. Res.]
  {10.1103/PhysRevResearch.5.013168}, 5, 013168

\bibitem[\protect\citeauthoryear{Kiuchi, Held, Sekiguchi  \& Shibata}{Kiuchi
  et~al.}{2022}]{Kiuchi:2022ubj}
Kiuchi K.,  Held L.~E.,  Sekiguchi Y.,   Shibata M.,  2022, \mn@doi [Phys. Rev.
  D] {10.1103/PhysRevD.106.124041}, 106, 124041

\bibitem[\protect\citeauthoryear{Kiuchi, Fujibayashi, Hayashi, Kyutoku,
  Sekiguchi  \& Shibata}{Kiuchi et~al.}{2023}]{Kiuchi:2022nin}
Kiuchi K.,  Fujibayashi S.,  Hayashi K.,  Kyutoku K.,  Sekiguchi Y.,   Shibata
  M.,  2023, \mn@doi [Phys. Rev. Lett.] {10.1103/PhysRevLett.131.011401}, 131,
  011401

\bibitem[\protect\citeauthoryear{Kiuchi, Reboul-Salze, Shibata  \&
  Sekiguchi}{Kiuchi et~al.}{2024}]{Kiuchi:2023obe}
Kiuchi K.,  Reboul-Salze A.,  Shibata M.,   Sekiguchi Y.,  2024, \mn@doi
  [Nature Astron.] {10.1038/s41550-024-02194-y}, 8, 298

\bibitem[\protect\citeauthoryear{Klion, Duffell, Kasen  \& Quataert}{Klion
  et~al.}{2021}]{Klion:2020efn}
Klion H.,  Duffell P.~C.,  Kasen D.,   Quataert E.,  2021, \mn@doi [Mon. Not.
  Roy. Astron. Soc.] {10.1093/mnras/stab042}, 502, 865

\bibitem[\protect\citeauthoryear{{Kondo}, {Sumi}, {Koshimoto}, {Rattenbury},
  {Suzuki}  \& {Bennett}}{{Kondo} et~al.}{2023}]{2023AJ....165..254K}
{Kondo} I.,  {Sumi} T.,  {Koshimoto} N.,  {Rattenbury} N.~J.,  {Suzuki} D.,
  {Bennett} D.~P.,  2023, \mn@doi [\aj] {10.3847/1538-3881/acccf9}, \href
  {https://ui.adsabs.harvard.edu/abs/2023AJ....165..254K} {165, 254}

\bibitem[\protect\citeauthoryear{{Korobkin} et~al.,}{{Korobkin}
  et~al.}{2021}]{Korobkin:2020spe}
{Korobkin} O.,  et~al., 2021, \mn@doi [\apj] {10.3847/1538-4357/abe1b5}, \href
  {https://ui.adsabs.harvard.edu/abs/2021ApJ...910..116K} {910, 116}

\bibitem[\protect\citeauthoryear{Kramida, {Yu.~Ralchenko}, Reader  \& {and NIST
  ASD Team}}{Kramida et~al.}{2021}]{NIST}
Kramida A.,  {Yu.~Ralchenko} Reader J.,   {and NIST ASD Team} 2021, {NIST
  Atomic Spectra Database (ver. 5.9). Available:
  {\tt{https://physics.nist.gov/asd}}. National Institute of Standards and
  Technology, Gaithersburg, MD.}

\bibitem[\protect\citeauthoryear{{Kulkarni}}{{Kulkarni}}{2005}]{Kulkarni:2005jw}
{Kulkarni} S.~R.,  2005, \mn@doi [arXiv e-prints]
  {10.48550/arXiv.astro-ph/0510256}, \href
  {https://ui.adsabs.harvard.edu/abs/2005astro.ph.10256K} {pp
  astro--ph/0510256}

\bibitem[\protect\citeauthoryear{{Kupka}, {Piskunov}, {Ryabchikova}, {Stempels}
   \& {Weiss}}{{Kupka} et~al.}{1999}]{1999A&AS..138..119K}
{Kupka} F.,  {Piskunov} N.,  {Ryabchikova} T.~A.,  {Stempels} H.~C.,   {Weiss}
  W.~W.,  1999, \mn@doi [\aaps] {10.1051/aas:1999267}, \href
  {https://ui.adsabs.harvard.edu/abs/1999A&AS..138..119K} {138, 119}

\bibitem[\protect\citeauthoryear{{Kurganov} \& {Tadmor}}{{Kurganov} \&
  {Tadmor}}{2000}]{2000JCoPh.160..241K}
{Kurganov} A.,  {Tadmor} E.,  2000, \mn@doi [Journal of Computational Physics]
  {10.1006/jcph.2000.6459}, \href
  {https://ui.adsabs.harvard.edu/abs/2000JCoPh.160..241K} {160, 241}

\bibitem[\protect\citeauthoryear{Kuroda}{Kuroda}{2010}]{Kuroda:2010zzb}
Kuroda K.,  2010, \mn@doi [Class. Quant. Grav.]
  {10.1088/0264-9381/27/8/084004}, 27, 084004

\bibitem[\protect\citeauthoryear{{Kurucz} \& {Bell}}{{Kurucz} \&
  {Bell}}{1995}]{1995all..book.....K}
{Kurucz} R.~L.,  {Bell} B.,  1995, {Atomic line list}

\bibitem[\protect\citeauthoryear{Kyutoku, Ioka  \& Shibata}{Kyutoku
  et~al.}{2013}]{Kyutoku:2013wxa}
Kyutoku K.,  Ioka K.,   Shibata M.,  2013, \mn@doi [Phys. Rev. D]
  {10.1103/PhysRevD.88.041503}, 88, 041503

\bibitem[\protect\citeauthoryear{Kyutoku, Ioka, Okawa, Shibata  \&
  Taniguchi}{Kyutoku et~al.}{2015}]{Kyutoku:2015gda}
Kyutoku K.,  Ioka K.,  Okawa H.,  Shibata M.,   Taniguchi K.,  2015, \mn@doi
  [Phys. Rev.] {10.1103/PhysRevD.92.044028}, D92, 044028

\bibitem[\protect\citeauthoryear{{Lamb} et~al.,}{{Lamb}
  et~al.}{2019}]{GRB160821BG}
{Lamb} G.~P.,  et~al., 2019, \mn@doi [\apj] {10.3847/1538-4357/ab38bb}, \href
  {https://ui.adsabs.harvard.edu/abs/2019ApJ...883...48L} {883, 48}

\bibitem[\protect\citeauthoryear{Lattimer \& Schramm}{Lattimer \&
  Schramm}{1974}]{Lattimer:1974slx}
Lattimer J.~M.,  Schramm D.~N.,  1974, \mn@doi [Astrophys. J.]
  {10.1086/181612}, 192, L145

\bibitem[\protect\citeauthoryear{Levan et~al.}{Levan
  et~al.}{2024}]{GRB230307AL}
Levan A.~J.,  et~al., 2024, \mn@doi [Nature] {10.1038/s41586-023-06759-1}, 626,
  737

\bibitem[\protect\citeauthoryear{Li \& Paczynski}{Li \&
  Paczynski}{1998}]{Li:1998bw}
Li L.-X.,  Paczynski B.,  1998, \mn@doi [Astrophys. J.] {10.1086/311680}, 507,
  L59

\bibitem[\protect\citeauthoryear{Lippuner \& Roberts}{Lippuner \&
  Roberts}{2015}]{Lippuner:2015gwa}
Lippuner J.,  Roberts L.~F.,  2015, \mn@doi [Astrophys. J.]
  {10.1088/0004-637X/815/2/82}, 815, 82

\bibitem[\protect\citeauthoryear{Lovelace, Duez, Foucart, Kidder, Pfeiffer,
  Scheel  \& Szilegyi}{Lovelace et~al.}{2013}]{Lovelace:2013vma}
Lovelace G.,  Duez M.~D.,  Foucart F.,  Kidder L.~E.,  Pfeiffer H.~P.,  Scheel
  M.~A.,   Szilegyi B.,  2013, \mn@doi [Class. Quant. Grav.]
  {10.1088/0264-9381/30/13/135004}, 30, 135004

\bibitem[\protect\citeauthoryear{{Margalit} \& {Piran}}{{Margalit} \&
  {Piran}}{2020}]{Margalit2020MNRAS}
{Margalit} B.,  {Piran} T.,  2020, \mn@doi [\mnras] {10.1093/mnras/staa1486},
  \href {https://ui.adsabs.harvard.edu/abs/2020MNRAS.495.4981M} {495, 4981}

\bibitem[\protect\citeauthoryear{Metzger et~al.,}{Metzger
  et~al.}{2010}]{Metzger:2010sy}
Metzger B.~D.,  et~al., 2010, \mn@doi [Mon. Not. Roy. Astron. Soc.]
  {10.1111/j.1365-2966.2010.16864.x}, 406, 2650

\bibitem[\protect\citeauthoryear{Nakar}{Nakar}{2007}]{Nakar:2007yr}
Nakar E.,  2007, \mn@doi [Phys. Rept.] {10.1016/j.physrep.2007.02.005}, 442,
  166

\bibitem[\protect\citeauthoryear{{Nakar} \& {Piran}}{{Nakar} \&
  {Piran}}{2011}]{Nakar2011Natur}
{Nakar} E.,  {Piran} T.,  2011, \mn@doi [\nat] {10.1038/nature10365}, \href
  {https://ui.adsabs.harvard.edu/abs/2011Natur.478...82N} {478, 82}

\bibitem[\protect\citeauthoryear{Nakar \& Piran}{Nakar \&
  Piran}{2017}]{Nakar:2016cih}
Nakar E.,  Piran T.,  2017, \mn@doi [Astrophys. J.]
  {10.3847/1538-4357/834/1/28}, 834, 28

\bibitem[\protect\citeauthoryear{Nativi, Bulla, Rosswog, Lundman, Kowal, Gizzi,
  Lamb  \& Perego}{Nativi et~al.}{2020}]{Nativi:2020moj}
Nativi L.,  Bulla M.,  Rosswog S.,  Lundman C.,  Kowal G.,  Gizzi D.,  Lamb
  G.~P.,   Perego A.,  2020, \mn@doi [Mon. Not. Roy. Astron. Soc.]
  {10.1093/mnras/staa3337}, 500, 1772

\bibitem[\protect\citeauthoryear{{Nishimura}, {Takiwaki}  \&
  {Thielemann}}{{Nishimura} et~al.}{2015}]{2015ApJ...810..109N}
{Nishimura} N.,  {Takiwaki} T.,   {Thielemann} F.-K.,  2015, \mn@doi [\apj]
  {10.1088/0004-637X/810/2/109}, \href
  {https://ui.adsabs.harvard.edu/abs/2015ApJ...810..109N} {810, 109}

\bibitem[\protect\citeauthoryear{Nissanke, Kasliwal  \& Georgieva}{Nissanke
  et~al.}{2013}]{Nissanke:2012dj}
Nissanke S.,  Kasliwal M.,   Georgieva A.,  2013, \mn@doi [Astrophys. J.]
  {10.1088/0004-637X/767/2/124}, 767, 124

\bibitem[\protect\citeauthoryear{{Paczynski}}{{Paczynski}}{1991}]{1991AcA....41..257P}
{Paczynski} B.,  1991, \actaa, \href
  {https://ui.adsabs.harvard.edu/abs/1991AcA....41..257P} {41, 257}

\bibitem[\protect\citeauthoryear{{Piskunov}, {Kupka}, {Ryabchikova}, {Weiss}
  \& {Jeffery}}{{Piskunov} et~al.}{1995}]{1995A&AS..112..525P}
{Piskunov} N.~E.,  {Kupka} F.,  {Ryabchikova} T.~A.,  {Weiss} W.~W.,
  {Jeffery} C.~S.,  1995, \aaps, \href
  {https://ui.adsabs.harvard.edu/abs/1995A&AS..112..525P} {112, 525}

\bibitem[\protect\citeauthoryear{{Pognan}, {Jerkstrand}  \& {Grumer}}{{Pognan}
  et~al.}{2022}]{Pognan2022MNRAS}
{Pognan} Q.,  {Jerkstrand} A.,   {Grumer} J.,  2022, \mn@doi [\mnras]
  {10.1093/mnras/stab3674}, \href
  {https://ui.adsabs.harvard.edu/abs/2022MNRAS.510.3806P} {510, 3806}

\bibitem[\protect\citeauthoryear{Rastinejad et~al.}{Rastinejad
  et~al.}{2022}]{GRB211211AR}
Rastinejad J.~C.,  et~al., 2022, \mn@doi [Nature] {10.1038/s41586-022-05390-w},
  612, 223

\bibitem[\protect\citeauthoryear{Rosswog}{Rosswog}{2005}]{Rosswog:2005su}
Rosswog S.,  2005, \mn@doi [Astrophys. J.] {10.1086/497062}, 634, 1202

\bibitem[\protect\citeauthoryear{Rosswog, Korobkin, Arcones, Thielemann  \&
  Piran}{Rosswog et~al.}{2014}]{Rosswog:2013kqa}
Rosswog S.,  Korobkin O.,  Arcones A.,  Thielemann F.,   Piran T.,  2014,
  \mn@doi [Mon. Not. Roy. Astron. Soc.] {10.1093/mnras/stt2502}, 439, 744

\bibitem[\protect\citeauthoryear{{Ryabchikova}, {Piskunov}, {Kurucz},
  {Stempels}, {Heiter}, {Pakhomov}  \& {Barklem}}{{Ryabchikova}
  et~al.}{2015}]{2015PhyS...90e4005R}
{Ryabchikova} T.,  {Piskunov} N.,  {Kurucz} R.~L.,  {Stempels} H.~C.,  {Heiter}
  U.,  {Pakhomov} Y.,   {Barklem} P.~S.,  2015, \mn@doi [\physscr]
  {10.1088/0031-8949/90/5/054005}, \href
  {https://ui.adsabs.harvard.edu/abs/2015PhyS...90e4005R} {90, 054005}

\bibitem[\protect\citeauthoryear{Shibata}{Shibata}{2015}]{Shibata2015}
Shibata M.,  2015, Numerical Relativity.
{WORLD} {SCIENTIFIC}, \mn@doi{10.1142/9692}, \url
  {https://doi.org/10.1142/9692}

\bibitem[\protect\citeauthoryear{Shibata \& Taniguchi}{Shibata \&
  Taniguchi}{2008}]{Shibata:2007zm}
Shibata M.,  Taniguchi K.,  2008, \mn@doi [Phys. Rev.]
  {10.1103/PhysRevD.77.084015}, D77, 084015

\bibitem[\protect\citeauthoryear{Shibata, Fujibayashi  \& Sekiguchi}{Shibata
  et~al.}{2021}]{Shibata:2021xmo}
Shibata M.,  Fujibayashi S.,   Sekiguchi Y.,  2021, \mn@doi [Phys. Rev. D]
  {10.1103/PhysRevD.104.063026}, 104, 063026

\bibitem[\protect\citeauthoryear{Shingles et~al.,}{Shingles
  et~al.}{2023}]{Shingles:2023kua}
Shingles L.~J.,  et~al., 2023, \mn@doi [Astrophys. J. Lett.]
  {10.3847/2041-8213/acf29a}, 954, L41

\bibitem[\protect\citeauthoryear{Steiner, Hempel  \& Fischer}{Steiner
  et~al.}{2013}]{Steiner:2012rk}
Steiner A.~W.,  Hempel M.,   Fischer T.,  2013, \mn@doi [Astrophys. J.]
  {10.1088/0004-637X/774/1/17}, 774, 17

\bibitem[\protect\citeauthoryear{Tanaka \& Hotokezaka}{Tanaka \&
  Hotokezaka}{2013}]{Tanaka:2013ana}
Tanaka M.,  Hotokezaka K.,  2013, \mn@doi [Astrophys. J.]
  {10.1088/0004-637X/775/2/113}, 775, 113

\bibitem[\protect\citeauthoryear{Tanaka, Hotokezaka, Kyutoku, Wanajo, Kiuchi,
  Sekiguchi  \& Shibata}{Tanaka et~al.}{2014}]{Tanaka:2013ixa}
Tanaka M.,  Hotokezaka K.,  Kyutoku K.,  Wanajo S.,  Kiuchi K.,  Sekiguchi Y.,
   Shibata M.,  2014, \mn@doi [Astrophys. J.] {10.1088/0004-637X/780/1/31},
  780, 31

\bibitem[\protect\citeauthoryear{Tanaka et~al.}{Tanaka
  et~al.}{2017}]{Tanaka:2017qxj}
Tanaka M.,  et~al., 2017, \mn@doi [Publ. Astron. Soc. Jap.]
  {10.1093/pasj/psx121}, 69, Publications of the Astronomical Society of Japan,
  Volume 69, Issue 6, 1 December 2017, 102, https://doi.org/10.1093/pasj/psx121

\bibitem[\protect\citeauthoryear{Tanaka et~al.}{Tanaka
  et~al.}{2018}]{Tanaka:2017lxb}
Tanaka M.,  et~al., 2018, \mn@doi [Astrophys. J.] {10.3847/1538-4357/aaa0cb},
  852, 109

\bibitem[\protect\citeauthoryear{Tanaka, Kato, Gaigalas  \& Kawaguchi}{Tanaka
  et~al.}{2020}]{Tanaka:2019iqp}
Tanaka M.,  Kato D.,  Gaigalas G.,   Kawaguchi K.,  2020, \mn@doi [Mon. Not.
  Roy. Astron. Soc.] {10.1093/mnras/staa1576}, 496, 1369

\bibitem[\protect\citeauthoryear{Tanvir, Levan, Fruchter, Hjorth, Wiersema,
  Tunnicliffe  \& de Ugarte~Postigo}{Tanvir et~al.}{2013}]{GRB130603BT}
Tanvir N.~R.,  Levan A.~J.,  Fruchter A.~S.,  Hjorth J.,  Wiersema K.,
  Tunnicliffe R.,   de Ugarte~Postigo A.,  2013, \mn@doi [Nature]
  {10.1038/nature12505}, 500, 547

\bibitem[\protect\citeauthoryear{{The LIGO Scientific Collaboration}, {the
  Virgo Collaboration}  \& {the KAGRA Collaboration}}{{The LIGO Scientific
  Collaboration} et~al.}{2024}]{LIGOScientific:2024elc}
{The LIGO Scientific Collaboration} {the Virgo Collaboration}  {the KAGRA
  Collaboration} 2024, arXiv e-prints, \href
  {https://ui.adsabs.harvard.edu/abs/2024arXiv240404248T} {p. arXiv:2404.04248}

\bibitem[\protect\citeauthoryear{Troja et~al.}{Troja
  et~al.}{2019}]{GRB160821BT}
Troja E.,  et~al., 2019, \mn@doi [Mon. Not. Roy. Astron. Soc.]
  {10.1093/mnras/stz2255}, 489, 2104

\bibitem[\protect\citeauthoryear{Troja et~al.}{Troja
  et~al.}{2022}]{GRB211211AT}
Troja E.,  et~al., 2022, \mn@doi [Nature] {10.1038/s41586-022-05327-3}, 612,
  228

\bibitem[\protect\citeauthoryear{Villar et~al.}{Villar
  et~al.}{2017}]{Villar:2017wcc}
Villar V.~A.,  et~al., 2017, \mn@doi [Astrophys. J.]
  {10.3847/2041-8213/aa9c84}, 851, L21

\bibitem[\protect\citeauthoryear{{Wanajo}, {Fujibayashi}, {Hayashi}, {Kiuchi},
  {Sekiguchi}  \& {Shibata}}{{Wanajo} et~al.}{2022}]{Wanajo:2022jgw}
{Wanajo} S.,  {Fujibayashi} S.,  {Hayashi} K.,  {Kiuchi} K.,  {Sekiguchi} Y.,
  {Shibata} M.,  2022, \mn@doi [arXiv e-prints] {10.48550/arXiv.2212.04507},
  \href {https://ui.adsabs.harvard.edu/abs/2022arXiv221204507W} {p.
  arXiv:2212.04507}

\bibitem[\protect\citeauthoryear{Waxman, Ofek, Kushnir  \& Gal-Yam}{Waxman
  et~al.}{2018}]{Waxman:2017sqv}
Waxman E.,  Ofek E.~O.,  Kushnir D.,   Gal-Yam A.,  2018, \mn@doi [Mon. Not.
  Roy. Astron. Soc.] {10.1093/mnras/sty2441}, 481, 3423

\bibitem[\protect\citeauthoryear{Wollaeger et~al.,}{Wollaeger
  et~al.}{2018}]{Wollaeger:2017ahm}
Wollaeger R.~T.,  et~al., 2018, \mn@doi [Mon. Not. Roy. Astron. Soc.]
  {10.1093/mnras/sty1018}, 478, 3298

\bibitem[\protect\citeauthoryear{Wu, Barnes, Martinez-Pinedo  \& Metzger}{Wu
  et~al.}{2019}]{Wu:2018mvg}
Wu M.-R.,  Barnes J.,  Martinez-Pinedo G.,   Metzger B.~D.,  2019, \mn@doi
  [Phys. Rev. Lett.] {10.1103/PhysRevLett.122.062701}, 122, 062701

\bibitem[\protect\citeauthoryear{{Wu}, {Ricigliano}, {Kashyap}, {Perego}  \&
  {Radice}}{{Wu} et~al.}{2022}]{Wu:2021ibi}
{Wu} Z.,  {Ricigliano} G.,  {Kashyap} R.,  {Perego} A.,   {Radice} D.,  2022,
  \mn@doi [\mnras] {10.1093/mnras/stac399}, \href
  {https://ui.adsabs.harvard.edu/abs/2022MNRAS.512..328W} {512, 328}

\bibitem[\protect\citeauthoryear{Yang et~al.,}{Yang et~al.}{2015}]{GRB060614Y}
Yang B.,  et~al., 2015, \mn@doi [Nature Commun.] {10.1038/ncomms8323}, 6, 7323

\bibitem[\protect\citeauthoryear{Zhu, Yang, Liu, Huang, Zhang, Li, Yu  \&
  Gao}{Zhu et~al.}{2020}]{Zhu:2020inc}
Zhu J.-P.,  Yang Y.-P.,  Liu L.-D.,  Huang Y.,  Zhang B.,  Li Z.,  Yu Y.-W.,
  Gao H.,  2020, \mn@doi [Astrophys. J.] {10.3847/1538-4357/ab93bf}, 897, 20

\bibitem[\protect\citeauthoryear{Zhu, Lund, Barnes, Sprouse, Vassh, McLaughlin,
  Mumpower  \& Surman}{Zhu et~al.}{2021}]{Zhu:2020eyk}
Zhu Y.~L.,  Lund K.,  Barnes J.,  Sprouse T.~M.,  Vassh N.,  McLaughlin G.~C.,
  Mumpower M.~R.,   Surman R.,  2021, \mn@doi [Astrophys. J.]
  {10.3847/1538-4357/abc69e}, 906, 94

\makeatother
\end{thebibliography}




\appendix

\section{Formulation}\label{app:form}
In this appendix, we describe the formulation of hydrodynamics equations in the spherical coordinates employed for the long-term evolution of ejecta. Throughout this appendix, the units of $c=1=G$ are employed where $G$ is the gravitational constant, unless otherwise mentioned. 
\subsection{Basic equations}
The basic equations for the numerical hydrodynamics employed in this work are formulated in the framework of the 3+1 decomposition of the spacetime~\citep[see, e.g.,][]{Shibata2015}. In the 3+1 form, the metric tensor $g_{\mu\nu}$ is decomposed as
\begin{align}
	ds^2=g_{\mu\nu}dx^\mu dx^\nu=-\alpha^2dt^2+\gamma_{ij}\left(dx^i+\beta^idt\right)\left(dx^j+\beta^jdt\right),
\end{align}
where $\mu$ and $\nu$ denote the spacetime indices, $i$ and $j$ denote the spatial indices, $\alpha$, $\beta^i$, and $\gamma_{ij}$ denote the lapse, shift, and spatial metric, respectively. We treat the matter as a perfect fluid and the energy-momentum tensor is given by 
\begin{align}
	T_{\mu\nu}=\rho hu_\mu u_\mu+Pg_{\mu\nu},
\end{align}
where $\rho$, $h$, $u^\mu$, and $P$ denote the rest-mass density, specific enthalpy, four velocity, and pressure, respectively. The Euler equation, energy equation, and continuity equation are given, respectively, by
\begin{align}
	\gamma_{\nu i}\nabla_\mu T^{\mu\nu}&=-\rho {\dot \epsilon}_{\rm esc} u_i\label{eq:eom}\\
	n_\nu \nabla_\mu T^{\mu\nu}&=-\rho {\dot \epsilon}_{\rm esc} n_\nu u^\nu\label{eq:eoe}\\
	\nabla_\mu \left(\rho u^\mu\right)&=0,\label{eq:eoc}
\end{align}
with the covariant derivative, $\nabla_\mu$. Here, $n_\nu=-\alpha\nabla_\nu t$, $\gamma_{\mu\nu}=g_{\mu\nu}+n_\mu n_\nu$, and ${\dot \epsilon}_{\rm esc}$ is the specific radioactive heating rate deposited in the form of neutrinos which entirely escape from the system without being thermalized. Note that the contribution of other radioactive decay channels that instantaneously thermalize (beta, alpha, and spontaneous fission) is taken into account by the same treatment as described in Appendix A of~\cite{Kawaguchi:2022bub}.


In the non-rotating black-hole spacetime with the isotropic coordinates, where ${\dot \alpha}=0$, $\beta^{i}=0$, and $K_{ij}=0$, the basic equations are simplified as
\begin{align}
	\partial_t \left(r^2{\rm sin}\theta {\tilde \rho}_*\right)&+\partial_r\left(r^2 {\rm sin}\theta {\tilde \rho}_* v^{(r)}\right)\nonumber\\
    &+\partial_\theta\left(r {\rm sin}\theta {\tilde \rho}_* v^{(\theta)}\right)\nonumber\\
    &+\partial_\phi\left(r {\tilde \rho}_* v^{(\phi)}\right)=0,
\end{align}
\begin{align}
	\partial_t \left(r^2{\rm sin}\theta {\tilde S}_{(r)}\right)&+\partial_r\left[r^2 {\rm sin}\theta\left( {\tilde S}_{(r)} v^{(r)}+P\alpha\sqrt{{\tilde \gamma}}\right)\right]\nonumber\\
    &+\partial_\theta\left(r {\rm sin}\theta {\tilde S}_{(r)} v^{(\theta)}\right)\nonumber\\
    &+\partial_\phi\left(r {\tilde S}_{(r)} v^{(\phi)}\right)\nonumber\\
	&=r^2{\rm sin}\theta\left[-{\tilde S}_0\partial_r \alpha-\frac{1}{2}\alpha\sqrt{{\tilde \gamma}}{\tilde S}_{(i)(j)}\partial_r {\tilde \gamma}^{(i)(j)}\right.\nonumber\\
	&+\frac{2}{r}\alpha\sqrt{{\tilde \gamma}} P +\frac{1}{r}\left({\tilde S}_{(\theta)}v^{(\theta)}+{\tilde S}_{(\phi)}v^{(\phi)}\right)\nonumber\\
    &\left.-\frac{\alpha}{hw} {\tilde S}_{(r)}{\dot \epsilon_{\rm esc}} \right],
\end{align}

\begin{align}
	\partial_t \left(r^3{\rm sin}\theta {\tilde S}_{(\theta)}\right)&+\partial_r\left(r^3 {\rm sin}\theta {\tilde S}_{(\theta)} v^{(r)}\right)\nonumber\\
    &+\partial_\theta \left[r^2 {\rm sin}\theta\left( {\tilde S}_{(\theta)} v^{(\theta)}+P\alpha\sqrt{{\tilde \gamma}}\right)\right]\nonumber\\
    &+\partial_\phi\left(r^2{\tilde S}_{(\theta)} v^{(\phi)}\right)\nonumber\\
	&=r^3 {\rm sin}\theta\left[\frac{1}{r}\alpha\sqrt{{\tilde \gamma}} P {\rm cot}\theta+\frac{1}{r}{\tilde S}_{(\phi)}v^{(\phi)}{\rm cot}\theta\right.\nonumber\\
    &\left.-\frac{\alpha}{hw} {\tilde S}_{(\theta)}{\dot \epsilon_{\rm esc}} \right],
\end{align}
\begin{align}
	\partial_t \left(r^3{\rm sin}^2\theta {\tilde S}_{(\phi)}\right)&+\partial_r\left(r^3 {\rm sin}^2\theta {\tilde S}_{(\phi)} v^{(r)}\right)\nonumber\\
    &+\partial_\theta\left(r^2 {\rm sin}^2\theta {\tilde S}_{(\phi)} v^{(\theta)}\right)\nonumber\\
    &+\partial_\phi\left[r^2 {\rm sin}\theta\left( {\tilde S}_{(\phi)} v^{(\phi)}+P\alpha\sqrt{{\tilde \gamma}}\right)\right]\nonumber\\
    &=-r^3{\rm sin}^2\theta \frac{\alpha}{hw} {\tilde S}_{(\phi)}{\dot \epsilon_{\rm esc}},
\end{align}

\begin{align}
	\partial_t\left( \alpha r^2{\rm sin}\theta {\tilde S}_0\right)&+\partial_r\left[\alpha r^2{\rm sin}\theta\left({\tilde S}_0v^{(r)}+P\sqrt{{\tilde \gamma}}v^{(r)}\right)\right]\nonumber\\
    &+\partial_\theta\left[\alpha r{\rm sin}\theta\left({\tilde S}_0v^{(\theta)}+P\sqrt{{\tilde \gamma}}v^{(\theta)}\right)\right]\nonumber\\
    &+\partial_\phi\left[\alpha r\left({\tilde S}_0v^{(\phi)}+P\sqrt{{\tilde \gamma}}v^{(\phi)}\right)\right]\nonumber\\
    &=-\alpha^2 r^2{\rm sin}\theta {\tilde \rho}_* {\dot \epsilon_{\rm esc}}.
\end{align}

Here, $K_{ij}$ denotes the extrinsic curvature, and the other variables which newly appear in the above equations are defined as follows:
\begin{align}
\sqrt{\gamma}&={\rm det}\left(\gamma_{ij}\right),\nonumber\\
\rho_*&=\rho w\sqrt{\gamma},\nonumber\\
w&=\alpha u^t,\nonumber\\
S_i&=\rho_* {\hat u}_i=\rho_* h u_i,\nonumber\\
S_0&=\rho_* {\hat e}=\rho_*\left(hw-\frac{P}{\rho w}\right),\nonumber\\
S_{ij}&=\rho hu_iu_j+P\gamma_{ij},\nonumber\\
v^i&=\frac{u^i}{u^t},\nonumber\\
\Lambda_{(r)}^r=1,~\Lambda_{(\theta)}^\theta=\frac{1}{r},&~\Lambda_{(\phi)}^\phi=\frac{1}{r{\rm sin}\theta},~\Lambda_{(i)}^j=0\,(i\ne j),\nonumber\\
{\tilde \gamma}_{(i)(j)}&=\Lambda_{(i)}^k\Lambda_{(j)}^l \gamma_{kl}=\psi^4 \delta_{(i)(j)},\nonumber\\
\sqrt{{\tilde \gamma}}&=\frac{1}{r^2{\rm sin}\theta}\sqrt{\gamma}=\psi^6,\nonumber\\
{\tilde \rho}_*&=\frac{1}{r^2{\rm sin}\theta} \rho_*,\nonumber\\
v^{(r)}=v^r,~v^{(\theta)}&=r v^\theta,~v^{(\phi)}=r {\rm sin}\theta\, v^\phi,\nonumber\\
{\tilde S}_{(r)}=\frac{1}{r^2{\rm sin}\theta} S_r,~{\tilde S}_{(\theta)}&=\frac{1}{r^3{\rm sin}\theta} S_\theta,~{\tilde S}_{(\phi)}=\frac{1}{r^3{\rm sin}^2\theta} S_\phi,\nonumber\\
{\tilde S}_{(i)(j)}&=\Lambda_{(i)}^k\Lambda_{(j)}^l S_{kl}.
\end{align}
$\psi$ denotes the conformal factor. The indices without the parenthesis denote the tensor components with respect to the coordinate basis, and the variables of indices with the parenthesis correspond to the orthonormal basis components in the spherical coordinates of the flat spacetime.

In this work, we numerically solve the set of these equations by employing a Kurganov-Tadmor scheme~\citep{2000JCoPh.160..241K} with a piecewise parabolic reconstruction for the quantities of cell interfaces and the minmod-like filter introduced in~\cite{2000JCoPh.160..241K} for the flux-limitter.

\section{Particle tracing}\label{app:pt}
To perform nucleosynthesis calculations, the Eulerian data obtained with a simulation have to be translated into Lagrangian evolution of the thermodynamical variables. For this purpose, the \textit{tracer particle method} is employed. In this appendix, we summarize our method of particle tracing and how we use the nucleosynthesis results for the HD simulation.

\subsection{Time evolution of particles}
Suppose we have a time series of three-velocity field $v^{i(n)}$, where $i(n)$ is the index of the spatial coordinates, at a time slice $t=t^{(n)}$. Here, the velocity field is defined at the spatial points $(x^1_j,x^2_k,x^3_l)$ discretely, where $j$, $k$, $l$ denote the grid points in the three-dimensional space. For a given spatial position of a particle $x^{i(n)}$ at the $n$th time slice, the coordinates of the particle in the $(n+1)$th time slice is solved with the so-called semi-implicit trapezoidal method \citep{Fujibayashi:2020qda,2015ApJ...810..109N}. In this method, we solve
\begin{align}
\frac{x^{i(n+1)}-x^{i(n)}}{t^{(n+1)}-t^{(n)}} = \frac{1}{2}\big(\dot{x}^{i(n)} + \dot{x}^{i(n+1)}\big)
\end{align}
for $x^{i(n+1)}$. Here, $\dot{x}^{i(n)}$ is the velocity of the particle at $n$th time slice defined with the tri-linear interpolation from the velocity field to the particle position. For the back-tracing of the ejecta particles, the time step $\Delta t = t^{(n+1)}-t^{(n)}$ is limited by the frequency of the outputs. Numerical accuracy of the particle tracing may be diagnosed with a timescale defined as
\begin{align}
\Delta t' = \min\bigg(\frac{x^1_{j+1}-x^1_{j}}{\dot{x}^{1(n)}}, \frac{x^2_{k+1}-x^2_{k}}{\dot{x}^{2(n)}}, \frac{x^3_{l+1}-x^3_{l}}{\dot{x}^{3(n)}}\bigg),
\end{align}
where the spatial coordinates of the particle are in the intervals $x^1_{j+1}$--$x^1_{j}$, $x^2_{k+1}$--$x^2_{k}$, and $x^3_{l+1}$--$x^3_{l}$ at $t=t^{(n)}$. $\Delta t'<\Delta t$ indicates that the particle position changes significantly within the time step $\Delta t$. In such a case, we perform several sub-steps between the $(n+1)$th and $n$th time slices. The velocity fields at time $t$ between $t^{(n+1)}$ and $t^{(n)}$ are linearly interpolated from the two time slices.

\subsection{Initial position of the particle}
The particles are distributed on a sphere with the radius $r_\mathrm{ext}=3\times10^8$\,cm. At a given time, the ejecta criterion is checked at $r=r_\mathrm{ext}$ with a set of the polar and azimuthal angles $(\theta_i, \phi_i)$, where $i$ is the index of the direction on which particles are placed. If the ejecta criterion is satisfied at a direction, a particle is placed at the angle. The particles are then traced backward in time.

The polar angles in the set $\{(\theta_i, \phi_i)\}$ are on a uniform grid in 0--$\pi/2$ with a spacing $\Delta \theta$. The azimuthal angles in the set are also on a uniform grid in 0--$2\pi$ but with a $\theta$-dependent spacing $\Delta \phi$, which decreases with the polar angle. The function of $\Delta \phi$ is determined to distribute the particles in an approximately uniform manner in the sphere.

The particles are placed repeatedly with a time interval $\Delta t_\textrm{p-set}$, which is determined by the average velocity of the ejecta at $r=r_\mathrm{ext}$ as
\begin{align}
\Delta t_\textrm{p-set} = \frac{r_\mathrm{ext}\Delta \theta}{\langle v^r\rangle},
\end{align}
where $\Delta \theta$ is the polar angle spacing of the particle-set locations and $\langle v^r\rangle$ is the average radial velocity of the particles located at given time. In this way, the particles are located uniformly in space, i.e., in similar radial and lateral distances.

The mass of each particle is assigned as
\begin{align}
m = {r_\mathrm{ext}}^2 \Delta\Omega \rho u^r \sqrt{-g}\Delta t_\textrm{p-set}.
\end{align}
This definition of the particle mass is consistent with the conserved mass flux at $r=r_\mathrm{ext}$. Therefore, the total mass of the particles converges to the ejecta mass of the NR simulation for increasing the number of the particles set on the sphere (i.e., $\Delta \theta \rightarrow0$ and thus $\Delta t_\textrm{p-set}\rightarrow0$).

\subsection{Nucleosynthesis}

In total, about 9600 tracer particles are generated with this method. Along the thermodynamical histories of the tracer particles, nucleosynthesis calculations are performed in the same manner as in \citet{Wanajo:2022jgw}. The initial value of $Y_e$ for the nucleosynthesis calculation is taken from the final value of each tracer particle, which is that shown in Figs.~\ref{fig:xy_prof}, \ref{fig:ye_prof}, and \ref{fig:noheat_prof}. We define the particles with $Y_e\leq0.08$ as those of dynamical ejecta. They are found to have very similar nucleosynthesis results with only a small difference in actinide abundances. Therefore, for the particles of the dynamical ejecta, we assign the same elemental abundance and radioactive heating rate, which are the mass-weighted average of those in \citet{Wanajo:2022jgw}. On the other hand, for the other  component ($Y_e>0.08$), we assign each particle with the abundance and heating rate obtained by the nucleosynthesis calculation along its thermodynamical history.

\subsection{Mapping particle data for HD simulation}
To take the radioactive heating into account in the HD simulation, and to obtain the final spatial distribution of elements, the nucleosynthesis results along with the tracer particles are used. For this purpose, the time and angular positions at which fluid elements are injected through $r=r_\mathrm{ext}$ are traced by advecting the three passive scalar variables. The injection time and angles are then converted to the tracer particles which reach the extraction radius at the similar times and angles.

In practice, we first construct a table of heating rates as functions of the time ($t$), and the injection time ($t_\mathrm{inj}$) and angles ($\theta_\mathrm{inj}$, $\phi_\mathrm{inj}$). 
The tracer particles are placed sparsely in terms of time and angles. Therefore, the properties of the tracer particles (abundance and radioactive heating rates) have to somehow be mapped onto the injection time and angle space.


For a given injection time and angles ($t_\mathrm{inj}$, $\theta_\mathrm{inj}$, $\phi_\mathrm{inj}$), we select a few tracer particles spatially closest to the (Cartesian) point $\vec{x}_\mathrm{inj} = (r_\mathrm{ext} \sin\theta_\mathrm{inj}\cos\phi_\mathrm{inj}, r_\mathrm{ext} \sin\theta_\mathrm{inj}\sin\phi_\mathrm{inj}, r_\mathrm{ext} \cos\theta_\mathrm{inj})$ at the time $t_\mathrm{inj}$. The location of a tracer particle at the time $t_\mathrm{inj}$, $\vec{x}_i(t_\mathrm{inj})$, is estimated as
\begin{align}
x_i(t_\mathrm{inj}) &= r_i(t_\mathrm{inj}) \sin\theta_i \cos\phi_i,\\
y_i(t_\mathrm{inj}) &= r_i(t_\mathrm{inj}) \sin\theta_i \sin\phi_i,\\
z_i(t_\mathrm{inj}) &= r_i(t_\mathrm{inj}) \cos\theta_i,
\end{align}
where $\theta_i$ and $\phi_i$ are the angles at which the particle is located on $r=r_\mathrm{ext}$, and $r_i(t_\mathrm{inj})$ is the radius of the $i$th particle estimated at $t=t_\mathrm{inj}$. The radius is estimated as
\begin{align}
r_i(t_\mathrm{inj}) = r_\mathrm{ext} + v^r_i (t_\mathrm{inj}-t_i),
\end{align}
where $v^r_i$ is the radial three velocity of the particle when it is located on $r=r_\mathrm{ext}$ at the time $t_i$. In short, we assume that tracer particles move only radially with the same velocity as what they have at $r=r_\mathrm{ext}$. The radioactive heating rate of the fluid elements injected at ($t_\mathrm{inj}$, $\theta_\mathrm{inj}$, $\phi_\mathrm{inj}$) is then estimated as
\begin{align}
\dot{q}(t) = \sum_{i\in \mathcal{P}} w_i \dot{q}_i(t),
\end{align}
where $\mathcal{P}$ is the set of the selected tracer particles, and $w_i$ is the weight that satisfies $\sum_i w_i = 1$. We define the weight as $w_i \propto 1/|\vec{x}_\mathrm{inj}-\vec{x}_i(t_\mathrm{inj})|$, i.e., the weight is inversely proportional to the distance between the injection point and the particle at the same time $t_\mathrm{inj}$.

\label{lastpage}
\end{document}